\documentclass[a4paper,11pt]{article}
\pdfoutput=1
\usepackage{jheppub}

\usepackage{graphicx}
\usepackage[figuresright]{rotating}
\usepackage{bm,amsmath,amssymb}
\usepackage[mathscr]{eucal}
\usepackage{multirow}
\usepackage{xcolor}
\definecolor{lcolor}{rgb}{0.,0.0,0.}
\definecolor{citcolor}{rgb}{0,0.,0.5}

\newcommand{\beq}{\begin{eqnarray}}
\newcommand{\eeq}{\end{eqnarray}}
\newcommand{\dis}{\displaystyle}
\def\dd{{\rm d}}

\newcommand{\bem}{\begin{multline}}
\newcommand{\eem}{\end{multline}}
\newcommand{\beg}{\begin{gather}}
\newcommand{\eeg}{\end{gather}}

\newcommand{\nn}{\nonumber\\}

\newcommand{\ben}{\begin{eqnarray*}}
\newcommand{\een}{\end{eqnarray*}}

\def\lnu{{\lambda_{K}}}
\def\lfr{{\lambda_{\rm fr}}}
\def\lk{{\lambda_{\kappa}}}
\def\lB{{\lambda_{B}}}

\newcommand{\secn}[1]{Section~1}
\newcommand{\appn}[1]{Appendix~1}

\long\def\comment#1{ }

\def\and{\quad\text{and}\quad}

\def\th{\theta}

\def\0{{\boldsymbol 0}}

\def\max{{\rm max}}

\newcommand{\erf}{\textrm{erf}}

\newcommand{\abar}{\bar{\alpha}}

\title{Dynamical grooming meets LHC data}
\author[a,b]{Paul Caucal, }
\emailAdd{pcaucal@bnl.gov}
\affiliation[a]{Institut de Physique Th\'{e}orique, Universit\'{e} Paris-Saclay, CNRS, CEA, F-91191, Gif-sur-Yvette, France}
\affiliation[b]{Physics Department, Brookhaven National Laboratory, Upton, NY 11973, USA}
\author[a]{Alba Soto-Ontoso, }
\emailAdd{alba.soto@ipht.fr}
\author[c,d]{Adam Takacs}
\emailAdd{adam.takacs@uib.no}
\affiliation[c]{Department of Physics and Technology, University of Bergen, Bergen 5020, Norway}
\affiliation[d]{Department of Astronomy and Theoretical Physics, Lund University, S-223 62 Lund, Sweden}

\abstract{In this work, we analyse the all-orders resummation structure of the momentum sharing fraction, $z_g$, opening angle, $\theta_g$, and relative transverse momentum, $k_{t,g}$, of the splitting tagged by the dynamical grooming procedure in hadronic collisions. We demonstrate that their resummation does non-exponentiate and it is free of clustering logarithms. Then, we analytically compute the probability distributions of ($z_g, \theta_g, k_{t,g}$) up to next-to-next-to-double logarithm accuracy (N$^2$DL) in the narrow jet limit, including a matching to leading order in $\alpha_s$. On the phenomenological side, we perform an analytic-to-parton level comparison with Pythia and Herwig. We find that differences between the analytic and the Monte-Carlo results are dominated by the infra-red regulator of the parton shower. Further, we present the first analytic comparison to preliminary ALICE data and highlight the role of non-perturbative corrections in such low-$p_t$ regime. Once the analytic result is corrected by a phenomenologically determined non-perturbative factor, we find very good agreement with the data.}

\begin{document}
\maketitle

%%%%%%%%%%%%%%%%%%
%%%%%%%%%%%%%%%%%%
\section{Introduction}
\label{sec:intro}
%%%%%%%%%%%%%%%%%%
%%%%%%%%%%%%%%%%%%
Jet physics aims at pinning down the microscopic properties of Quantum Chromodynamics (QCD~\cite{Sterman:1977wj}. In the context of heavy-ion physics, the modification of jets with respect to their vacuum counterparts is regarded as an experimental evidence for the formation of a dense, thermal medium, namely the Quark-Gluon plasma~\cite{Connors:2017ptx}. 

Nowadays, most of the efforts in the field from a theoretical point of view, both from an analytic perspective and with machine learning tools (see Ref.~\cite{Marzani:2019hun} for a review), are directed towards studying the space-time structure of a jet by characterising its radiation pattern through jet substructure observables, i.e.\ constructed from one (e.g.\ $z_g$~\cite{Larkoski:2014wba,Larkoski:2015lea}), or a few branchings (e.g.\ N-subjettiness~\cite{Thaler:2010tr} or the Lund jet plane~\cite{Dreyer:2018nbf,Lifson:2020gua}) at most. In this paper we focus on the former category where typically the one branching that defines the observable is selected in a region of phase space where perturbative QCD calculations are applicable, that is, far away from the soft and wide angle sector. This tagging task is handled by so-called `grooming methods' through which the hard and collinear core of the jet is isolated. The way this general goal is achieved differs from one groomer to the other, e.g. Modified Mass Drop Tagger(mMDT)~\cite{Butterworth:2008iy} or its extension Soft Drop (SD)~\cite{Larkoski:2014wba} selects the splitting whose momentum sharing fraction obeys $z\!>\!z_{\rm cut}\theta^\beta$, while trimming~\cite{Krohn:2009th} first reclusters a jet into subjets with a smaller radius $R_{\rm sub}$ and then keep only those subjets whose $p^{\rm subjet}_t\!>\!z_{\rm cut} p_t^{\rm jet}$. In the previous expressions ($z_{\rm cut}, \beta $ and $R_{\rm sub}$) are free parameters that need to be tuned with Monte-Carlo simulations to achieve an optimal performance~\cite{Aad:2020elc}. These methodological differences leave their imprint into the analytic behavior of the observables that they define~\cite{Dasgupta:2013ihk,Dasgupta:2013via}. For example, as we shall see in more detail in what follows, due to the presence of an explicit $z_{\rm cut}$ in the Soft Drop grooming condition this method is free of non-global logarithms in the resummation function. This fact has enable to push the accuracy of the calculation of Soft Drop groomed observables up to next-to-leading log accuracy in $p+p$~\cite{Marzani:2017mva,Marzani:2017kqd,Kang:2018jwa,Dasgupta:2012hg,Kang:2019prh,Baron:2020xoi,Anderle:2020mxj,Kang:2018vgn} and even next-to-next-to-leading log in $e^++e^-$~\cite{Kardos:2020gty,Kardos:2018kth,Frye:2016aiz}. It is then clear that the usefulness of a given grooming method should not be judged only on the basis of its resilience to non-perturbative physics, but also on its analytic structure from a pQCD point of view. This paper aims at deepening our analytic understanding of jet substructure observables as defined by a novel grooming technique that has been recently introduced and dubbed `dynamical grooming' (DyG)~\cite{Mehtar-Tani:2019rrk,Mehtar-Tani:2020oux,Soto-Ontoso:2020ola}.

The dynamical grooming method consists in identifying the `hardest' branching in a jet tree as a proxy for the physical jet scale. The hardness measure is given by
\beq
\label{eq:hardness}
\kappa^{(a)}=\frac{1}{p_{t,{\rm jet}}}z(1-z)p_t\theta^{a}
\eeq
where $a$ is a continuous free parameter that has to be larger than zero in order to guarantee collinear safety. For certain values of $a$ in Eq.~\eqref{eq:hardness}, the hardness measure translates into familiar kinematical quantities, e.g $\kappa^{(1)}\!=\!k_t$, where $k_t$ is the transverse momentum of the splitting, or $\kappa^{(2)}\!=\!m^2$, with $m$ being the branching mass. In addition, we define $\theta\!=\!\Delta R/R$ where $\Delta R$ is the angular separation between the sub-jets and $R$ corresponds to the cone size. The hardest splitting is obtained after re-clustering the jet sample with Cambridge/Aachen algorithm~\cite{Dokshitzer:1997in} and finding the node with the largest $\kappa$ in the clustering sequence. In fact, it can be proven through analytical arguments~\cite{Nason:2004rx} that it is sufficient to look for the hardest splitting along the primary Lund plane of the jet, i.e. following the branch with the larger transverse momentum at each de-clustering step.\footnote{We have numerically checked that our results are robust if we look for the hardest splitting in the whole tree and not only on the primary branch.}
First steps towards the calculation from first-principles in perturbative QCD of the probability distribution of the momentum sharing fraction, $z_g$, the mass and the relative transverse momentum, $k_{t,g}$, of the hardest splitting were presented in the original dynamical grooming paper in the resummation region, i.e.\ when $z_g(k_{t,g})\ll 1$~\cite{Mehtar-Tani:2019rrk}. Interestingly, it was found that similarly to the Soft Drop case, the $z_g$ distribution pertains to a special class of jet observables known as Sudakov safe~\cite{Larkoski:2013paa,Larkoski:2015lea}. Together with the modified leading-log calculation of DyG jet substructure observables, a Monte-Carlo study of the impact of non-perturbative physics was presented in Ref.~\cite{Mehtar-Tani:2019rrk}. An overall similar performance than Soft Drop was shown, but with a remarkable resilience to hadronization in some cases like the $z_g$ distribution as tagged by $a\!=\!1$. This novel idea has triggered the interest of the ALICE collaboration that has recently conducted some preliminary measurements on the $z_g,\theta_g$~\cite{Mulligan:2020cnp}, and $k_{t,g}$~\cite{Ehlers:2020piz} distributions at $\sqrt s\!=\!5.02$~TeV in the jet transverse momentum bin of $60\!<\!p^{ch}_t\!<\!80$ GeV. As we will see, the low $p_t$ reach of the ALICE detector challenges the analytic description of such data set given that non-perturbative effects are sizeable. In addition, first steps towards the experimental use of DyG in heavy-ion collisions were reported in Ref.~\cite{Mulligan:2020tim}.

From an analytic point of view the purposes of this paper are multifold: (i) understand the resummation structure of dynamical grooming observables and propose a definition for their logarithmic accuracy which circumvents their non-exponentiating nature (double log, next-to-double log, etc) (ii) advance the resummation of $z_g$ and $k_{t,g}$ from modified-leading logarithm~\cite{Mehtar-Tani:2019rrk} to next-to-next-to double logarithmic accuracy,\footnote{For a precise definition of N$^p$DL accuracy, see Eq.~\eqref{eq:Logarithmic_expansion} and the discussion below.} as well as presenting for the first time the resummation of $\theta_g$, (iii) highlight the absence of clustering logarithms in dynamically groomed observables, (iv) perform a fixed-order matching for all three dynamically groomed observables. This last point is not trivial for the pair of Sudakov safe observables and we propose a novel method to match the resummed and fixed-order distributions.  All these ingredients are contained in Section~\ref{sec:vacuum}. After a few sanity checks on the analytic side, in Sec.~\ref{sec:analytic} we compare our results to Monte-Carlo simulations at parton level in a high-$p_t$ setup, where non-perturbative effects are mild. Next, in Sec.~\ref{sec:data-to-theory}, we present the first comparison between an analytic calculation and the preliminary ALICE data for ($z_g,\theta_g$ and $k_{t,g}$). In addition, Monte-Carlo studies with different general purpose event generators are performed showing the impact of different details in the definition of the dynamical grooming method in Appendices~\ref{sec:appendix-b},~\ref{sec:appendix-c}. The discriminating power of these type of jet substructure observables with respect to different hadronization models and parton showers are shown in App.~\ref{sec:appendix-e}.

%%%%%%%%%%%%%%%%%%%%%%%%%%%%%%%%%%%%
%%%%%%%%%%%%%%%%%%%%%%%%%%%%%%%%%%%%
\section{Theoretical analysis of dynamically groomed observables}
\label{sec:vacuum}
%%%%%%%%%%%%%%%%%%%%%%%%%%%%%%%%%%%%
%%%%%%%%%%%%%%%%%%%%%%%%%%%%%%%%%%%%

In this section, we present the all-order perturbative calculation of dynamically groomed observables in the $\kappa^{(a)}\!\ll\!1$ region\footnote{At low enough values of $\kappa$ the calculation is dominated by non-perturbative effects. Therefore, strictly speaking our resummation is valid when $\kappa_{{\rm NP}}\!\ll\!\kappa\!\ll\!1$.} and their matching to fixed order results applicable when $\kappa^{(a)}\!\sim\!1$. In the soft limit, $z\!\ll \!1$ and the $(1\!-\!z)$ factor can be removed from Eq.~\eqref{eq:hardness}. Furthermore, as the hardest splitting takes place along the primary branch, we neglect momentum degradation such that $p_{t}\!=\!p_{t,{\rm jet}}$. Therefore, instead of Eq.~\eqref{eq:hardness}, the definition $\kappa^{(a)}\!=\! z\theta^a$ is adopted throughout this section, and we sometimes omit the $a$ superscript to lighter the notation.\footnote{One can rigorously prove that $1\!-\!z$ corrections are beyond our targeted accuracy (see Appendix~\ref{sec:appendix-b}).}

%%%%%%%%%%%%%%%%%%%%%%%%%%%%%%%%%%%%
\subsection{Double-logarithmic estimation and basic properties}
\label{sec:dla}
%%%%%%%%%%%%%%%%%%%%%%%%%%%%%%%%%%%%
We shortly revisit the baseline calculation performed in Ref.~\cite{Mehtar-Tani:2019rrk}. In the $\kappa\!\ll\!1$ limit, the two-dimensional probability distribution of a splitting, with kinematic variables ($z,\theta$), to be hardest in the clustering sequence is given by
\beq
\label{eq:prob-dist}
\dis\frac{\dd^2 \mathcal P_i(z,\theta|a)}{\dd \theta\dd z}=\widetilde{P}_i(z,\theta)\Delta_i(\kappa|a)\,,
\eeq
where $i$ indicates the flavor of the jet initiating parton. The two ingredients entering the right-hand side of the previous equation are actually connected through
\begin{align}
\label{eq:sudakov_general}
\ln \Delta_i(\kappa|a) &= -\displaystyle\int_0^{1} \dd z' \displaystyle\int_0^1 \dd\theta' \widetilde{P}_i(z',\theta') \Theta\left(z'\theta'^a-\kappa^{(a)}\right)\,.
\end{align}
In physical terms, the branching kernel, $\widetilde{P}(z,\theta)$, represents the probability of a splitting with ($z,\theta$) to occur, while $\Delta(\kappa|a)$ is the so-called Sudakov form factor and vetoes all harder emissions, i.e.\ those with $\kappa'\!>\!\kappa$. From Eq.~\eqref{eq:sudakov_general}, it is easy to see that $a\!>\!0$ is required to regulate the collinear singularity. The normalised probability distribution to measure an observable $\kappa^{(b,c)}\!=\!z^b\theta^c$ on the $\kappa^{(a)}$ tagged splitting is given by
\begin{align}
\label{eq:hardness-general}
\left.\frac{1}{\sigma} \frac{\dd\sigma}{\dd \kappa^{(b,c)}} \right|_a &= \int_0^{1} \dd\theta \int_0^1 \dd z \,  \mathcal P_i(z,\theta|a)\delta\big(z^b\theta^c - \kappa^{(b,c)}\big)\,,
\end{align}
where a sum over flavors including the proper quark/gluon fraction is implicit. The observables that we focus on are obtained from Eq.~\eqref{eq:hardness-general} by setting: $(b\!=\!1, c\!=\!0)$ for $z_g$, $(b\!=\!0, c\!=\!1)$ for $\theta_g$, and $(b\!=\!1, c\!=\!1)$ for $k_{t,g}$.

We start by considering branchings in the soft-collinear limit ($z\!\ll\!1$ and $\theta\!\ll\!1$) that generate terms with powers of $\alpha_s\ln^2(\kappa^{(b,c)})$ in Eq.~\eqref{eq:prob-dist}. That is, we achieve double logarithmic accuracy (DLA) in the language of logarithmic resummation, as we will see below. The soft-collinear limit of the branching kernel reads
\beq
\widetilde{P}_i(z,\theta)=\dis\frac{\alpha_s}{\theta\pi}P_i(z)\,,
\eeq
where $P_i$ is the leading-order Altarelli-Parisi splitting function that, in this approximation, is given by
\beq
P_{i}(z)= \displaystyle\frac{2C_i}{z}\,,
\eeq
with $C_i$ being the color factor of the jet initiator parton; $C_i\!=\!C_A$ for gluons, and $C_F$ for quarks. The running of the strong coupling is beyond DLA and therefore we fix to its value at the jet scale, i.e.\ $\alpha_s\!\equiv\!\alpha_s(p_{t,{\rm jet}}R)$. In this limit, the Sudakov reduces to
\begin{align}
\label{eq:sudakov_dla}
 \ln \Delta_i(\kappa|a)&= -\displaystyle\int_{\kappa}^{1} \dd z' \displaystyle\int_{(\kappa/z')^{1/a}}^1 \frac{\dd\theta'}{\theta'} \dis\frac{\alpha_s}{\pi} \displaystyle\frac{2C_i}{z} = -\displaystyle\frac{\bar\alpha}{a} \ln^2\kappa\,,
\end{align}
where $\bar{\alpha}\!=\!C_i\alpha_s/\pi$. By plugging Eq.~\eqref{eq:sudakov_dla} into Eq.~\eqref{eq:hardness-general}, we obtain the momentum sharing fraction of the tagged splitting
\begin{align}
\label{eq:zg_dla}
\dis\frac{1}{\sigma}\dis\frac{\dd\sigma}{\dd z_g}= \dis\frac{1}{z_g}\dis\sqrt{\frac{\bar\alpha\pi}{a}}\left[{\rm{erf}}\left(\dis\sqrt{\frac{\bar\alpha}{a}}\ln z_g\right)+1\right]\,,
\end{align}
its opening angle
\begin{align}
\label{eq:thetag_dla}
\dis\frac{1}{\sigma}\dis\frac{\dd\sigma}{\dd \theta_g} = \frac{1}{\theta_g}\dis\sqrt{\bar{\alpha}\pi a}\left[{\rm{erf}}\left(\sqrt{\bar{\alpha}a}\ln(\theta_g)\right)+1\right]\,,
\end{align}
and its relative transverse momentum
\begin{align}
\label{eq:ktg_dla}
\dis\frac{1}{\sigma}\dis\frac{\dd\sigma}{\dd k_{t,g}} = \frac{1}{k_{t,g}}\dis\frac{\sqrt{\bar{\alpha}\pi a}}{a-1}\left[{\rm{erf}}\left(\sqrt{\frac{\bar{\alpha}}{a}}\ln(k_{t,g})\right)-{\rm{erf}}\left(\sqrt{\bar{\alpha}a}\ln(k_{t,g})\right)\right]\,.
\end{align}

%%%%%%%%%%%%%%%%%%
\paragraph{Location of the peak.} 
%%%%%%%%%%%%%%%%%%
An important feature of Eqs.~\eqref{eq:zg_dla}--\eqref{eq:ktg_dla} is the value at which they are cut off. Its location can be obtained by taking the derivative of e.g. Eq.~\eqref{eq:thetag_dla}\begin{align}
 \frac{\dd}{\dd \theta_g}\left(\frac{1}{\sigma}\dis\frac{\dd\sigma}{\dd \theta_g}\right)&= \sqrt{\bar{\alpha}a}\frac{1}{\theta_g^2}\left[-\sqrt{\pi}\left(1-\erf(\sqrt{x})\right)+\frac{2\sqrt{x}\exp(-x)}{\ln(1/\theta_g)}\right]\,,
\end{align}
where $x\equiv\abar a\ln^2(1/\theta_g)$. Then, the maximum value of the distribution, $\theta_{\rm max}$, satisfies the implicit equation
\begin{equation}
 \frac{2\sqrt{x}\exp(-x)}{\sqrt{\pi}(1-\erf(\sqrt{x}))}=\ln\left(\frac{1}{\th_{\rm max}}\right)\,.
\end{equation}
If $\theta_{\rm max}\ll1$, the left hand side can be approximated by its asymptotic behaviour ($x\rightarrow\infty$)
\begin{equation}\label{eq:x-exp}
 \frac{2\sqrt{x}\exp(-x)}{\sqrt{\pi}(1-\erf(\sqrt x))}\simeq 2x\,,
\end{equation}
such that 
\begin{equation}\label{eq:theta-cut}
\ln\left(\frac{1}{\th_{\rm max}}\right)=\frac{1}{2a\abar}+\mathcal{O}(1)\,.
\end{equation}
This equation indicates that the smaller the value of $a$, the deeper the tagged splitting is on the angular ordered shower, i.e.\ at smaller angles. In other words, at fixed $\theta_g$, larger values of $a$ lead to a bigger Sudakov suppression. Therefore, the distribution shifts to larger $\theta_g$ for larger $a$. Thus, Eq.~\eqref{eq:theta-cut} confirms and provides an analytic explanation for the result reported in Ref.~\cite{Mehtar-Tani:2019rrk} on the location of the tagged branching in the jet tree using Pythia~\cite{Sjostrand:2007gs} simulations. Notice that, in order to solve the implicit equation for the peak position, we have assumed that $\th_{\rm max}\ll1$. This approximation holds for not too large values of $a$. Otherwise, the smallness of $\abar$ can be compensated by $a$ in the product $a\abar$ appearing in Eq.~\eqref{eq:theta-cut} and $\th_{\rm max}\sim 1$. 

Following similar steps for the maximum of the momentum sharing fraction, we obtain\footnote{Notice that this equation can be also obtained by applying the $a\!\mapsto\!1/a$ transformation in Eq.~\eqref{eq:theta-cut}}
\begin{equation}\label{eq:z-cut}
\ln\left(\frac{1}{z_{\rm max}}\right)=\frac{a}{2\abar}+\mathcal{O}(1)\,.
\end{equation}
Again, the previous expression is valid as long as $a$ is not too small. 

Finally, in the case of $k_{t,g}$, we find that
\begin{equation}\label{eq:kt-cut}
\ln\left(\frac{1}{k_{t,\rm max}}\right)=\frac{a^{{\rm sgn}(a-1)}}{2\bar\alpha}+\mathcal{O}(1)\,.
\end{equation}
This analytic estimate confirms the ordering observed numerically in Figure 9 of Ref.~\cite{Mehtar-Tani:2019rrk}, i.e.\ the $a\!=\!0.1$ curve is peaked at a smaller $k_t$ than the $a\!=\!2$ case is, being $a\!=\!1$ the curve peaking at the largest value.

%%%%%%%%%%%%%%%%%%%%%%%%%%%%%%%%%%%%
\paragraph{Infra-red and collinear safety.}
%%%%%%%%%%%%%%%%%%
The first step towards boosting the accuracy of our calculation is to analyse the IRC (un)safety of the observables that we are dealing with. As we have already mentioned, and was shown in Ref.~\cite{Mehtar-Tani:2019rrk}, dynamically groomed observables are collinear unsafe for $a\!\leq\!0$. For $a\!>\!0$, while $k_{t,g}$ is a \textit{standard} IRC safe observable, both $z_g$ and $\theta_g$ are Sudakov safe only~\cite{Larkoski:2015lea} . This means that the all-order resummation encompassed in the Sudakov form factor regulates the singularities that appear at each order in $\alpha_s$ when $\theta_g\to0$ or $z_g\to0$. Notice that for $\theta_g$, this behavior represents a stark difference with respect to Soft Drop grooming, where this observable is, in fact, IRC safe~\cite{Larkoski:2014wba}. This can be understood as a result of the $z_{\rm cut}$ that appears in the Soft Drop condition and regulates the soft singularity. In turn, Dynamical Grooming does not introduce any sharp cut-off on the radiation phase-space and thus nothing forbids the hardest splitting to be in the soft ($z\sim 0$) region. 

A well-known consequence of Sudakov safety~\cite{Larkoski:2013paa,Larkoski:2015lea} is that the $z_g$ and $\theta_g$-distributions have an ill-defined expansion in (integer) powers of $\alpha_s$. To illustrate this fact, we introduce the cumulative distribution, that defines the probability to measure an observable below a certain value $\nu$, i.e.
\begin{equation}
\Sigma(\nu) = \dis\int_0^\nu \dd \nu'\dis\frac{1}{\sigma}\dis\frac{\dd \sigma}{\dd \nu'}\,.
\end{equation}
The $k_{t,g}$ cumulative distribution at DLA reads
\begin{align}\label{eq:ktg-sigma}
\Sigma(k_{t,g})=\frac{1}{a-1}\Bigg[&a\exp\left(-\frac{\bar\alpha}{a}\ln^2(k_{t,g})\right)-\exp\left(-\bar\alpha a\ln^2(k_{t,g})\right)\nn
&+\sqrt{\pi a\bar\alpha}\ln(k_{t,g})\left[{\rm erf}\left(\sqrt{\frac{\bar\alpha}{a}}\ln k_{t,g}\right)-{\rm erf}\left(\sqrt{\bar\alpha a}\ln k_{t,g}\right)\right]\Bigg]\,,
\end{align}
and its expansion in $\alpha_s$ (or equivalently in $\bar\alpha$) 
\begin{equation}
	\Sigma(k_{t,g})=1-\bar\alpha\ln^2\left(\frac{1}{k_{t,g}}\right)+\frac{1+a+a^2}{6a}\bar\alpha^2\ln^4\left(\frac{1}{k_{t,g}}\right)+\mathcal O(\bar\alpha^3)\,.
\end{equation}
From the previous expression, it is clear that $\Sigma(k_{t,g})$ admits an analytic expansion in $\bar\alpha$, as it is expected for an IRC safe observable. 
 
In contrast, the $\theta_g$-cumulative distribution is
\begin{equation}\label{eq:Sigma-thetag}
 \Sigma(\th_g)=\exp\left(-\bar{\alpha} a \ln^2\left(\frac{1}{\th_g}\right)\right)-\sqrt{\bar{\alpha}\pi a}\ln\left(\frac{1}{\th_g}\right)\left[{\rm{erf}}\left(\dis-\sqrt{\bar\alpha a}\ln \left(\frac{1}{\th_g}\right)\right)+1\right]\,,
 \end{equation}
and its expansion in powers of $\bar{\alpha}$ is given by
\begin{equation}
\label{eq:Sigma-thetag-series}
 \Sigma(\th_g)=1-\sqrt{\bar{\alpha}\pi a}\ln\left(\frac{1}{\th_g}\right)+\bar{\alpha}a\ln^2\left(\frac{1}{\th_g}\right)+\mathcal{O}(\bar{\alpha}^2)\,.
\end{equation}
In this case, the resumming function is not analytic as the second term in Eq.~\eqref{eq:Sigma-thetag-series} is of order $\sqrt{\bar{\alpha}}$. The non-analyticity on the dynamically groomed $\theta_g$ is caused uniquely by the $\sqrt{\bar{\alpha}}$ term. That is, all other powers of $\bar\alpha$ appearing in Eq.~\eqref{eq:Sigma-thetag-series} are integer and the function
\begin{equation}\label{eq:Sigma-zg}
 \Sigma(\th_g)+\sqrt{\bar{\alpha}\pi a}\ln\left(\frac{1}{\th_g}\right)
\end{equation}
is, in fact, analytic. The same arguments apply to $z_g$ where again one can utilise the $a\mapsto 1/a$ transformation, to confirm that
\begin{equation}
\label{eq:Sigma-zg-series}
\Sigma(z_g)+\sqrt{\frac{\bar{\alpha}\pi}{a}}\ln\left(\frac{1}{z_g}\right)
\end{equation}
has an analytic dependence on $\bar\alpha$ at any perturbative order. Interestingly, this $\alpha_s$-expansion of $z_g$ is remarkably different from its Soft Drop counterpart when $\beta\!>\!0$. For Soft Drop, the expansion is driven by $\alpha_s^{n/2}$ terms where the integer $n\!\geq\!1$~\cite{Larkoski:2015lea}. Whether this is a purely mathematical statement, or an explanation in physical terms exists, is beyond our degree of understanding and further work is required to clarify it.  

To sum up, in this section we have shown that the opening angle and momentum sharing fraction of the splitting tagged by dynamical grooming are unconventional observables from a pQCD point of view. The Sudakov safety of the $z_g$ and $\theta_g$ distributions leads to an ambiguous definition of the logarithmic accuracy in their resummation, as was noted in Ref.~\cite{Larkoski:2013paa}. Furthermore, the standard matching to fixed-order calculations is not trivial due to the non-analyticity of the resummed result. In this context, a careful definition of logarithmic accuracy in the resummation is required and will be provided next.

%%%%%%%%%%%%%%%%%%%%%%%%%%%%%%%%%%%%
\subsection{Revisiting the meaning of accuracy: from IRC to Sudakov safe observables}
\label{accuracy}
%%%%%%%%%%%%%%%%%%%%%%%%%%%%%%%%%%%%

We start by considering a general resummed formula for an IRC safe distribution obtained with dynamical grooming. Following our previous notation, we denote $\kappa^{(b,c)}$ the observable that we measure on the splitting whose hardness, $\kappa^{(a)}\!=\!z\theta^a$, is the largest in the shower. The cumulative distribution to measure $\kappa^{(b,c)}\!\ll\!1$ reads
\begin{equation}
\label{eq:master-form}
\Sigma(\kappa^{(b,c)})=\mathcal \dis\int_0^1\dd z \dis\int_0^1\dd \theta\,\widetilde{P}(z,\theta)\Delta(\kappa|a)\Theta(\kappa^{(b,c)}- z^b\theta^c)\,,
\end{equation}
where we have omitted the flavour index for simplicity. An important comment regarding the values of $(a,b,c)$ is in order. Only when $b\!\leq\!1$ and $c\!\leq\!a$, the hierarchy $\kappa^{(a)}\!\leq\!\kappa^{(b,c)}\!\leq\!1$ is satisfied and thus $\Delta(\kappa|a)$ can be accurately computed through resummation techniques. Any other combination of $(a,b,c)$ leads to a situation in which $\kappa^{(b,c)}\!\ll\!1$ does not necessarily imply $\kappa^a\! \ll \!1$ such that fixed order contributions to $\Delta(\kappa|a)$ become relevant. 

Having these constraints in mind, we derive the Sudakov safe distributions of $z_g\!\equiv\!\kappa^{(1,0)}$ and $\theta_g\!\equiv\!\kappa^{(0,1)}$ as two limits of Eq.~\eqref{eq:master-form}, i.e.
\begin{equation}
\label{eq:resum-zg}
\Sigma(z_g)=\lim_{c\to0}\Sigma(\kappa^{(1,c)})\,,
\end{equation}
and
\begin{equation}
\label{eq:resum-thetag}
\Sigma(\theta_g)=\lim_{b\to0}\Sigma(\kappa^{(b,1)})\,.
\end{equation}
The key point is that we define the accuracy of $z_g$ and $\theta_g$ through the accuracy of the IRC safe distribution $\Sigma(\kappa^{(b,c)})$. For instance, we shall state that $\Sigma(z_g)$ is known at DLA, if $\Sigma(\kappa^{(b,c)})$ is known at the same degree of accuracy for all $c\!>\!0$, or at least in the neighbourhood of $c\!=\!0$. Our prescription to define the accuracy of Sudakov safe observables follows the spirit of Ref.~\cite{Larkoski:2013paa}. However, instead of defining the accuracy of the Sudakov-safe observable by marginalization of an IRC safe double differential distribution, we exploit the IRC safety of the $\kappa^{(b,c)}$ observable itself.  It's important to realise that the perturbative expansion of the Sudakov safe observables is only defined after taking first the appropriate limit on $\Sigma(\kappa^{(b,c)})$ as given by Eqs.~\eqref{eq:resum-zg},\eqref{eq:resum-thetag}. If these steps are taken in reverse order, i.e.\ expanding $\Sigma(\kappa^{(b,c)})$ in powers of $\alpha_s$ first and subsequently taking the limit of $b(c)\to 0$, one can show that the correct $\alpha_s$-expansion, given by Eqs.~\eqref{eq:Sigma-thetag-series}--\eqref{eq:Sigma-zg-series} at DLA, is not recovered. In short, these two operations do not commute.

The perturbative expansion of $\Sigma(\kappa^{(b,c)})$ can be written as
\begin{equation}\label{eq:Logarithmic_expansion}
 \Sigma(\kappa^{(b,c)})= \sum_{n=0}^\infty \alpha_s^n\sum_{m=0}^{2n}c_{nm}\,\ln^m(\kappa^{(b,c)})\,,
\end{equation}
where the $c_{nm}$ coefficients have to be determined. Then, we adopt the following convention~\cite{Dasgupta:2018nvj,Hamilton:2020rcu}: the logarithmic accuracy of $\Sigma(\kappa^{(b,c)})$ is said to be N$^p$DL if the $c_{nm}$ coefficients are known for all $n$ and $2n-p\leq m\leq 2n$. Notice that in many other jet substructure calculations it is customary to define the logarithmic accuracy at the level of $\ln\Sigma$ instead of on the cumulative distribution itself. The reason why we use $\Sigma(\kappa^{(b,c)})$ is because, in general, due to the marginalisation procedure stated in Eq.~\eqref{eq:master-form} the resummation of DyG observables does not exponentiate~\cite{Catani:1992ua,Banfi:2004yd} as it clear from Eq.~\eqref{eq:ktg-sigma}. This no exponentiation property is part of other jet substructure observables such as subjet multiplicities. Yet, there is a specific case for which it does: when $b\!=\!1$ and $c\!=\!a$. That is, when the kinematic variable used for tagging coincides with the measured observable. For instance, select the splitting with the largest $k_t$ in the shower, and compute its $k_t$-distribution. In this case, the cumulative distribution is simply the Sudakov form factor, i.e.
\begin{equation}
 \Sigma(\kappa^{(1,a)})=\Delta(\kappa|a)\,
\end{equation}
that is equivalent to the plain distribution. 

A natural question at this point is how does one relate the $c_{nm}$ coefficients with the accuracy of $\widetilde{P}(z,\theta)$ and $\Delta(\kappa|a)$. In other words, which are the relevant terms that one needs to include in the branching kernel and in the Sudakov form factor in order to reach a given accuracy? To answer this question we rely on the exponentiate property of $\Delta(\kappa|a)$, to write its logarithmic structure in the traditional form\footnote{Notice that, in contrast to some cases in the literature, the $g$-functions contain both collinear and soft, non-global terms, i.e. we do not write a separate $\mathcal S$ factor as in~\cite{Banfi:2004yd}.}~\cite{Banfi:2004yd}
\begin{equation}
\label{sudakov-log}
\Delta(\kappa|a)=\left(1+\displaystyle\sum_{n\geq1}\alpha_s^nC_n\right)e^{\ln(\kappa)g_1(x)+g_2(x)+\alpha_sg_3(x)+\mathcal{O}(\alpha_s^{n+2}\ln^n \kappa)}\,,
\end{equation}
with $C_n$ being constant coefficients, $g_i$ analytic functions
\begin{equation}
g_i(x)=\dis\sum_{i=1}^\infty g_{ij}x^j\,,
\end{equation}
and $x\equiv \alpha_s\ln \kappa$. In the N$^p$LL type of counting, the resuming function $g_1$ would be referred as LL, $g_2$ as NLL and so on. Our targeted accuracy is N$^2$DL in the rest of the paper, with the possibility of keeping sub-leading terms. After expanding Eq.~\eqref{sudakov-log} in powers of $\alpha_s$ we realize that one has to account for the following $g_{nm}$ coefficients at the corresponding level of accuracy
\begin{align}
& \text{DL} (p=0): g_{11}\,,\\
& \text{NDL} (p=1): g_{11}, g_{12}, g_{21}\,,\\ 
& \text{N$^2$DL} (p=2): g_{11}, g_{12}, g_{13}, g_{21}, g_{22}, C_1 \,.
\end{align}

The $g_{11}$ was already computed in Sec.~\ref{sec:dla} where we accounted for soft and collinear emissions only
\begin{equation}
g_{11} = -\dis\frac{C_i}{a\pi}\,.
\end{equation}
The other coefficients and their physical interpretation are provided in the following section up to N$^2$DL. Given that the constant $C_1$ term is related to the interplay between the resummation and fixed-order calculations, we postpone its discussion to Sec.~\ref{sub:matching-ktg} and neglect it in the resummation-related part. 

Turning to the terms that are needed in $\widetilde{P}(z,\theta)$, we start by working out the plain case ($b\!=\!1$ and $c\!=\!a$). The exponentiation property of the resummation, in this particular case, leads to a one-to-one mapping between the terms in the Sudakov and in the branching kernel. More concretely, following Eq.~\eqref{eq:master-form} one gets
\begin{equation}
-\dis\int_0^1 \dd z \dis\int_0^1 \dd \theta \widetilde{P}(z,\theta)\Theta(z\theta^a-\kappa) = \ln (\kappa)g_1(\alpha_s\ln\kappa)+g_2(\alpha_s\ln\kappa)+\cdots\,,
\end{equation}
that reduces in the N$^2$DL case to
\begin{equation}
\label{pbranch-n2dl}
-\dis\int_0^1 \dd z \dis\int_0^1\dd \theta \widetilde{P}(z,\theta)\Theta(z\theta^a-\kappa) = \ln(\kappa)(g_{11}x+g_{12}x^2+g_{13}x^3) +g_{21}x+g_{22}x^2\,,
\end{equation}
where, again, $x\!\equiv\!\alpha_s\ln \kappa$. The previous equation, derived exploiting the exponentiation property of the plain case, is sufficient to reach N$^2$DL for all values of ($b,c$). Using Eq.~\eqref{pbranch-n2dl} with any $b$ and $c$ may produce power suppressed or sub-leading ($p\!\ge\!3$) logarithmic corrections in front of $\alpha_s^n\ln^{2n-2}(\kappa)$ terms, that are nevertheless negligible in the resummation region. The physical insight behind the 'universality' of Eq.~\eqref{pbranch-n2dl} relates to the fact that $\widetilde{P}(z,\theta)$ is just a probability to have a splitting with a given $z$ and $\theta$. Thus, the branching kernel should be a priori independent of both $a$ and the observable we measure on this branching.

%%%%%%%%%%%%%%%%%%%%%%%%%%%%
\subsection{$k_{t,g}$ at LO+N$^2$DL accuracy}
\label{sec:n2dl}
%%%%%%%%%%%%%%%%%%%%%%%%%%%%
After this rather formal discussion, we would like to shed light on our statements through an explicit calculation. Namely, we compute the IRC safe $k_{t,g}$ distribution in the small jet radius limit at N$^2$DL accuracy on the resummation side and include its matching to a fixed-order calculation at leading order, thus achieving a solid analytic description for all values of $k_{t,g}$.  
\subsubsection{Resummation}
From the general formula given by Eq.~\eqref{eq:master-form} it is straightforward to calculate the cumulative $k_{t,g}$ distribution by setting $b\!=\!c\!=\!1$. It reads,
\begin{equation}\label{eq:master-form-ktg}
\Sigma(k_{t,g})=\mathcal \dis\int_0^1\dd z \dis\int_0^1 \dd\theta \widetilde{P}(z,\theta)\Delta(\kappa|a)\Theta(k_{t,g}-z\theta)
\end{equation}
such that the differential cross section is
\begin{equation}
\displaystyle\frac{1}{\sigma_0}\dis\frac{\dd\sigma}{\dd k_{t,g}}=\dis\frac{\dd \Sigma(k_{t,g})}{\dd k_{t,g}}
\end{equation}
where $\sigma_0$ represents the Born level total cross-section. In what follows, we calculate the necessary $g_{nm}$ coefficients that enter in the Sudakov form factor and the branching kernel, see Eqs.~\eqref{sudakov-log},~\eqref{pbranch-n2dl}, and organise them according to the underlying physical effect.  

\paragraph{Hard-collinear emissions.} Due to its simplicity, the first term that we add to our calculation is the one arising from including hard-collinear corrections ($z\!\sim\!1$, $\theta\!\ll\!1$) in the splitting function. This amounts to take into account the finite part of the splitting functions as follows:
\begin{equation}\label{Phardcoll}
P^{({\rm h-c})}_i(z)=\frac{2C_i}{z}\Theta\left(e^{-B_i}-z\right)\,,
\end{equation}
where $B_q\!=\!2/C_F$, $B_g\!=\!11/12-n_fT_r/(3C_A)$, $T_r\!=\!1/2$, and we fix the number of flavors to $n_f\!=\!5$. The analytic integration of the new finite piece that appears both in the Sudakov and the branching kernel is useful to illustrate the point about sub-leading terms that appear naturally in the calculation. In fact, 
\begin{equation}
 - \dis\int_0^1 \dd z\dis\int_0^1\frac{\dd \theta}{\theta}\frac{\alpha_s}{\pi}\dis P^{({\rm h-c})}_i(z)\Theta(z\theta^a-\kappa) = -\frac{\alpha_sC_i}{\pi a}\left(B_i+\ln(\kappa)\right)^2\,.
 \end{equation}
From the previous equation one can easily read off the $g_{21}$ coefficient
\begin{equation}
g_{21}=-\dis\frac{2C_iB_i}{a\pi}\,,
\end{equation}
while the term proportional to $B_i$ and no $\ln(\kappa)$ dependence is sub-leading, although might be large when $a\!\ll\!1$. Strictly speaking, this latter contribution is not required to reach N$^2$DL in our calculation, but we will check its numerical impact by the end of this section. Notice that since there is no soft singularity for flavor switching splittings, they contribute as a power correction to $\kappa^{(a)}$ in our Sudakov form factor and we do not include them here. This argument is valid as long as $\kappa\!\ll\!1$ along the lines of the role played by $y_{\rm{cut}}$ in App.~B of Ref.~\cite{Dasgupta:2013ihk}. 

%%%%%%%%%%%%%%%%%%
\paragraph{Running coupling.} 
%%%%%%%%%%%%%%%%%%
Up to now, we have fixed the coupling in order to achieve compact, fully analytic expression. However beyond DLA, the running of the coupling has to be taken into account. At 1-loop in perturbation theory it is given by:
\begin{align}
\label{eq:alphas-1loop}
 \alpha_s^{\rm 1\ell}(k_t)&=\dis\frac{\alpha_s}{1+2\beta_0\alpha_s\ln\left(\frac{k_t}{Q}\right)}\\
 &=\alpha_s\left[1-2\beta_0\alpha_s\ln\left(\frac{k_t}{Q}\right)+4\beta_0^2\alpha_s^2\ln^2\left(\frac{k_t}{Q}\right)\right]+\mathcal{O}(\alpha_s^4)\,,
\end{align}
with the reference value $\alpha_s\!\equiv\!\alpha_s(Q)$ is set at the jet scale $Q\!\equiv\!p_{t,\rm jet}R$, $\beta_0\!=\!(11C_A\!-\!4n_fT_r)/(12\pi)$ and $k_t\!=\!z\theta p_{t,{\rm jet}}$. 

Next, we integrate analytically the branching kernel with the 1-loop running coupling 
\begin{align}
\label{eq:rc-terms-1loop}
 -\dis\int_0^1 \dd z\dis\int_0^1\dd \theta& \dis\frac{\alpha_s^{\rm 1\ell}(k_t) }{\pi\theta}P_i(z)\Theta(z\theta^a-\kappa)=\nn
 &\ln\kappa\left(g_{11}\alpha_s\ln\kappa+\dis\frac{2C_i\beta_0(1+a)}{3a^2\pi }\alpha_s^2\ln^2\kappa
-\dis\frac{2C_i\beta^2_0(1+a+a^2)}{3a^3\pi }\alpha^3_s\ln^3\kappa\right)\nn
 &+g_{21}\alpha_s\ln(\kappa)+\dis\frac{2B_iC_i\beta_0}{a^2\pi }\alpha_s^2\ln^2\kappa+\mathcal{O}(\textrm{N}^3\textrm{DL})\,.
 \end{align}
Note that in the previous expression we have only kept the relevant terms up to N$^2$DL, as indicated by the $\mathcal{O}(\textrm{N}^3\textrm{DL})$ notation. Now, we can identify the terms corresponding to the soft and collinear piece of the splitting function to be
 \begin{align}
 g_{12}& = \dis\frac{2\beta_0C_i(1+a)}{3a^2\pi} \,, \\
 g_{13}& = -\dis\frac{2\beta^2_0C_i(1+a+a^2)}{3a^3\pi}.
 \end{align}
 while the hard-collinear correction results into
  \begin{equation}
 g^{1}_{22} = \dis\frac{2B_iC_i\beta_0}{a^2\pi}.
 \end{equation}
 In the last equation, the upper subscript in the coefficient indicates that this is not the only term that contributes to the $g_{22}$ coefficient, i.e.\ $g_{22}\!=\!\sum_i g^i_{22}$.

To achieve N$^2$DL accuracy, we need to go to the next order in the running coupling. We work in the CMW scheme~\cite{Catani:1990rr} which enables to include also the 2-loop contribution of the splitting functions in the soft limit. Then, the running of the coupling at two loops is given by
\begin{align}
\label{eq:2loop-rc-def}
\alpha_s^{\rm 2\ell}(k_t)=\dis\frac{\alpha_s}{1+2\alpha_s\beta_0\ln(\frac{k_t}{Q})}-\dis\frac{\beta_1\alpha_s^2}{\beta_0}\dis\frac{\ln(1+2\alpha_s\beta_0\ln(\frac{k_t}{Q}))}{[1+2\alpha_s\beta_0\ln(\frac{k_t}{Q})]^2}+\dis\frac{K}{2\pi}\dis\frac{\alpha_s^2}{[1+2\alpha^2_s\beta_0\ln(\frac{k_t}{Q})]^2}\,,
\end{align}
with $\beta_1\!=\!(17C_A^2\!-\!5C_An_f\!-\!3C_f)/(24\pi^2)$ and $K\!=\!(67/18\!-\!\pi^2/6)C_A\!-\!5n_f/9$. Again, we can integrate the branching kernel with the 2-loop running coupling to identify another contribution to the $g_{22}$ coefficient:
  \begin{equation}
 g^{2}_{22} = -\dis\frac{KC_i}{2a\pi^2}\,.
 \end{equation}
 
An important aspect is that the domain of applicability of Eqs.~\eqref{eq:alphas-1loop} and \eqref{eq:2loop-rc-def} is restricted to perturbative scales, i.e.\ above $\mu_{\rm fr}\!\sim\!1$~GeV. Hence, in order to avoid the divergence appearing at such scale, a freezing of the coupling, i.e.
\begin{equation}
\label{eq:alpha-s-ir}
\alpha_s^{\rm IR}(k_t) = \alpha_s(\mu_{\rm fr})\Theta(\mu-k_t)
\end{equation}
is implemented. We would like to point out that this choice is completely ad-hoc and one could think of replacing Eq.~\eqref{eq:alpha-s-ir} by a more general functional form and systematically study its impact on jet substructure observables. We will investigate this possibility in future studies.

%%%%%%%%%%%%%%%%%%
\paragraph{Soft emissions at large angles.}
%%%%%%%%%%%%%%%%%%
The dynamically groomed $k_{t,g}$ pertains to the category of so-called non-global observables,\footnote{This also applies to $z_g$ and $\theta_g$.} i.e.\ it is sensitive to a certain region of the radiation phase space. As such it is affected by a particularly complex class of logarithms known as non-global logs~\cite{Dasgupta:2001sh,Banfi:2010pa,Delenda:2012mm}. A typical configuration that can give rise to these contributions is a collection of large angle gluons outside the jet which subsequently radiate softer gluons inside. In order to understand how this topology contributes to the $k_{t,g}$ distribution, we first calculate the lowest $\mathcal{O}(\alpha_s^2)$ term coming from such configurations. For illustrative purposes, we start with the calculation of the leading non-global logarithm in $e^+e^-$ annihilation, in which the color structure of the event is simpler. We discuss the straightforward generalization to $p+p$ collisions in the following paragraph. Once again, we rely on the small-$R$ limit and sketch how to lift this approximation in the next section. 

The calculation of the non-global contribution at $\mathcal{O}(\alpha_s^2)$ is standard: one calculates the cross-section for two correlated gluon emissions strongly ordered in energy, with the first emission outside the jet and the second inside. For dynamical grooming, one can rely on the fact that the gluon inside the jet is necessarily the hardest, since it is the only one at this order. Thus, the double differential distribution for having a dynamically groomed $z_g$ and $\th_g$ value from a non-global configuration initiated by a $q\bar{q}$ dipole is:\
\begin{align}\label{eq:NG-ev}
 \frac{1}{\sigma_0}\frac{\dd^2\sigma^{\rm NG}}{\dd z_g\dd \cos(R_g)}&=4C_FC_A\left(\frac{\alpha_s}{2\pi}\right)^2\int_0^{p_T}\frac{\dd\omega_1}{\omega_1}\int_0^{\omega_1}\frac{\dd\omega_2}{\omega_2}\int_{-1}^{1}\dd\cos R_1\int_{-1}^{1}\dd\cos R_2\,\Omega(\cos R_1,\cos R_2)\nn
 &\Theta(\cos(R)-\cos(R_1))\Theta(\cos(R_2)-\cos(R))\delta\left(z_g-\frac{\omega_2}{p_T}\right)\delta(\cos(R_2)-\cos(R_g))\nn
 &=4C_FC_A\left(\frac{\alpha_s}{2\pi}\right)^2\frac{1}{z_g}\ln\left(\frac{1}{z_g}\right)\Theta(\cos(R_g)-\cos(R))\nn
 &\hspace{4cm}\times\int_{-1}^{\cos(R)}\dd\cos R_1\,\Omega(\cos R_1,\cos R_g)\,,
\end{align}
where the first $\Theta$-function in the second line enforces the first gluon to be outside the jet, while the second $\Theta$-function constrains the tagged emission to be inside. Note that this is the real term only.\footnote{In principle the lower bound in the integration range of $\cos R_1$ depends on the the jet selection. However, in the small $R$ approximation (see the discussion thereafter), those corrections are power of $R$ suppressed.} The function $\Omega$ is the azimuthal average of the real cross-section for correlated double gluon emission from a quark~\cite{Marzani:2019hun}:
\begin{equation}
 \Omega(\cos R_1,\cos R_2)=\frac{2}{(\cos(R_2)-\cos(R_1))(1-\cos(R_1))(1+\cos(R_2))}\,.
\end{equation}
The $R_g\!=\!\th_g R$ integral in Eq.~\eqref{eq:NG-ev} is non-singular in the collinear limit, so that one can perform the two angular integrals exactly to get the leading term in the soft and $R\to0$ limit:
\begin{equation}\label{eq:non-global-as2}
 \frac{1}{\sigma_0}\frac{\dd\sigma^{\rm NG}}{\dd z_g}=2C_FC_A\left(\frac{\alpha_s}{2\pi}\right)^2\frac{1}{z_g}\ln\left(\frac{1}{z_g}\right)\frac{\pi^2}{3}\,.
\end{equation}
In the previous equation, the soft singularity when $z_g\rightarrow0$ induces a single log contribution which has to be taken into account at N$^2$DL as part of the $g_2$ function in Eq.~\eqref{sudakov-log}. 

In $p+p$ collisions, the situation is \textit{a priori} more involved. Each Born level partonic configuration needs to be broken into distinct hard dipoles. However, as shown in Ref.~\cite{Dasgupta:2012hg}, only the dipoles involving the measured jet matter in the small $R$ limit (i.e.\ neglecting terms proportional to $\theta^n$), and all such contributions are enhanced by the same $\pi^2/3$ factor as in the $e^+e^-$ result in Eq.~\eqref{eq:non-global-as2}. Consequently, the non-global contribution to the resummed distributions factorize according to the flavour of the jet, in the same way as the collinear piece calculated above. In other words, by imposing the small $R$ limit we can use the $e^+e^-$ result from Eq.~\eqref{eq:non-global-as2} for the $p+p$ case. By doing so, we can extract the last piece of the $g_{22}$ coefficient, namely
\begin{equation}
\label{eq:non-global-piece}
g^3_{22}=-\dis\frac{\pi^2}{3}\dis\frac{C_iC_A}{(2\pi)^2}\,.
\end{equation} 

At this point let us summarize the main ingredients obtained so far in a compact way that shall facilitate the reproducibility of our results. At N$^2$DL accuracy, and in the small-$R$ limit, the Sudakov form factor given by Eq.~\eqref{sudakov-log} involves the coefficients provided in Table~\ref{table:sudakov}.
\begin{table}[ht]
\centering\small
\begin{tabular}{|c|c|}
\hline
$g_{nm}$ & Physical origin \\
\hline% inserts table %heading\hline
$g_{11}=-\dis\frac{C_i}{a\pi}$ &Soft and collinear\\[1ex]
$g_{12}=\dis\frac{2\beta_0C_i(1+a)}{3a^2\pi}$ & Soft and collinear + $\alpha_s^{\rm 1\ell}(k_t)$\\[1ex]
$g_{13}= -\dis\frac{2\beta^2_0C_i(1+a+a^2)}{3a^3\pi}$ & Soft and collinear + $\alpha_s^{\rm 1\ell}(k_t)$\\[1ex]
$g_{21}= -\dis\frac{2C_iB_i}{a\pi}$ & Hard and collinear\\[1ex]
$g_{22}= \dis\frac{2C_iB_i\beta_0}{a^2\pi}-\dis\frac{KC_i}{2a\pi^2}-\dis\frac{\pi^2}{3}\dis\frac{C_iC_A}{(2\pi)^2} $& Hard and collinear +$\alpha_s^{\rm 1\ell}(k_t)$, $\alpha_s^{\rm 2\ell}(k_t)$, non-global soft \\[1ex]
\hline
\end{tabular}
\caption{\label{table:sudakov} Relevant coefficients for the Sudakov form factor at N$^2$DL accuracy.}
\end{table}

Equivalently, using Eq.~\eqref{pbranch-n2dl} we arrive to the following, non unique expression for the branching kernel
\begin{align}
\label{eq:pbranch}
\widetilde P(z,\theta)&=\dis\frac{2\alpha_sC_i}{\pi z\theta}(1-2\alpha_s\beta_0\ln(\mu_K z\theta)+4\alpha_s^2\beta_0^2\ln^2(\mu_K z\theta))+\displaystyle\frac{2\alpha_s C_i B_i}{\pi\theta}(1-2\alpha_s\beta_0\ln(\mu_K \theta)) \nn
&+\dis\frac{K}{2\pi}\dis\frac{2C_i\alpha_s^2}{\pi z\theta}-2C_iC_A\left(\dis\frac{\alpha_s}{2\pi}\right)^2\dis\frac{\pi^2}{3}\dis\frac{\ln(\mu_K z)}{z}\,.
\end{align}
In order to estimate the uncertainty of the resummation, we have introduced the dimensionless multiplicative factor $\mu_K$ that will be varied between $0.5$ and $2$. A subtle issue\footnote{We are grateful to Gregory Soyez for pointing this out.} concerning this $\mu_K$ variation is that the first non-trivial correction that arises after integrating over ($z,\theta$) Eq~\eqref{eq:pbranch} is given by $\propto\alpha_s^2\ln^2(\kappa)\ln(\mu_K)$. This is of the same order as the corresponding $K$-term, i.e. $\propto\alpha_s^2\ln^2(\kappa)K$. To overcome the non-desirable variation of a $g_{nm}$ coefficient, we vary $\mu_K$ under the condition that the $K$ term is constant, i.e.\ $K$ is shift to $K+4\pi\beta_0\ln(\mu_K)$ in the calculation (and similarly when considering varition of the renormalization scale $Q$).

Note that the approximations made to derive the coefficients inside the Sudakov factor $\Delta(\kappa|a)$ and the branching kernel $\widetilde P$ lead to a differential cross-section which is not necessarily normalized to the Born jet cross-section. To restore the correct normalization, one can simply divide the cumulative distribution Eq.~\eqref{eq:master-form-ktg} by $\Sigma(1)$. This overall normalization factor is a non-logarithmic correction which does not spoil our targeted accuracy. 

Lastly, we would like to draw the reader's attention to the fact that multiple gluon emissions were not considered in this calculation. Generically speaking, at leading logarithmic accuracy, a single emission dominates jet substructure observables. This strong ordering might be broken beyond leading-log, like in the jet mass case, such that an arbitrary number of emissions give comparable contributions to the final measured value. The region of phase space for which this happens has to be determined on an observable basis thus increasing the complexity of analytic calculations. In the dynamical grooming case, multiple emissions do not have to be considered for the observables computed in this paper. This property is a direct consequence of how the method is built. That is, dynamically groomed observables in tagging mode are not additive but defined on the hardest emission and thus it is the only one that contributes to all orders in the resummation. 

%%%%%%%%%%%%%%%%%%
\paragraph{N$^2$DL and N$^2$DL'.}
%%%%%%%%%%%%%%%%%%
Insofar, we have provided the minimal set of $g_{nm}$ coefficients that lead us to N$^2$DL accuracy. For that purpose we have neglected all terms that are not logarithmically enhanced. In order to gauge the impact of these sub-leading contributions, we will also provide results with the `complete' branching kernel, i.e.
\begin{equation}\label{eq:Pbranch-full}
 \widetilde P(z,\theta)=\left[\frac{2\alpha_s^{\rm 2\ell}(\mu_K z\theta Q)C_i}{\pi z \theta}-2C_iC_A\frac{\pi^2}{3}\left(\frac{\alpha_s}{2\pi}\right)^2\frac{\ln(\mu_K z)}{z}\right]\Theta\left(e^{-B_i}-z\right)\,,
\end{equation}
and the Sudakov $\Delta(\kappa|a)$ calculated \textit{exactly} from this complete branching kernel whose explicit expression can be found in App.~\ref{sec:appendix-a}. Note that the resulting differential cross-section is then normalized by construction. The running of the coupling is neglected in the non-global term for simplicity, adding it would enable to account for part of the full resummation of the non-global soft function \cite{Dasgupta:2001sh}.  We will refer to this resummation as N$^2$DL', where the prime indicates that the resummation actually includes some of the sub-leading logarithmic corrections with $p\!\ge\!3$. That said, we emphasize that in all rigour, both ways of doing the resummation --- either `minimally' using Eq.~\eqref{eq:pbranch} and the coefficients in Tab.~\ref{table:sudakov} in the Sudakov or with the complete branching kernel given by Eq.~\eqref{eq:Pbranch-full} --- reach the same N$^2$DL logarithmic level accuracy, and not more.\footnote{Indeed, a complete N$^3$DL resummation would require at least the first term in the analytic expansion of the $g_3$ (NNLL) function inside $\Delta(\kappa|a)$.} Therefore, we will use this resummation scheme freedom to leverage our uncertainty.

%%%%%%%%%%%%%%%%%%
\paragraph{Beyond N$^2$DL and the small-$R$ limit}
%%%%%%%%%%%%%%%%%%
Before we move on to the fixed-order section, we would like to sketch which steps have to be taken in order to extend the calculation that we have just presented. 

In the first place, if the small-$R$ constraint is lifted, one has to account for process dependent terms that enter the calculation as a power series in the jet radius. Physical scenarios that lead to such contributions involve soft and large angle emissions that end up being clustered in the reconstructed jet. For example, a splitting originated from the initial state partons can be tagged by Dynamical Grooming and induce single logarithmic terms suppressed by powers of the jet radius $R$ in the resummation. The difficulty with soft emissions at large angles comes from the fact that such emissions have a complicated color structure which depend on the full Born level event and not only on the Casimir factor of the measured jets. In order to handle such corrections, which are expected to be important for $R\!\sim\!1$, one could decide to rely on the large $N_c$ limit and decompose each Born processes into different colour flows, as done in Ref.~\cite{Dasgupta:2001sh,Lifson:2020gua} in the context of the Lund plane density. Then, each color flow corresponds to a superposition of hard dipoles, which can radiate a soft large angle gluon into the measured jet. In practical terms, adding these contributions would promote the jet flavour dependence of $\widetilde P(z,\theta)$ and $\Delta_i(\kappa|a)$ to a color flow one. Once these new terms are taken into account N$^2$DL accuracy is reached beyond the small jet radius limit, but in the large $N_c$ approximation. If one does not resort to the large $N_c$ limit, one has to deal with matrix formulae in color space, as in Ref.~\cite{Dasgupta:2012hg}. It is however unclear if the simple structure of Eq.~\eqref{eq:master-form-ktg} remains when the exponentiation has a matrix form and it deserves a dedicated study.

From the non-global logarithms side, their full resummation is required if a higher accuracy in the resummation is intended. This is a complicated task for the $\kappa^{(b,c)}$ observable in $p+p$ collisions, even in the large $N_c$ limit. With the latter approximation, one could resort to the same numerical method as in Ref.~\cite{Dasgupta:2012hg} (see also Ref.~\cite{Kang:2019prh} for the $\th_g$ distribution defined with Soft Drop).

%%%%%%%%%%%%%%%%%%
\subsubsection{Matching to fixed-order}
\label{sub:matching-ktg}
%%%%%%%%%%%%%%%%%%

In order to produce reliable predictions when $k_{t,g}\!=\!\mathcal{O}(1)$ and to achieve N$^2$DL via the $C_1$ term, the resummed distribution obtained in the previous section needs to be matched with a fixed-order calculation. Several matching schemes are available in the literature. For our purposes, it is clearly desirable to have a matching scheme satisfying the two following conditions: (i) the matching scheme should produce `for free' the $C_1$ term, (ii) the matching scheme should preserve the fixed order endpoint of the distribution at $k_{tg,\rm max}\!=\!0.5$. Two possible matching schemes that satisfy these requirements are the multiplicative and the $\log(R)$ matching~\cite{Catani:1992ua,Banfi:2010xy}. In what follows, we shall use multiplicative matching at leading order ($\mathcal{O}(\alpha_s)$) and discuss how to extend it to next-to-leading order ($\mathcal{O}(\alpha^2_s)$). 

As the colour structure of the resummation is tremendously simplified within our targeted accuracy, i.e.\ it only depends on the jet flavor. The matching formula can be decomposed accordingly as follows~\cite{Banfi:2010xy}:
\begin{align}
\label{x-match}
 \Sigma^{\rm LO+N^2DL}(k_{t,g})=\dis\frac{1}{\sigma_0+\sigma_1}&\left\{\dis\sum_{i=q,g}\tilde\Sigma^{\rm N^2DL}_i(k_{t,g})\left(1+\frac{\Sigma^{\rm LO}_{i}(k_{t,g})-\tilde\Sigma^{\rm N^2DL}_{i,1}(k_{t,g})}{\sigma_{0,i}}\right)\right. \\ \nonumber
 &\hspace{8cm}+\Sigma^{\rm LO}_{\rm else}(k_{t,g})\Bigg\}\,.
\end{align}
We proceed to describe the ingredients entering the previous equation, except the meaning of the last term that will become clear later on. First, $\sigma_0$ and $\sigma_1$ are the inclusive dijet cross-section at leading order and next-to-leading order respectively. The $\tilde\Sigma_i$ is the resummed cumulative $k_{t,g}$ cross-section for $i$-jets, that shares the same endpoint, $k_{tg,\max}$, as the fixed order distribution. At a given accuracy, this is achieved through the following transformation:\footnote{When the resummed distribution has an endpoint different from $1$, Eq.~\eqref{endpoint-resum} needs to be modified accordingly. In particular, when using the calculation of $\Sigma$ at N$^2$DL$'$, the $+1$ inside the logarithm is replaced by $\exp(B_q)$.}
\begin{equation}\label{endpoint-resum}
 \tilde\Sigma_{i}(k_{t,g})=\sigma_{0,i}\Sigma_i\left[\exp\left(-\log\left(\frac{1}{k_{t,g}}-\frac{1}{k_{tg,\rm max}}+1\right)\right)\right]\,.
\end{equation}
Notice that, besides the shift in the endpoint, we have multiplied the resummed cumulative distribution by $\sigma_{0,i}$ in order to ensure that $\tilde\Sigma^{\rm N^2DL}$ and $\Sigma^{\rm LO}$ have the same units. Following up with the pieces entering Eq.~\eqref{x-match}, $\tilde\Sigma^{\rm N^2DL}_{i,1}$ is the $\mathcal{O}(\alpha_s)$ term in the expansion of $\tilde\Sigma^{\rm N^2DL}_i$, while $\Sigma^{\rm LO}_{i}$ is the leading order distribution defined as
\begin{equation}
 \Sigma^{\rm LO}_{i}(k_{t,g})=\sigma_{1,i}-\int_{k_{t,g}}^{1}\,\dd k_{t,g}'\frac{\dd \sigma^{\rm LO}_i}{\dd k_{t,g}'}\,,
\end{equation}
Regarding the normalization, we find that, by construction, $\Sigma^{\rm LO+N^2DL}(k_{tg,\rm max})\!=\!1$. 

One can check that the limiting behavior of the matched distribution is correct. Indeed, Eq.~\eqref{x-match} gives back the LO distribution for $k_{t,g}\sim k_{tg,\rm max}$. In turn, when $k_{t,g}\ll 1$ the distribution behaves like
\begin{equation}
\label{eq:limit-resum}
 \Sigma^{\rm LO+N^2DL}(k_{t,g})\simeq\frac{1}{\sigma_0+\sigma_1}\sum_{i=q,g}\tilde\Sigma^{\rm N^2DL}_i(k_{t,g})(1+\alpha_sC_{1,i})\,,
\end{equation}
where the $C_1$ term is given by its standard definition:
\begin{equation}\label{eq:def-C1}
\alpha_sC_{1,i}=\lim\limits_{k_{t,g}\rightarrow0}\frac{\Sigma^{\rm LO}_{i}(k_{t,g})-\tilde\Sigma^{\rm N^2DL}_{i,1}(k_{t,g})}{\sigma_{0,i}}\,.
\end{equation}
Thus, Eq.~\eqref{eq:limit-resum} shows that the matched distribution reproduces the resummed result in its regime of validity. In our calculation, $C_1$ is a constant up to $R^2$-suppressed single logarithmic contributions.

In practice, the LO $k_{t,g}$ differential cross-section and the LO and NLO jet cross-sections $\sigma_0$ and $\sigma_1$ are obtained using MadGraph5~\cite{Alwall:2014hca} (in fixed order mode) with CT10nlo PDF set~\cite{Lai:2010vv}. The factorisation scale for the PDF convolution is set to $\mu_F Q$, with $\mu_F$ a dimensionless factor introduced to estimate the uncertainty relative to this prescription. For a given jet selection $[p_{t,\rm min},p_{t,\rm max}]$, a unique generation cut is imposed in the fixed order calculation. Namely, the sum of the transverse momenta of the partons is required to be larger than $p_{t,\rm min}$ and one asks for at least one jet with $p_t\!>\!p_{t,\rm min}/4$. We have checked that the resulting cross-sections are insensitive to the precise value of these cuts. The reference value of the strong coupling at the jet scale, $\alpha_s(\mu_R Q)$, is evaluated in the $\overline{\rm{MS}}$ scheme. The $\mu_R$ is a dimensionless factor used to gauge the uncertainty with respect to the renormalization scale. When the $p_t$ selection is broad, such as in the ATLAS set-up detailed in the following section, the $p_t$ range is divided into smaller bins in which the inclusive jet and $k_{tg}$ cross-sections are calculated. The extension of Eq.~\eqref{x-match} in this case is straightforward.

The last ingredient in Eq.~\eqref{x-match} involves the decomposition according to the flavour of the jet. This is done in an IRC safe way for both the $k_{t,g}$ differential and inclusive jet cross-section. More concretely, at LO the jets have at most two constituents. Then, when the jet has zero or one net flavour, the jet is tagged as a gluon or quark jet, respectively. Otherwise, whenever the jet is multi-flavored, i.e.\ it contains two (anti)-quarks of different flavor, it pertains to what we call the 'else' category. The LO $k_{t,g}$ differential cross-section for these multi-flavored jets goes to zero at small $k_{t,g}$ and contributes to the full match result via the $\Sigma^{\rm LO}_{\rm else}(k_{t,g})$ term in Eq.~\eqref{x-match}.

Finally, as for the resummation part, we would like to comment on how to further extend the matching procedure to higher accuracy. In this case, the equivalent of the multiplicative matching formula Eq.~\eqref{x-match} at NLO can be found in Ref.~\cite{Banfi:2010xy} and it involves the NLO $k_{t,g}$ differential cross-section. The latter can be obtained by generating 3-jet events at NLO with MadGraph, or any other code dedicated to matrix element calculations. Even if there is no conceptual difficulty in promoting our matching to NLO, we postpone it for further studies given that its quantitative impact on the resulting distributions could be as sizeable as the missing power of $R$ suppressed terms on the resummation. Therefore, we believe that these two endeavors should be pursued simultaneously and must be included for refining our phenomenological studies presented in Sec.~\ref{sec:numerics}.

%%%%%%%%%%%%%%%%%%
\subsubsection{Results}
\label{sec:sanity-checks}
%%%%%%%%%%%%%%%%%%
Once the analytic framework has been presented, we proceed to show some numeric results for high-$p_t$ jets ($800\!<\!p_t\!<\!1000$~GeV) at top LHC energy $\sqrt s\!=\!13$~TeV with cone size $R\!=\!0.4$. The central value of the following curves is obtained with $\mu_F\!=\!\mu_R\!=\!1$. Further, the error bars are obtained by varying a factor of two the following parameters in the calculation: factorization and renormalization scales through the 7-point rule~\cite{Cacciari:2003fi}, the parameter $\mu_K$ that controls the scale at which the strong coupling runs (see Eq.~\eqref{eq:pbranch}) and the freezing scale $\mu_{\rm fr}$ used to avoid the Landau pole in Eq.~\eqref{eq:alpha-s-ir}. Then, we combine the various uncertainties by taking the envelope of all distributions.

\begin{figure}
\centering
  \includegraphics[width=\textwidth]{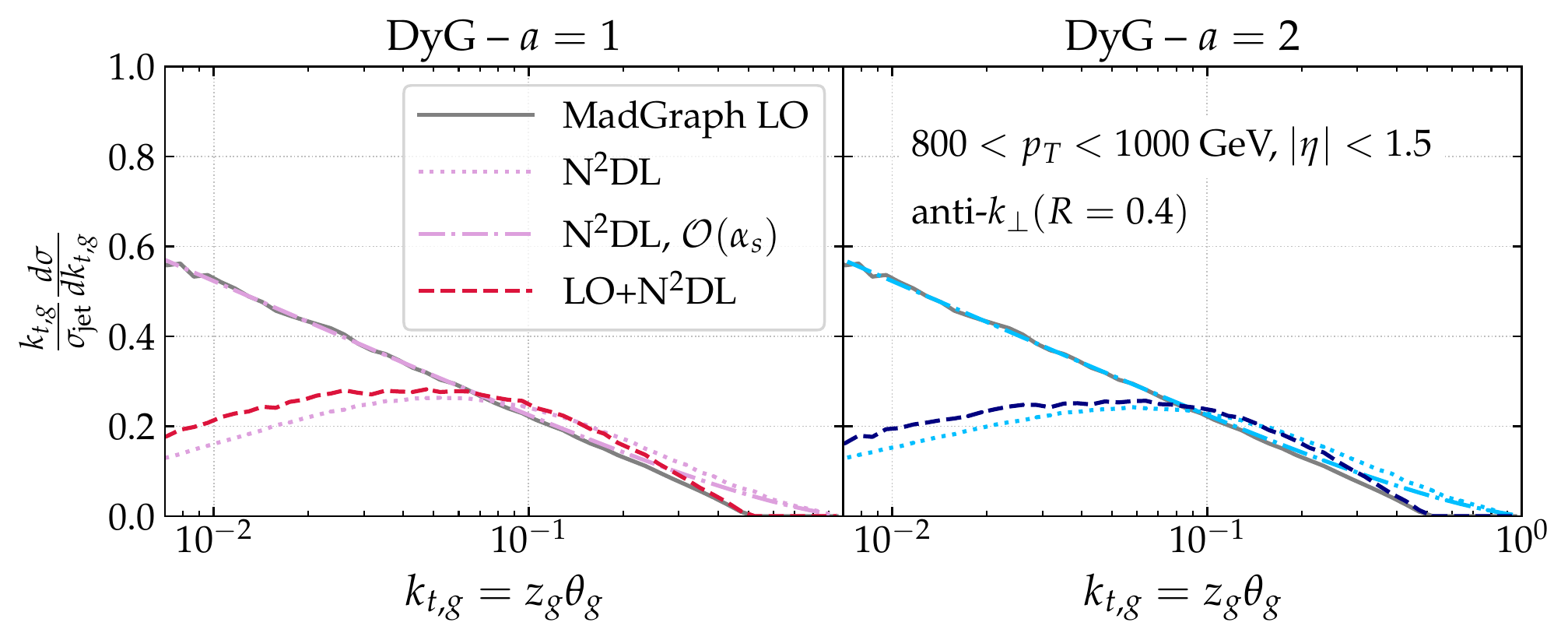}
  \caption{The $k_{t,g}$-distribution computed in different ways: at leading-order with MadGraph, resumed at N$^2$DL as given by Eq.~\eqref{pbranch-n2dl} and Tab.~\ref{table:sudakov}, first order expansion of the resumed result, and the matched distribution (see Eq.~\eqref{x-match}) for $a\!=\!1$ (left) and $a\!=\!2$ (right). The normalization factor $\sigma_{\rm jet}$ reduces to $\sigma_0\!+\!\sigma_1$ for the resummed and matched distributions and to $\sigma_0$ in the other two cases.}  
  \label{fig:ktg-analytics-matching}
\end{figure}

In Fig.~\ref{fig:ktg-analytics-matching} we present four distributions of $k_{t,g}$: (i) the fixed-order result, (ii) the resummed result at N$^2$DL, (iii) the $\mathcal O(\alpha_s)$ expansion of the latter and (iv) the matched distribution at LO+N$^2$DL. It is clear from this figure that the matching procedure works as expected, i.e.\ the LO+N$^2$DL recovers the N$^2$DL result at small $k_{t,g}$, while it tends towards the leading-order curve in the opposite regime. In addition, the endpoint of the resummation is shifted by the matching procedure to the fixed-order one at $k_{t,g}\!=\!0.5$.  By comparing the fixed-order result and the first term in the $\alpha_s$-expansion of the resummed result, we can get a hint on the size of the $\mathcal{O}(R^n)$ logarithmically enhanced terms that we have so far neglected. In fact, the difference between the $\mathcal{O}(\alpha_s)$ term of the N$^2$DL curve and the exact leading order result converges towards a constant at small $k_{t,g}$. This indicates that these power suppressed terms enter with a small coefficient in the cumulative distribution and can be safely neglected for the setup studied in this work. All the previous statements hold for both values of $a$. In particular, the fixed order result is independent of $a$ because there is only one splitting tagged.

\begin{figure}
\centering
  \includegraphics[width=\textwidth]{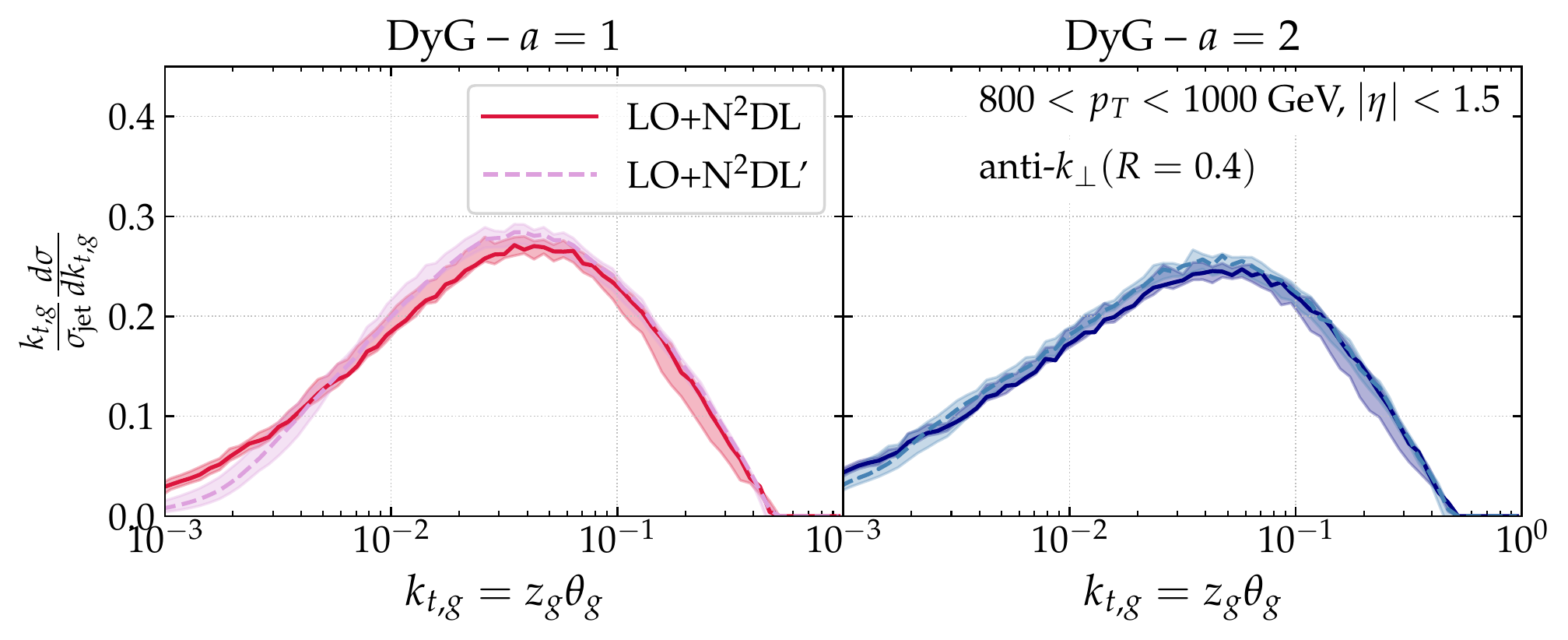}
  \caption{The $k_{t,g}$-distribution with a minimal N$^2$DL resummation (see Eq.~\eqref{eq:pbranch} and Tab.~\ref{table:sudakov}), and including sub-leading contributions N$^2$DL' (see Eq.~\eqref{eq:Pbranch-full} and App.~\ref{sec:appendix-a}). Both curves are normalized to $\sigma_0\!+\!\sigma_1$. 
  }
  \label{fig:ktg-analytics-ndl}
\end{figure}

Next, we compare in Fig.~\ref{fig:ktg-analytics-ndl} the two prescriptions to perform the resummation that we have discussed above, i.e.\ keeping uniquely the logarithmically enhanced terms at N$^2$DL or including sub-leading corrections (N$^2$DL'). In the large $k_{t,g}$ regime, we observe no difference between the LO+N$^2$DL and the LO+N$^2$DL' as it is expected since in this limit the fixed-order contribution dominates the matched result. This is no longer the case for $k_{t,g}\!\ll\!1$, where details of the resummation structure do matter. An important remark is that the discrepancy between the two curves diminishes when increasing the parameter $a$ that determines the hardness condition in the grooming algorithm. We attribute this to the $a-$scaling of the $g_{nm}$ parameters that, as one can see in Tab.~\ref{table:sudakov}, satisfies $g_{nm}\sim 1/a$. Hence, the larger the value of $a$ is, the smaller the coefficients in front of the higher order terms are and the narrower the difference between N$^2$DL and N$^2$DL'  becomes. In the phenomenological section, we will include the differences between N$^2$DL and N$^2$DL' as part of our uncertainty band given that, from a logarithmic counting point of view, there is no preferred option.

%%%%%%%%%%%%%%%%%%
\subsection{$z_g,\theta_{g}$ at LO+N$^2$DL accuracy}
%%%%%%%%%%%%%%%%%%

As discussed at length in Sec.~\ref{sec:dla}, the momentum sharing fraction, $z_g$, and opening angle, $\theta_g$, of the splitting tagged by dynamical grooming are Sudakov safe observables. In Eqs.~\eqref{eq:resum-zg}--\eqref{eq:resum-thetag}, we defined their distribution as the limit of the IRC safe $k_{t,g}$ distribution. This allows us to follow the same logic as in the previous section to obtain the resummation part of their distribution. In turn, the fixed-order result is not even well defined, as shown in Eqs.~\eqref{eq:Sigma-zg-series} and \eqref{eq:Sigma-thetag-series}, and thus the matching strategy differs to that presented in Sec.~\ref{sub:matching-ktg}. In what follows, we provide the necessary ingredients to reach LO+N$^2$DL accuracy in the small-$R$ limit.

%%%%%%%%%%%%%%%%%%%%%%%%%%%%%%
\subsubsection{Boundary logarithms for the $\th_g$ distribution}
\label{sec:boundary_logs}
%%%%%%%%%%%%%%%%%%%%%%%%%%%%%%
In the case of $z_g$, the resummation proceeds in exactly the same fashion as for $k_{t,g}$. In turn, for $\theta_g$, another source of logarithmic enhancement appears, caused by the interplay between the anti-$k_\perp$ algorithm~\cite{Cacciari:2008gp} used to cluster the jet, and the C/A algorithm to decluster it in the dynamical grooming procedure. These so-called clustering boundary logarithms~\cite{Lifson:2020gua} are of the form:
\begin{equation}
 \alpha_s^2\frac{1}{z}\ln\left(\frac{1}{z}\right)\ln\left(\frac{R}{R-R_g}\right)\,,
\end{equation}
where the double logarithmic enhancement becomes important when $R_g\rightarrow R$ ($\theta_g\!\equiv\!R_g/R$ close to 1).

Boundary logarithms arise from a non-global configuration where the first emission is outside the anti-$k_\perp$ jet, while the second is inside and almost collinear to the first one. In this situation, C/A algorithm would cluster these two emissions together, as part of the jet. As such, their leading contribution to the branching kernel can be obtained from the same calculation as the one done in Eq.~\eqref{eq:NG-ev}, but focusing in the regime where $\theta_1\!\simeq\!\theta_2\!\simeq\!R\!\ll\!1$~\cite{KhelifaKerfa:2011zu,Lifson:2020gua}, such that:
\begin{equation}
  \Omega(\th_1,\th_2)\simeq\frac{4}{\th_1^2(\th_1^2-\th_2^2)}\,,
\end{equation}
and
\begin{align}
	\frac{1}{\sigma_0}\frac{\dd^2\sigma^{\rm NG, \rm cl}}{\dd z_g\dd\th_g}&\simeq 4C_FC_A\left(\frac{\alpha_s}{2\pi}\right)^2\frac{1}{z_g}\ln\left(\frac{1}{z_g}\right)R^2\int_{R}^\infty\dd\th_1\th_1\th_g\,\Omega(\theta_1,\th_g R)\nn
	&= 8C_RC_A\left(\frac{\alpha_s}{2\pi}\right)^2\frac{1}{z_g}\ln\left(\frac{1}{z_g}\right)\frac{1}{\th_g}\ln\left(\frac{1}{1-\th_g^2}\right)\label{eq:leading-clust-exact}\\
	&\simeq 8C_RC_A\left(\frac{\alpha_s}{2\pi}\right)^2\frac{1}{z_g}\ln\left(\frac{1}{z_g}\right)\frac{1}{\th_g}\ln\left(\frac{1}{1-\th_g}\right)\,.\label{eq:leading-clust}
\end{align}
The upper boundary in the $\th_1$ integral can be safely sent to $\infty$ since the integral is dominated by the region $\th_1\!\simeq\!\th_2\simeq R$. To get the last line, we have kept the dominant contribution at $\th_g\sim1$. 

As stated above, this $\mathcal{O}(\alpha_s^2)$-contribution is enhanced by two soft logarithms of the type $\ln(z_g)/z_g$ and one boundary logarithm $\ln(1\!-\!\th_g)$. However, the logarithmic divergence associated with $\th_g\!\sim\!1$ is integrable in a neighbourhood of $1$. Consequently, as long as one deals with an observable in which the angle $R_g\sim R$ is integrated out between some lower bound and $1$ (such as $z_g$ or $k_{t,g}$ distributions), the boundary divergence is harmless. More precisely, its integral over $\th_g$ should give back the soft single logarithmic divergence that was part of our treatment of non-global configurations. This argument also applies to the Sudakov factor, since vetoing all emissions with hardness larger than $\kappa$ translates into the following integral
\begin{equation}\label{eqclust-kappa}
 \int_0^1\dd z'\int_0^1\dd \th'\, \frac{1}{\sigma_0}\frac{\dd^2\sigma^{\rm NG, \rm cl}}{\dd z'\dd \th'}\Theta(z'\th'^a-\kappa)\,,
\end{equation}
where it is clear that $\th'$ is always marginalized in the neighbourhood of $1$. In other words, the dynamical grooming procedure lowers the singularity associated with boundary logarithms from double- to single-log, and this single-log term is already taken into account by the coefficient $g_{22}^3$ in the Sudakov.

From that perspective, the $\th_g$ distribution is peculiar since the $R_g\rightarrow R$ logarithmic divergence from Eq.~\eqref{eq:leading-clust} is not integrated out. To effectively include boundary logarithms for the $\th_g$ distribution, we replace the last term in the branching kernel $\widetilde P(z,\th)$ given in Eq.~\eqref{eq:pbranch} by 
\begin{equation}
-2C_iC_A\left(\dis\frac{\alpha_s}{2\pi}\right)^2\dis\frac{\pi^2}{3}\dis\frac{\ln z}{z}\to-8C_iC_A\left(\dis\frac{\alpha_s}{2\pi}\right)^2\dis\frac{\ln z}{z}\dis\frac{1}{\theta_g}\ln\left(\dis\frac{1}{1-\theta_g} \right)\,\Theta(\th_g-\bar\theta)\,,\label{eq:thg-BL}
\end{equation}
with $\bar\theta$ defined such that 
\begin{equation}
 \int_{\bar\theta}^1\frac{\dd \th_g}{\th_g}\ln\left(\frac{1}{1-\th_g}\right)=\frac{\pi^2}{12}\,.
\end{equation}
Numerically, one finds $\bar\th\!\simeq\!0.66$. The step function guarantees that the single logarithmic term from soft non-global configurations is correctly accounted for within our targeted accuracy and without double counting.\footnote{The spirit of the step function is essentially the same as in our treatment of hard collinear emissions via the effective splitting function given by Eq.~\eqref{Phardcoll} Another way of including boundary logarithms without double counting is to use directly Eq.~\eqref{eq:leading-clust-exact} (without step function) since $-\int_0^1\dd\th\ln(1\!-\!\th^2)/\th\!=\!\pi^2/12$.} Such a constraint is also physically expected since boundary logarithms come from the region where $\th_g\!\sim\!1$ by definition. 

Finally, we would like to discuss how this new logarithmic divergence affects the logarithmic counting provided in Sec.~\ref{accuracy}. For the $\th_g$ distribution, we have found two sources of logarithmic enhancement that are either of the form $\ln(\th_g)$ or $\ln(1\!-\!\th_g)$. Since the veto factor in $\Delta(\kappa|a)$ suppresses boundary logarithms, there is only one power of $\ln(1\!-\!\th_g)$ that appears in the $\alpha_s$ expansion of the $\th_g$ distribution and it comes from the $\alpha_s^2$ result given by Eq.~\eqref{eq:thg-BL}. In order to have the correct logarithms at N$^2$DL in front of this single power of $\ln(1\!-\!\th_g)$, it is enough to solely include the first one-loop correction in the running coupling $\alpha_s^2\!\to\!\alpha_s^2(1\!-\!4\alpha_s\beta_0\log(z))$ and the hard-collinear correction at fixed coupling $1/z\!\to\! \Theta(e^{-B_i}\!-\!z)/z$ in Eq.~\eqref{eq:thg-BL}.

%%%%%%%%%%%%%%%%%%%%%%%%%%%%%%
\subsubsection{Comparison between the resummation structure of Soft Drop and Dynamical Grooming}
%%%%%%%%%%%%%%%%%%%%%%%%%%%%%%
The idea of studying the momentum sharing fraction and opening angle of a given splitting in the shower was originally proposed in Ref.~\cite{Larkoski:2014wba}. In this work, the splitting at issue was selected through the Soft Drop procedure, that is, the first branching in the de-clustering sequence that satisfies $z\!>\!z_{\rm cut}\theta^\beta$. These observables, ($z_g,\theta_g$), have been measured experimentally~\cite{Aad:2019vyi,Mulligan:2020cnp} and resummed to modified-leading log~\cite{Tripathee:2017ybi} and next-to-leading log accuracy~\cite{Kang:2019prh}, respectively. An important comment at this point is that Soft Drop observables do exponentiate and, therefore, a N$^p$LL counting applies. Hence, strictly speaking, an apples-to-apples comparison on the resummation structure for Soft Drop and Dynamical Grooming does not exist. 

We have identified one major simplification in the resummation structure of $\theta_g$ when it is defined through the dynamical grooming procedure instead of with Soft Drop: dynamical grooming is free of clustering logarithms. Let us briefly recap how these contributions arise for two correlated emissions~\cite{Delenda:2012mm}. Consider the emission of a gluon, $p_1$, off a hard quark $p_0$ together with a secondary emission, $p_2$, off $p_1$. These two emissions have commensurate angles $\theta_{01}\!\sim\!\theta_{02}$, while their energies (and thus transverse momentum) are strongly ordered $z_2\!\ll\! z_1$ ($k_{t,2}\!\ll\! k_{t,1}$). The C/A algorithm will miss-cluster the secondary gluon as a primary if $\theta_{02}\!<\!\theta_{12}$, with $\theta_{12}$ being the relative distance between the two emissions. Then, if $p_2$ is a real emission it will trigger the Soft Drop condition, even though $z_1\!\gg\!z_2$, and consequently $\theta_{02}\!=\!\theta_g$. In turn, if $p_2$ is virtual, the tagged splitting would be $p_1$ and $\theta_1\!=\!\theta_g$. This mismatch between the real and virtual contributions lead to a tower of logarithms at NLL that were numerically computed, in the large $N_c$-limit, in Ref.~\cite{Kang:2019prh} for $\theta_g$, as defined by Soft Drop, and also discussed in the context of the Lund plane in Ref.~\cite{Lifson:2020gua}. In the Dynamical grooming case, even if some secondary emissions can be `wrongly' pushed by the C/A algorithm into the primary Lund plane, it will never be the hardest given that $z_1\!\gg\!z_2$ and, therefore, $\kappa_1\!=\!z_1\theta_1^a$ will be larger than $\kappa_2\!=\!z_2\theta_2^a$ even if the angles are commensurate. The only effect of these emissions on DyG, beyond N$^2$DL, would be a small contribution to the $p_t$ degradation of the primary branch. 

From the non-global logarithms side, that affect not only $z_g$ but also $k_{t,g}$, we have shown in Sec.~\ref{sec:n2dl}, that they are proportional to $\ln(z_g)$ for Dynamical grooming. In the Soft Drop case, the soft singularity is cured by the definition of the grooming condition. That is, non-global logs enter in the Soft Drop calculation as $\propto\ln(z_{\rm cut}\theta^\beta)$ and thus have a smaller impact that in the DyG option.

Therefore, the cleanest grooming procedure from a theoretical point of view in order to avoid the resummation of both non-global and clustering logarithms at NLL would be to combine the two methods. Then, the grooming procedure would be a two-step process: first, one removes all emissions with $z\!<\!z_{\rm{cut}}$ and then one looks for the hardest one in the Dynamical Grooming sense. This possibility will be further studied in an upcoming publication~\cite{us}.

%%%%%%%%%%%%%%%%%%%%%%%%%%%%%%
\subsubsection{Matching to fixed-order}
\label{sec:matching-sudakov}
%%%%%%%%%%%%%%%%%%%%%%%%%%%%%%
The first leading order matching scheme for Sudakov safe observables was proposed in Ref.~\cite{Larkoski:2013paa}. It is based on constructing a $n$-dimensional IRC safe distribution, that we dub `IRC safe companion', and re-defining the Sudakov safe observable by an appropriate marginalization. In our case, the 2-dimensional IRC safe distribution would be $\dd^2\sigma/\dd z_g\dd \theta_g$, that can be interpreted as the \textit{joint} probability distribution for having a tagged branching with momentum sharing fraction $z_g$ and (normalised) opening angle $\th_g$. Although $z_g$ and $\theta_g$ are Sudakov safe observables by themselves, measuring them simultaneously, i.e. $z_g$ in a given bin of $\th_g$ or vice versa, restores IRC safety. 

To define a matching scheme for a Sudakov safe observable, one then rely on the matching of the IRC safe companion. Such matching can be done for instance in a multiplicative way,
\begin{equation}\label{eq:x-matching}
 \dd^2\sigma_i^{\rm LO+N^2DL}=\frac{\dd^2\sigma_i^{\rm LO}\times\dd^2\sigma_i^{\rm N^2DL}}{\dd^2\sigma_{i,1}^{\rm N^2DL}}\,.
\end{equation}
This formula guarantees that $\dd^2\sigma_i^{\rm LO+N^2DL}$ has exactly the same $\mathcal{O}(\alpha_s)$ coefficient as the LO result and reproduces the resumed calculation in the kinematic region enhanced by large logarithms. Notice also that at LO, $\dd^2\sigma_i^{\rm LO}$ coincides with the primary Lund plane density. We would like to point out that this matching scheme applies to jets with a given flavour $i$. As we shall see, this simplifies the resulting formula as our resummed result depends on the jet's flavour via the Casimir factor of the jet. At LO, such a decomposition is trivial, as explained in Sec.~\ref{sub:matching-ktg}. However, beyond LO, this requires to determine in an IRC safe manner the jet's flavour. This can be done using, for instance, the flavour-$k_t$ clustering algorithm~\cite{Banfi:2006hf}.

Once $\dd^2\sigma_i^{\rm LO+N^2DL}$ is known, the $z_g$ or $\th_g$ distributions are computed by marginalization. In the case of $z_g$, for example, this amounts to  
\begin{equation}\label{eq:zg-matching}
 \frac{1}{\sigma}\frac{\dd \sigma^{\rm LO+N^2DL}}{\dd z_g}=\frac{1}{\sigma}\sum_{i=q,g}\dis\int_0^1\dd z\int_0^1\dd \th\,\frac{\dd^2\sigma_i^{\rm LO+N^2DL}}{\dd z\dd \th}\delta(z-z_g)\,,
\end{equation}
where $\sigma$ is the inclusive jet cross-section. An important feature of Sudakov safe observables is hidden behind the apparent simplicity of Eq.~\eqref{eq:zg-matching}. In fact, not all matching schemes for the IRC safe companion lead to a well-defined integral after marginalization. For instance, choosing an additive matching,
\begin{equation}\label{eq:-match}
  \dd^2\sigma_i^{\rm LO+N^2DL}=\left[\dd^2\sigma_i^{\rm LO}-\dd^2\sigma_i^{\rm N^2DL,(1)}\right]+\dd^2\sigma_i^{\rm N^2DL}\,,
\end{equation}
induces a collinear divergent term (the one inside the bracket) which is not cured by the Sudakov. On the contrary, the integral in Eq.~\eqref{eq:zg-matching} is well defined because the Sudakov factor in $\dd^2\sigma_i^{\rm N^2DL}$ shields the $\th\!=\!0$ logarithmic divergence. 

We now turn the concrete implementation of Eq.~\eqref{eq:zg-matching} used in this paper. In principle we could compute $\dd^2\sigma_i^{\rm LO}$ using MadGraph as in our matched calculation of $k_{t,g}$. However, we decide here to take another path that we find more enlightening from the physics point of view and, at the same time, easier to implement numerically. Namely, in the small jet radius limit that we consider throughout this paper, it is possible to provide an explicit analytic expression for $\dd^2\sigma_i^{\rm LO}$. Up to powers of $\th_g$ corrections, it reads
\begin{equation}
\label{eq:match-lo-zg-theta-g}
 \frac{\dd^2\sigma_i^{\rm LO}}{\dd z_g\dd \th_g}\simeq\sigma_{0,i}\frac{2\alpha_sC_i}{\pi}\frac{1}{\th_g}\overline P_i(z_g)+\mathcal{O}(\th_g^n)\,,
\end{equation}
where $\overline P_i$ is the symmetrized splitting function of a parton $i$: $\overline P_i(z)\!=\!P_i(z)\!+\!P_i(1-z)$, summed over all decay channels. Matching our resummed distribution to this form of the LO result is then straightforward as it amounts to replace $2C_i\Theta(e^{-B_i}\!-\!z)/z$ in $\widetilde P(z,\th)$ (Eq.~\eqref{eq:Pbranch-full}) by $2C_i\overline P_i(z)$.\footnote{See also Refs.~\cite{Larkoski:2015lea,Tripathee:2017ybi} for a similar trick in the calculation of the Soft Drop $z_g$ distribution.} 

\begin{figure}
\centering
  \includegraphics[scale=0.6,page=1]{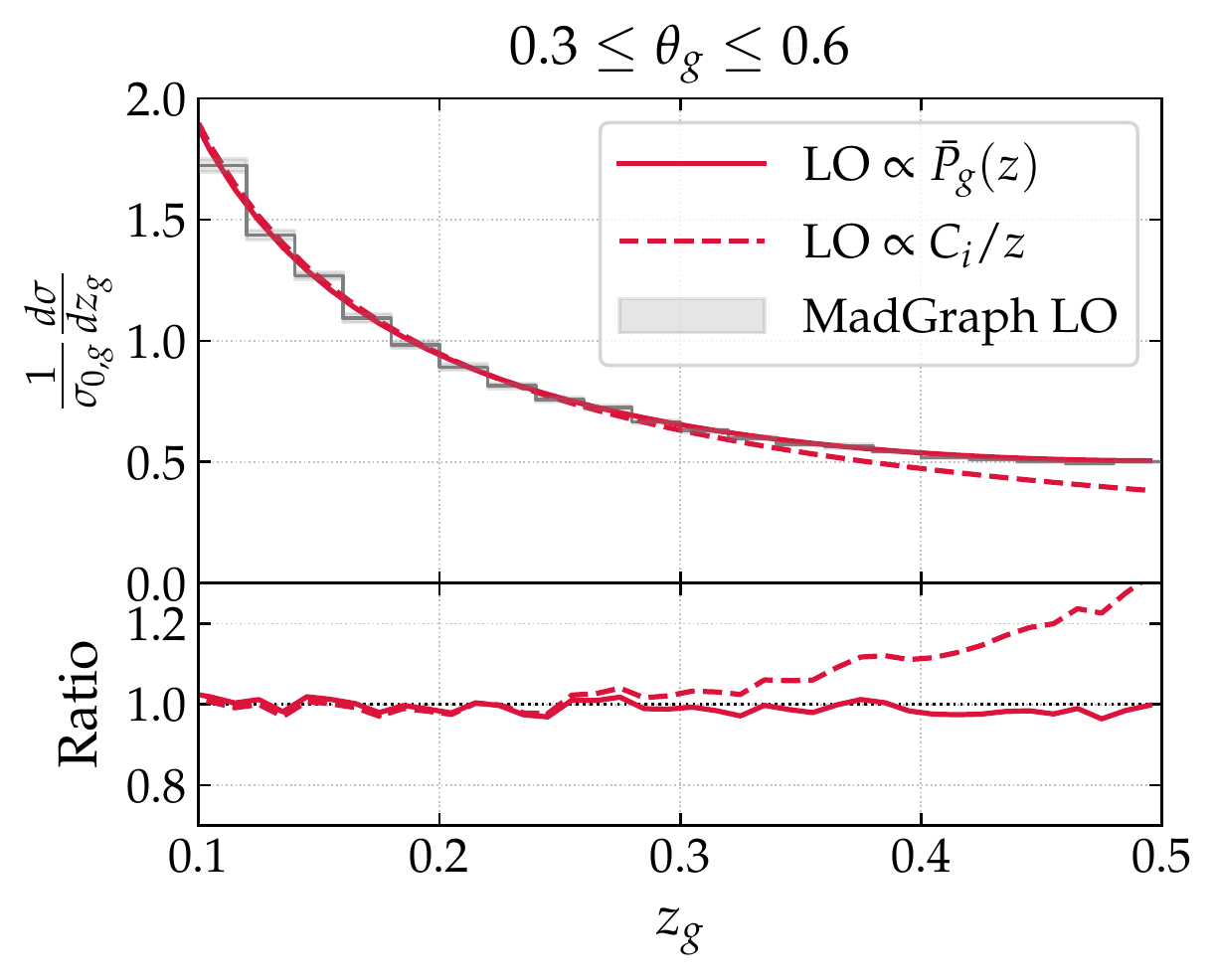} \includegraphics[scale=0.6,page=2]{plot-zgthg.pdf}
  \caption{Left: $z_g$ differential distribution at leading order in $\alpha_s$ for the gluon channel in a bin of $\theta_g$ computed in three different ways: with MadGraph (gray), using the splitting function as a proxy for the leading order result, see Eq.~\eqref{eq:match-lo-zg-theta-g}, either with the leading order expression of $P_i(z)$ (solid, red) or taking the soft limit (dashed, red). Right: same as left panel but for $\theta_g$.
  }
  \label{fig:madgraph-vs-zg}
\end{figure}

In Fig.~\ref{fig:madgraph-vs-zg}, we show a comparison between the exact result of $\dd^2\sigma_i^{\rm LO}/\dd z_g\dd \th_g$ obtained through MadGraph and the approximation given by Eq.~\eqref{eq:match-lo-zg-theta-g}. In addition, we compare these two options with the one that we get after replacing the full symmetrized splitting function by its soft limit in Eq.~\eqref{eq:match-lo-zg-theta-g}. More concretely, we have computed $\dd^2\sigma_i^{\rm LO}$ for the gluon channel in the three ways that we have just mentioned and show the $z_g$ and $\theta_g$-projections for two bins of $\theta_g$ and $z_g$, respectively. We see that Eq.~\eqref{eq:match-lo-zg-theta-g} matches the exact leading order result while the soft limit of the splitting function is not enough to accurately reproduce the MadGraph output throughout the whole range of $z_g$. Similar conclusions can be drawn by analysing the $\theta_g$-projection of $\dd^2\sigma_i^{\rm LO}$. Both for $z_g$ and $\theta_g$, the deviation of Eq.~\eqref{eq:match-lo-zg-theta-g} to the exact result remains below $5\%$. We have deliberately chosen a low $p_t$ bin to ensure that using $\overline P_i$ as a proxy for $\dd^2\sigma_i^{\rm LO}$ is valid in the regime in which the ALICE measurement has been recorded.  

We decide to normalize the $z_g$ and $\theta_g$ distributions to the Born level cross-section, $\sigma_0$, in contrast to the $k_{t,g}$ case where the NLO correction to the inclusive jet cross section, $\sigma_1$, was taken into account. For $k_{tg}$, the reason we included this correction is to account for the $C_1$ term, but in principle, a LO matched cross-section can safely be normalized by the Born cross-section without spoiling the targeted accuracy. For Sudakov safe cross-sections such as $z_g$ and $\th_g$, the question of finding a matching scheme which makes possible the inclusion of the NLO correction to $\sigma$ in a consistent way with respect to the resummation counterpart is in fact closely related to the $C_1$ problem that we proceed to tackle.

%%%%%%%%%%%%%%%%%%%%%%%%%%%%%%%%%
\paragraph{The $C_1$ term in Sudakov safe observables.} 
%%%%%%%%%%%%%%%%%%%%%%%%%%%%%%%%%
The impossibility to perturbatively expand Sudakov safe observables in powers of $\alpha_s$ invalidates the definition of the $C_1$ term given by Eq.~\eqref{eq:def-C1}. Up to now, the question on whether a `$C_1$-like' contribution to the resummation exists for this type of observables has not even been addressed in the literature. In this paper, we would like to outline a new, dedicated matching scheme for Sudakov safe observables. The main difference with respect to the original proposal by the authors in Ref.~\cite{Larkoski:2013paa} is to rely on an IRC safe cross-section which is one-dimensional. This IRC safe companion is built from the Sudakov safe distribution with an additional cut on the kinematic variable that is integrated out. More explicitly, for the dynamically groomed $z_g$ distribution, one defines the IRC safe cumulative distribution $\Sigma(z_g|\th_{\rm cut})$ using the same grooming procedure, but with an additional cut-off on the angle of the splitting, $\th_g$, that is denoted $\th_{\rm cut}$. Then, our matching formula is:
\begin{equation}\label{eq:new-match}
 \Sigma^{\rm LO+N^2DL}(z_g)=\Sigma^{\rm LO+N^2DL}(z_g|\th_{\rm cut})+\Sigma^{\rm N^2DL}(z_g)-\Sigma^{\rm N^2DL}(z_g|\th_{\rm cut})\,,
\end{equation}
with $\Sigma^{\rm LO+N^2DL}(z_g|\th_{\rm cut})$ defined using multiplicative matching as in Eq.~\eqref{x-match}.\footnote{The generalization of Eq.~\eqref{eq:new-match} to $\theta_g$ and beyond leading order is straightforward.} Notice that the normalization of $\Sigma^{\rm LO+N^2DL}(z_g)$ to $\sigma_0\!+\!\sigma_1$ is ensured. At first sight, this formula seems to depend on the value of $\th_{\rm cut}$. Yet, it is not the case provided that $\th_{\rm cut}$ is low enough. To understand this, recall that the Sudakov form factor in $\Sigma^{\rm N^2DL}(z_g)$ provides a natural cut-off, $\theta_c$, on the angular integration of the branching kernel that scales at DLA with $\th_c\!\simeq\!\exp(-1/\sqrt{\abar a})$ see Eq.~\eqref{eq:theta-cut} (also in Ref.~\cite{Mehtar-Tani:2019rrk}). Thus, if $\th_{\rm cut}$ is chosen smaller than $\th_c$, we expect that
\begin{equation}
\label{eq:new-match-1}
 \Sigma^{\rm N^2DL}(z_g)= \Sigma^{\rm N^2DL}(z_g|\th_{\rm cut})
\end{equation}
for $z_g\!\gtrsim\!\th_{\rm cut}^a$. Consequently, far from the resummation region, the dominant term in Eq.~\eqref{eq:new-match} is $\Sigma^{\rm LO+N^2DL}(z_g|\th_{\rm cut})$ which correctly captures the large $z_g\sim 0.5$ domain. On the contrary, in the small $z_g$ limit (but not smaller than $\th_{\rm cut}^a$), we obtain:
\begin{equation}
\label{eq:new-match-2}
 \Sigma^{\rm LO+N^2DL}(z_g)\simeq \Sigma^{\rm N^2DL}(z_g)+\alpha_s C_1(\th_{\rm cut})\Sigma^{\rm N^2DL}(z_g)\,.
\end{equation}
The second term in the previous equation is a correction to our resummed formula, which looks like a $C_1$ term. It depends on the value of $\th_{\rm cut}$ as a reminiscence of the non IRC safety of the $z_g$ distribution. To see that, one notices that up to a constant factor, $C_1(\th_{\rm cut})\!\propto\!\ln(\th_{\rm cut})$. Since $\th_{\rm cut}$ cannot be larger than $\th_c$, the $C_1$ correction is actually of order $\mathcal{O}(\alpha_s/\sqrt{\alpha_s a})\!=\!\mathcal{O}(\sqrt{\alpha_s/a})$, at least. There are two interesting features in this scaling behavior. First, the appearance of the square root of $\alpha_s$ is characteristic of Sudakov safe quantities. Second, we observe how $C_1$ can become sizeable for $a\ll1$. The latter point reminds us that introducing and ad-hoc parameter, $\th_{\rm cut}$, in the matching scheme comes with some associated difficulties. In short, from the resummation point of view, one would like $\th_{\rm cut}$ to be as small as possible such that Eq.~\eqref{eq:new-match-1} holds. However, the smallness of $\th_{\rm cut}$ can lead to a sizeable $C_1$ correction in Eq.~\eqref{eq:new-match-2}, thus spoiling the correct asymptotic limit. A clear trade-off exists and the concrete value of $\th_{\rm cut}$ in the proposed matching scheme and its applicability to phenomenological applications deserve further investigation. 

\subsubsection{Results}
\begin{figure}
\centering
  \includegraphics[scale=0.6,page=1]{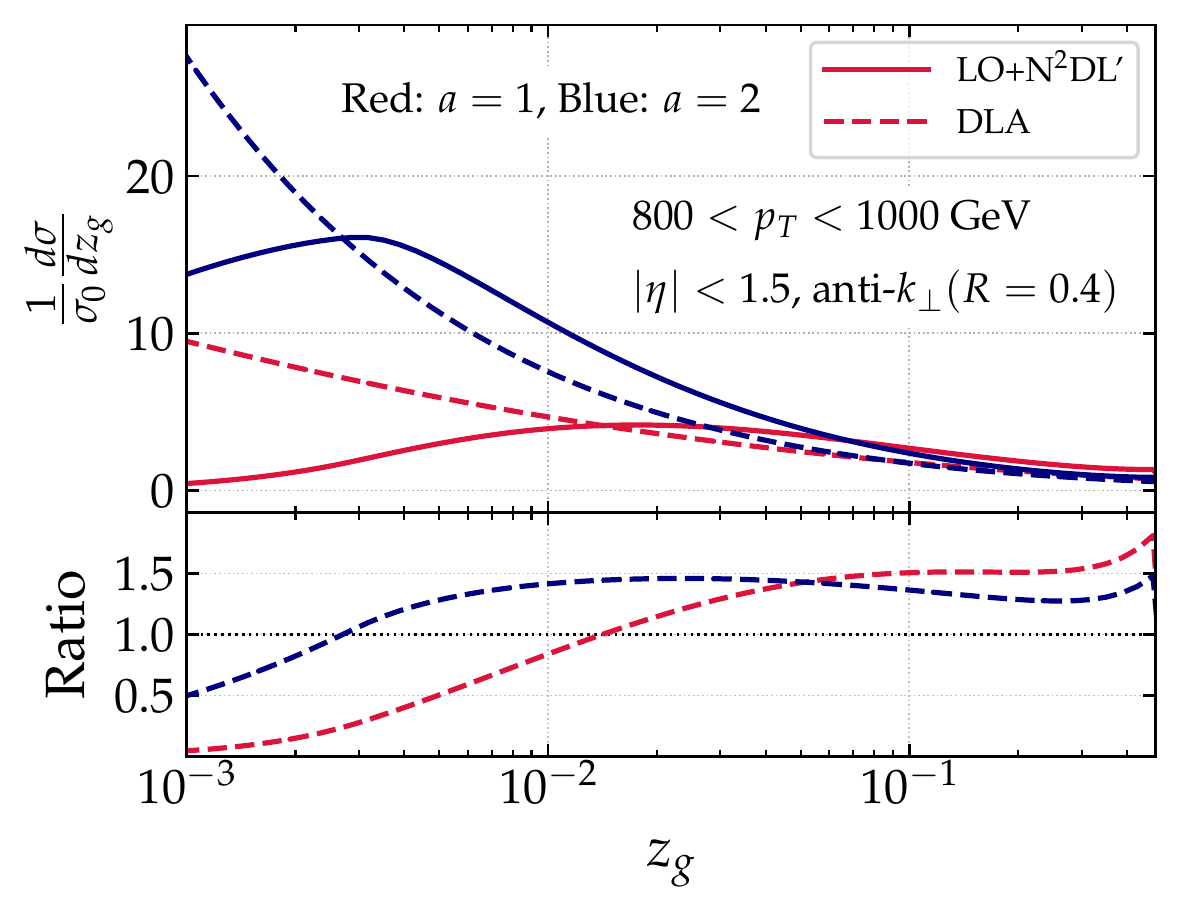} \includegraphics[scale=0.6,page=2]{plot-zgthg-final.pdf}
  \caption{Left: $z_g$-distribution for $a\!=\!1$ (red) and $a\!=\!2$ (blue) at two different accuracies, DLA and LO+N$^2$DL' and their ratio. Right: $\theta_g$-distribution for $a\!=\!1$ (red) and $a\!=\!2$ (blue) with and without boundary logarithms in the LO+N$^2$DL' result and their ratio.
  }
  \label{fig:zg-thetag-analytics}
\end{figure}

Following the reasoning of the $k_{t,g}$ section, we would like to highlight some features of the $z_g$ and $\theta_g$ analytic distributions before moving on to the comparison against Monte-Carlo simulations and ALICE preliminary data. In the left panel of Fig.~\ref{fig:zg-thetag-analytics}, we quantify the difference between the double-log calculation of the $z_g$ distribution and the LO+N$^2$DL'. The purpose of this figure is to highlight the deviation of the $z_g$-distribution from the $1/z$ behavior when higher orders in the resummation are included. Indeed, we have shown in Eq.~\eqref{eq:z-cut}, the $z_g$-distribution has a dynamically generated cut-off at $z_{\rm cut}\!\sim\!e^{-a/\bar{\alpha}}$. For $z\!>\!z_{\rm cut}$, it was shown in Ref.~\cite{Mehtar-Tani:2019rrk} that the distribution falls off like the soft limit of the Altarelli--Parisi splitting function. We observe that NDL and N$^2$DL contributions such as the running of the strong coupling or the presence of non-global logarithms induce an almost $50\%$ difference with respect to the DLA result. This should be taken into account when interpreting the experimental data specially when searching for modifications in heavy-ion measurements~\cite{Sirunyan:2017bsd,Adam:2020kug}.

In the right panel of Fig.~\ref{fig:zg-thetag-analytics}, we asses the impact of the boundary logarithms in the $\theta_g$ distribution that were discussed in Sec.~\ref{sec:boundary_logs}. As expected, they only matter at large angles and diverge when $\theta_g\to1$. Their contribution amounts to a $10\!-\!20\%$ and is therefore mandatory to include them if a theory-to-data comparison is aimed. 

%%%%%%%%%%%%%%%%%%%%%%%%%%%%%
%%%%%%%%%%%%%%%%%%%%%%%%%%%%%
\section{Phenomenology at LHC energies}
\label{sec:numerics}
%%%%%%%%%%%%%%%%%%%%%%%%%%%%%
%%%%%%%%%%%%%%%%%%%%%%%%%%%%%

The analytic calculations that we have presented above rely mainly on two approximations: the narrow jet limit and the use of the Altarelli--Parisi splitting function in the matching as a proxy for the leading order result in the case of $z_g$ and $\theta_g$. In order to evaluate the goodness of such approximations, we test our analytic results against parton level simulations in realistic experimental conditions together with the available experimental data. The results are presented for two values of $a$ in the dynamical grooming condition: $a\!=\!1$ and $a\!=\!2$. The reason why we do not consider smaller values of $a$ and, in particular, $a\!=\!0.1$ as done in the ALICE measurement, is because non-perturbative phenomena, beyond the reach of our analytic pQCD calculation, notably affect dynamically groomed observables when $a\!<\!1$. In addition, we utilise the N$^2$DL' prescription on the resummation side.  

%%%%%%%%%%%%%%%%%%
\subsection{Analytics vs. Monte-Carlo parton level}
\label{sec:analytic}
%%%%%%%%%%%%%%%%%%%%%%%%%%%%%
In this section, we compare our analytic calculation for ($k_{t,g}, z_g, \theta_g$) to parton level simulations of dijet events with Pythia8.235~\cite{Sjostrand:2007gs} and Herwig7.1.2~\cite{Bellm:2015jjp}. For the latter we use both the default angular-ordered shower that we denote `Herwig7-AO'~\cite{Gieseke:2003rz} and the dipole-type shower, `Herwig7-Dip', based on Ref.~\cite{Platzer:2009jq}. Given that non-perturbative effects are reduced when going to larger $p_t$, we study an experimental setup, that lies within the ATLAS capabilities~\cite{Aad:2019vyi}, where the comparison to pQCD calculations are deemed to be cleaner. The centre of mass energy is set to $\sqrt s\!=\!13$~TeV, jets are clustered with the anti-k$_\perp$ algorithm with $R\!=\!0.4$ and re-clustered with Cambridge/Aachen using FastJet3.3.1~\cite{Cacciari:2011ma}. The analysis is performed on those jets that satisfy: $800\!<\!p_t\!<\!1000$~GeV and $|\eta|\!<\!1.5$. For the Monte-Carlo studies, we used the DyG condition given by Eq.~\eqref{eq:hardness}.

\begin{figure}
\centering
  \includegraphics[width=\textwidth]{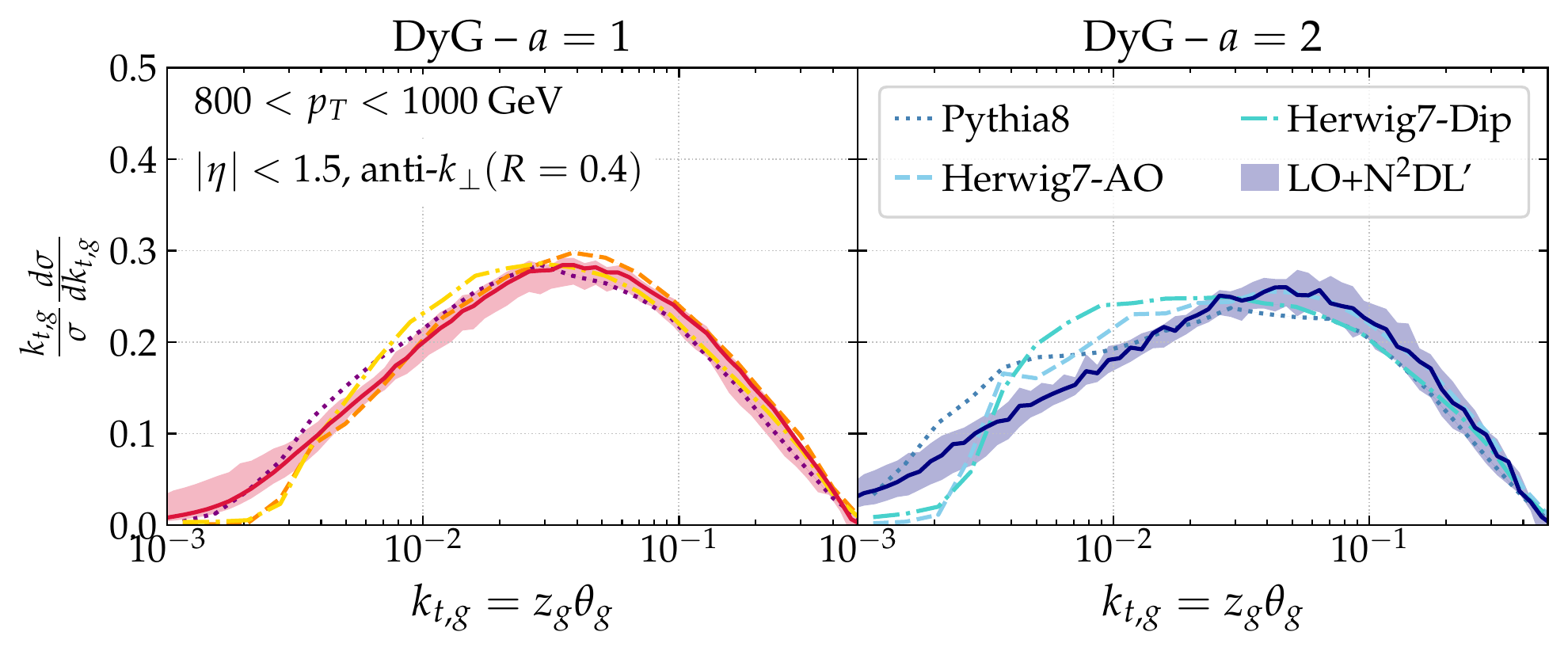}
  \caption{Theory to parton level comparison of $k_{t,g}$ in the ATLAS-like scenario for $a\!=\!1$ (left) and $a\!=\!2$ (right) in the dynamical grooming condition, see Eq.~\eqref{eq:hardness}.
  }
  \label{fig:ktg-mc-parton-highpt}
\end{figure}

We show the comparison between our analytic result and parton level Monte-Carlo simulations with Pythia8 and Herwig7 for the relative transverse momentum of the dynamically groomed splitting in Fig.~\ref{fig:ktg-mc-parton-highpt}. A crucial point to understand the fixed-order dominated regime, i.e.\ $k_{t,g}\gtrsim10^{-1}$, is that in the default setting of both Monte-Carlos, the parton shower starts off a leading-order $2\!\to\!2$ matrix element.\footnote{The $\alpha_s$ counting might be misleading at this point. Notice that what we refer to as LO in the analytic result is actually a NLO contribution in the sense that it enters at order $\alpha_s$, i.e. we consider $p+p\to jj$ at NLO.} Therefore, the fixed-order contribution to the $k_{t,g}$ distribution is exactly zero for these event generators. Hence, at large $k_{t,g}$, an exact agreement between our analytic result, dominated by the exact NLO matrix element, and the Monte-Carlos, where $k_{t,g}$ is exclusively generated by the parton shower, is not expected. Nevertheless, both event generators use at the very least the leading order Altarelli--Parisi splitting functions. As we have discussed in Sec.~\ref{sec:matching-sudakov}, the use of the full splitting function in the resummation (or, similarly, in the parton shower) effectively reproduces the fixed-order result in the narrow jet limit. Then, part of the higher-order corrections to the Born-level process are incorporated through the splitting function in the parton shower. This can explain the nice agreement between the analytic result and the Monte-Carlo curves for $k_{t,g}\!\gtrsim\!10^{-1}$.

On the resummation side, both showers in Herwig are in relatively good agreement and notably differ from Pythia. This is, a priori, rather counterintuitive based on the nature of the three parton showers that we are evaluating. The dipole-style Herwig shower and the Pythia one use a Catani-Seymour like~\cite{Catani:1996vz} dipole map,  transverse momentum ordering and implement a local recoil scheme. In turn, Herwig7-AO evolves through $1\!\to\!2$ splittings by means of a generalised angular variable and employs a global recoil scheme. Based on these general arguments, one would expect Herwig7-Dip and Pythia showers to deliver somewhat similar results. The opposite behavior observed in Fig.~\ref{fig:ktg-mc-parton-highpt} points out to a more general Pythia-to-Herwig difference rather than to the showers themselves. We identify the scale at which the QCD shower is stopped to be the source of this discrepancy. In fact, in the default setting, Pythia imposes a relatively low infra-red cut-off of $0.5$~GeV, while Herwig uses a more conservative scale of $\sim 1$~GeV that is common to both showers~\cite{Bellm:2016rhh}. Then, more phase-space is available for radiation in the Pythia case and this leads to the differences observed on the low $k_{t,g}$ side in Fig.~\ref{fig:ktg-mc-parton-highpt}. Thus, we conclude that the small-$k_{t,g}$ part of the differential distribution is sensitive to the way the infra-red is handled and thus to hadronisation. This point will be further emphasized in the following section. Moreover, any higher order term contained in the Monte-Carlo and not present at N$^2$DL in the resummation, e.g. energy-momentum conservation, would affect the low $k_{t,g}$ regime.

In what concerns the comparison between MCs and the analytic result, an enhancement at low $k_{t,g}$ values appears. A very similar trend is observed in Fig.~\ref{fig:pythia-recoil} of App.~\ref{sec:appendix-b} where we evaluate the impact of removing the $1\!-\!z$ in the hardness variable $\kappa$ (see Eq.\eqref{eq:hardness}) for the Monte-Carlo results. We remind the reader that this factor is a sub-leading, non-logarithmic correction in our analytic calculation at N$^2$DL accuracy. However, this mismatch in the $\kappa$ definition on the analytics and the Monte-Carlos amounts to a $\sim 10\%$ difference on the low $k_{t,g}$ regime and is, therefore, partly responsible for the bump at $k_{t,g}\!\sim\!10^{-3}$.
 
\begin{figure}
\centering
  \includegraphics[width=\textwidth]{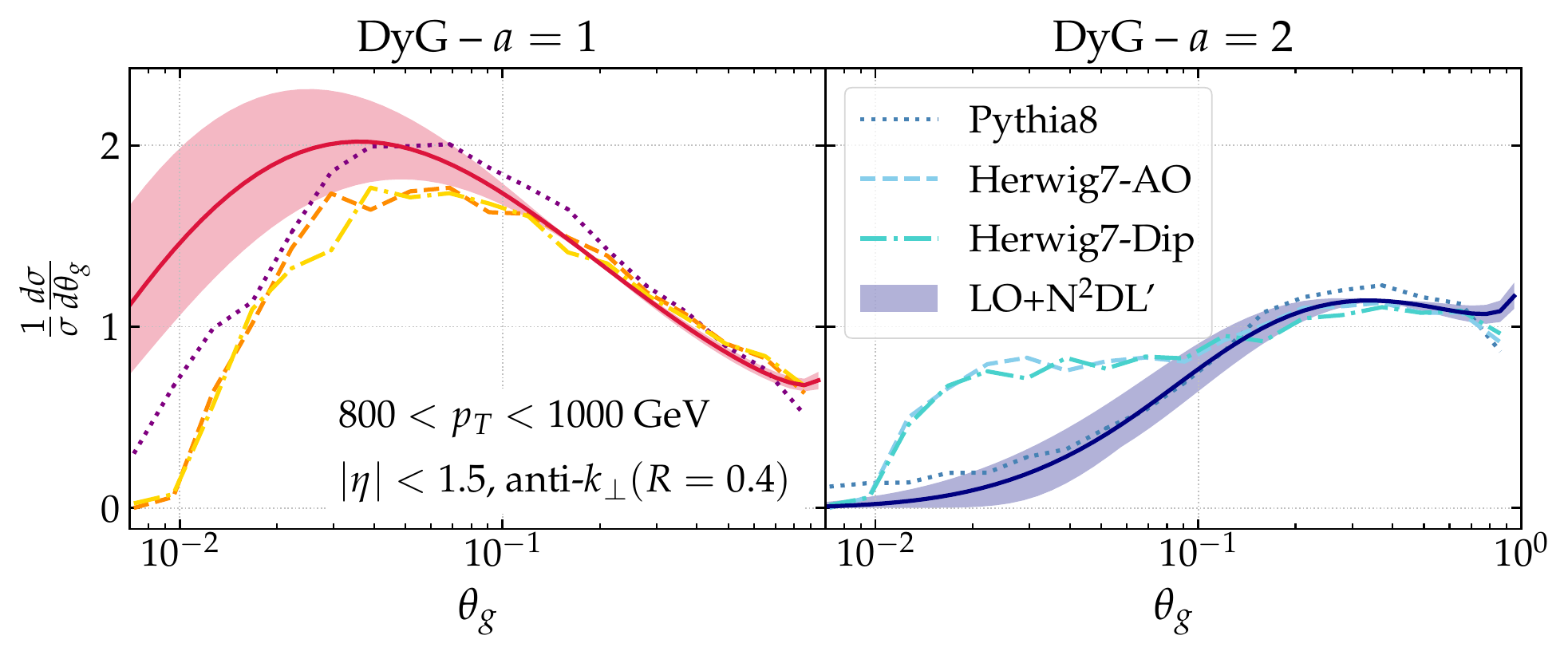}
  \caption{Same as Fig.~\ref{fig:ktg-mc-parton-highpt} but for $\theta_g$.
  }
  \label{fig:thetag-mc-parton-highpt}
\end{figure}

The distribution of the opening angle $\theta_g$ is displayed in Fig.~\ref{fig:thetag-mc-parton-highpt}. On the Monte-Carlo side, we again observe a strong sensitivity to the momentum scale at which the shower is cut-off. In fact, no significant differences are observed between the angle tagged by Herwig7-AO and Herwig7-Dip, thus indicating a strong dominance of the choice of IR-scale. In particular, we have checked that the small angle bump for $a\!=\!2$ disappears if the infra-red scale is lowered down in Herwig. We have pinned down two other sources for the analytic-to-Monte-Carlo discrepancy. The first one concerns again the $1\!-\!z$ factor in the definition of $\kappa$. In Fig.~\ref{fig:pythia-recoil} in App.~\ref{sec:appendix-b}, we quantify this effect and observe that the small $\theta_g$ region can be distorted by $\sim 20\!-\!40\%$, depending on the value of $a$. The enhancement of $\theta_g$ at large angles in the MCs with respect to the analytic curve could be explained by the $\mathcal{O}(\th_g^n)$ terms that we have neglected all along our calculation, both on the resummation side and also on the matching procedure where power suppressed terms in the fixed order result were ignored. On the other hand, it is not guaranteed that the branching kernels implemented in the Monte-Carlos recover the exact soft and large angle limit. Then, we conclude that the disagreement between the analytic calculation of the $\theta_g$-distribution and the parton shower results can be understood as a result of the choice of the infra-red scale, the finite $z$ corrections in the $\kappa$ definition and the jet clustering procedure, being the first one the strongest effect. 

\begin{figure}
\centering
  \includegraphics[width=\textwidth]{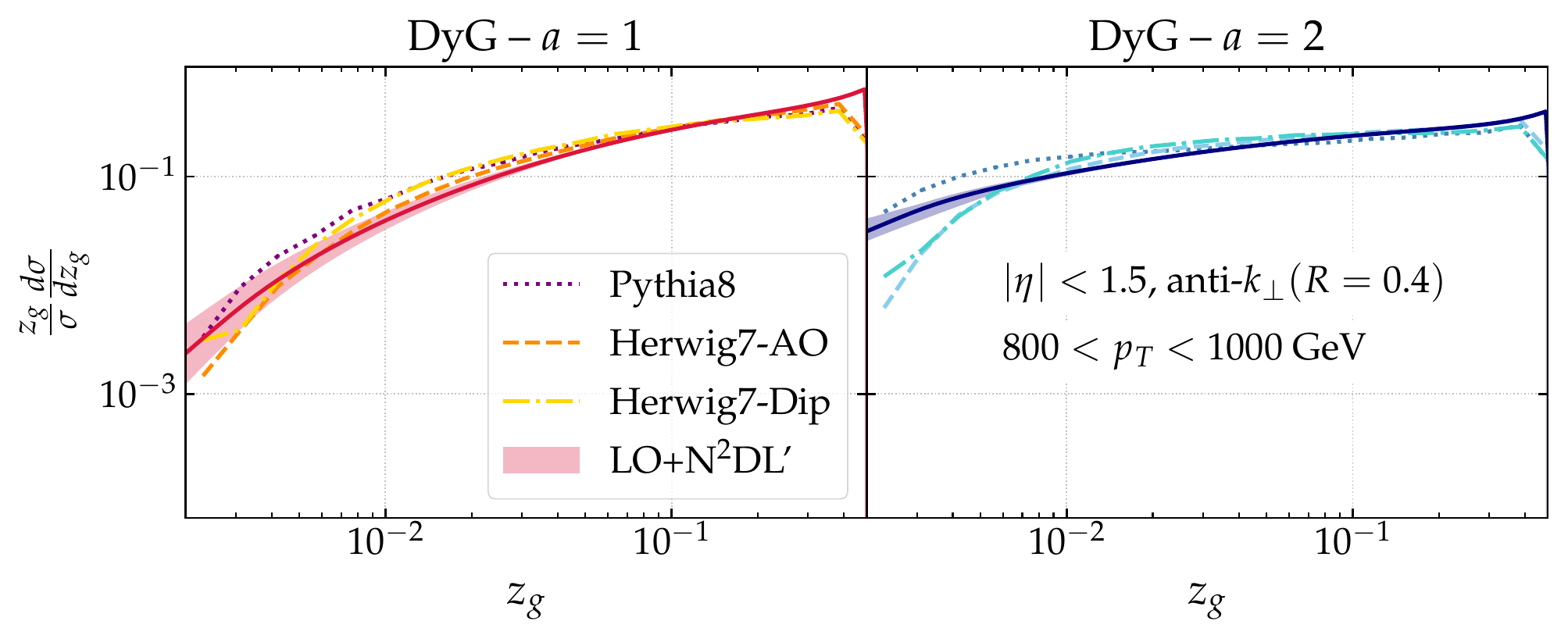}
  \caption{Same as Fig.~\ref{fig:thetag-mc-parton-highpt} but for $z_g$.
  }
  \label{fig:zg-mc-parton-highpt}
\end{figure}

Finally, the momentum sharing splitting fraction $z_g$ is presented in Fig.~\ref{fig:zg-mc-parton-highpt}. We clearly observe the presence of the dynamically generated cut-off that separates the fall-off of the distributions from the flattening. The latter starts earlier for $a\!=\!2$ given that $z_{\rm cut}$ is smaller in this case, see Eq.\eqref{eq:z-cut}. The agreement between the theory calculation and the Monte-Carlos is reasonable in the intermediate regime of $10^{-2}\!<\!z_g\!<\!10^{-1}$. Outside this interval, the recoil factor in the hardness definition is responsible for both the depletion at large $z_g$ in the MC's with respect to the analytic as well as for the excess at small-$z$, as can be seen in App.~\ref{sec:appendix-b}. In addition, the reduced phase space for emissions at infra-red scales in Herwig as compared to Pythia is manifest and further studied in App.~\ref{sec:appendix-d}. 

%%%%%%%%%%%%%%%%%%%%%%%%%%%%
\subsection{Comparison to preliminary ALICE data}
\label{sec:data-to-theory}
%%%%%%%%%%%%%%%%%%%%%%%%%%%%
In this section, we scrutinise our analytic calculation against preliminary measurements of DyG observables~\cite{Mulligan:2020cnp,Ehlers:2020piz}. The experimental analysis is performed at $\sqrt s\!=\!5.02$~TeV on jets clustered with the anti-k$_\perp$ algorithm with $R\!=\!0.4$. An important aspect is that only charged jets that satisfy $60\!<\!p^{ch}_t\!<\!80$~GeV and $|\eta|\!<\!0.5$ are considered. In the analytic calculation, no distinction is made between charged and neutral particles, thus a method to translate the charged $p_t$ bin of the data into its full counterpart has to be designed. A few possibilities exist to tackle this problem. One could be to identify via Monte-Carlo simulations the transverse momentum bin that, after subtracting the neutral component, yields most of the jets in the $60\!<\!p_t\!<\!80$~GeV interval. We have carried out this exercise and found that 80\% of the jets that fulfil $64.5\!<\!p_t \!<\!102.5$~GeV, fall in the $60\!<\!p^{ch}_t\!<\!80$~GeV category. Another option is to absorb this $p_t$-shift from charged to full jets into a non-perturbative factor that also accounts for the effect of hadronization, initial state radiation and multi-parton interactions. This latter is the approach followed in this paper (see also Ref.~\cite{Lifson:2020gua}) and works as follows. We perform the analytic calculation in the same $p_t$-bin as where the experimental measurement is carried out, i.e.\ $60\!<\!p_t\!<\!80$~GeV. Then, to construct the non-perturbative factor two samples have to be generated with Monte-Carlo. The first one includes all non-perturbative effects and only charged particles are clustered. Then, the dynamical grooming analysis is performed on the charged jets that satisfy $60\!<\!p^{ch}_t\!<\!80$~GeV. The second sample is generated as parton level events. Again, we select jets in the same $p_t$-bin as the theoretical calculation, i.e. $60\!<p_t\!<80$~GeV, without any charge selection. Then, our non-perturbative factor is defined as the ratio of the charged hadron level sample and the parton level one, both computed in the same transverse momentum bin. These results are plotted in App.~\ref{sec:appendix-d}, where further details on the role of the infra-red cutoff in the parton shower are provided. Finally, the theoretical results are multiplied by this phenomenological parameter. Then, the theoretical error band includes both the uncertainty of the non-perturbative factor and the analytic uncertainties characterised in the end of Sec.~\ref{sec:sanity-checks}. We label these results as `LO+N$^2$DL'+NP'.

\begin{figure}
\centering
  \includegraphics[width=\textwidth]{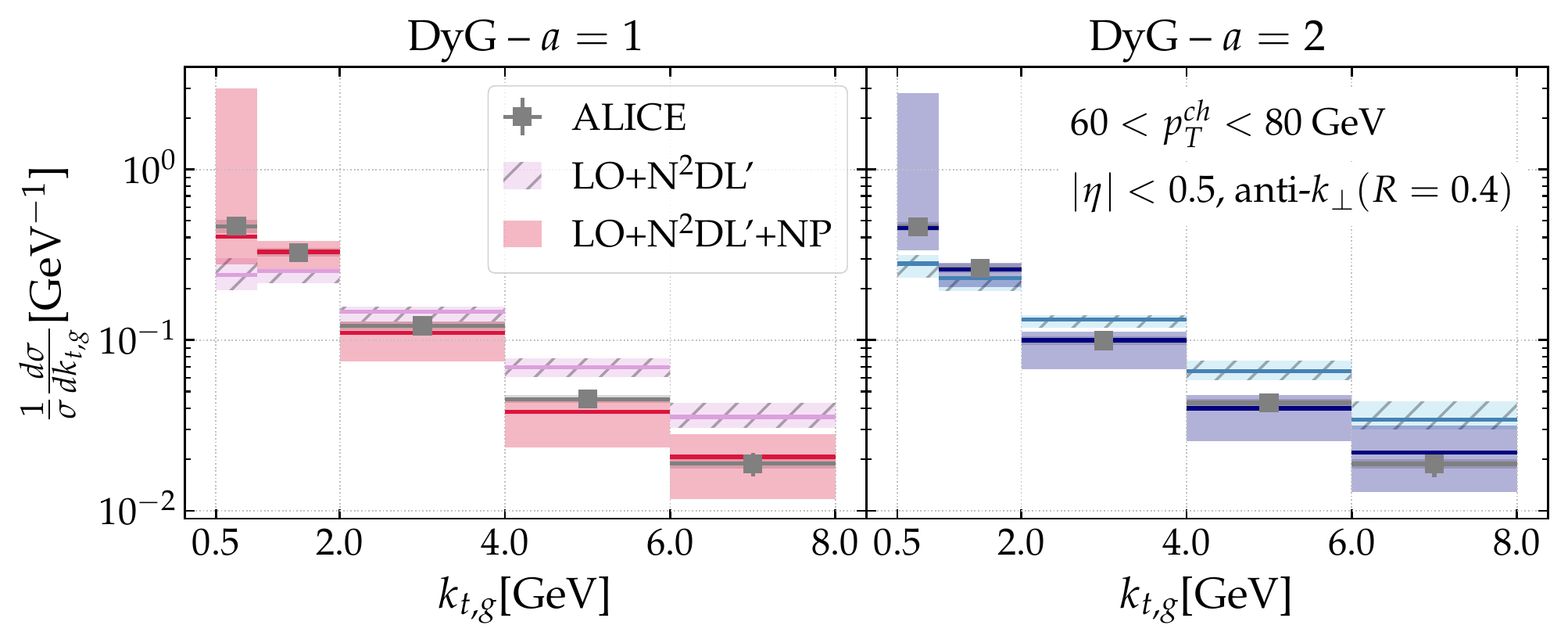}
  \caption{Comparison between the analytic result obtained in this paper and the preliminary ALICE data~\cite{Ehlers:2020piz} of $k_{t,g}$.}
  \label{fig:ktg-theory-alice}
\end{figure}

Like in the previous section, we start the discussion with the $k_{t,g}$ distribution shown in Fig.~\ref{fig:ktg-theory-alice}. To begin with, an important remark is that a mismatch exists between how the $k_{t,g}$ is defined in the analytic calculation, i.e.\ $k_{t,g}\!=\!z_gR_g$,\footnote{Notice that in the previous section we have used $k_{t,g}\!=\!z_g\theta_g$ and now we replace $\theta_g$ by $R_g\!=\!\theta_gR$ to follow ALICE convention.} and in ALICE's measurement, where $k_{t,g}\!=\!z_g\sin(R_g)p_t$. As we have already mentioned, $p_t$ degradation is ignored in our calculation because the hardest branching is located on the primary Lund plane at our degree of accuracy. Then, to accommodate the $p_t$ dependence of the experimental definition we simply multiply our analytic result by the lower bound of the $p_t$ bin, $60$~GeV in this case. We have checked that changing this factor by any other value within the explored $p_t$-bin leads to variations that are well covered by our uncertainty bands. The functional form of the angular dependence of the two $k_{t,g}$ definitions is a bit more delicate. This is so because considering $\sin(R_g)$ instead of $R_g$ brings additional power-corrections in the calculation that we have so far neglected based on our narrow jet approximation. Besides this fact, we observe in Fig.~\ref{fig:ktg-theory-alice} that the data points are only described by the theoretical calculation if the non-perturbative factor, displayed in Fig.~\ref{fig:np-factors}, is included.
In particular, its role is most prominent for the first bin and generates a large uncertainty. This is yet another manifestation of the different methods that Pythia and Herwig employ to regulate the infra-red sector in the shower. We also provide the Monte-Carlo to data comparison in App.~\ref{sec:appendix-e} and find that all three explored setups result into $10\!-\!20\%$ deviations with respect to the data both for $a\!=\!1$ and $a\!=\!2$. Therefore, we conclude that the agreement between the analytic result presented in this paper and ALICE data is satisfactory in spite of the low $p_t$ selection where hadronization effects are very large. 

\begin{figure}
\centering
  \includegraphics[width=\textwidth]{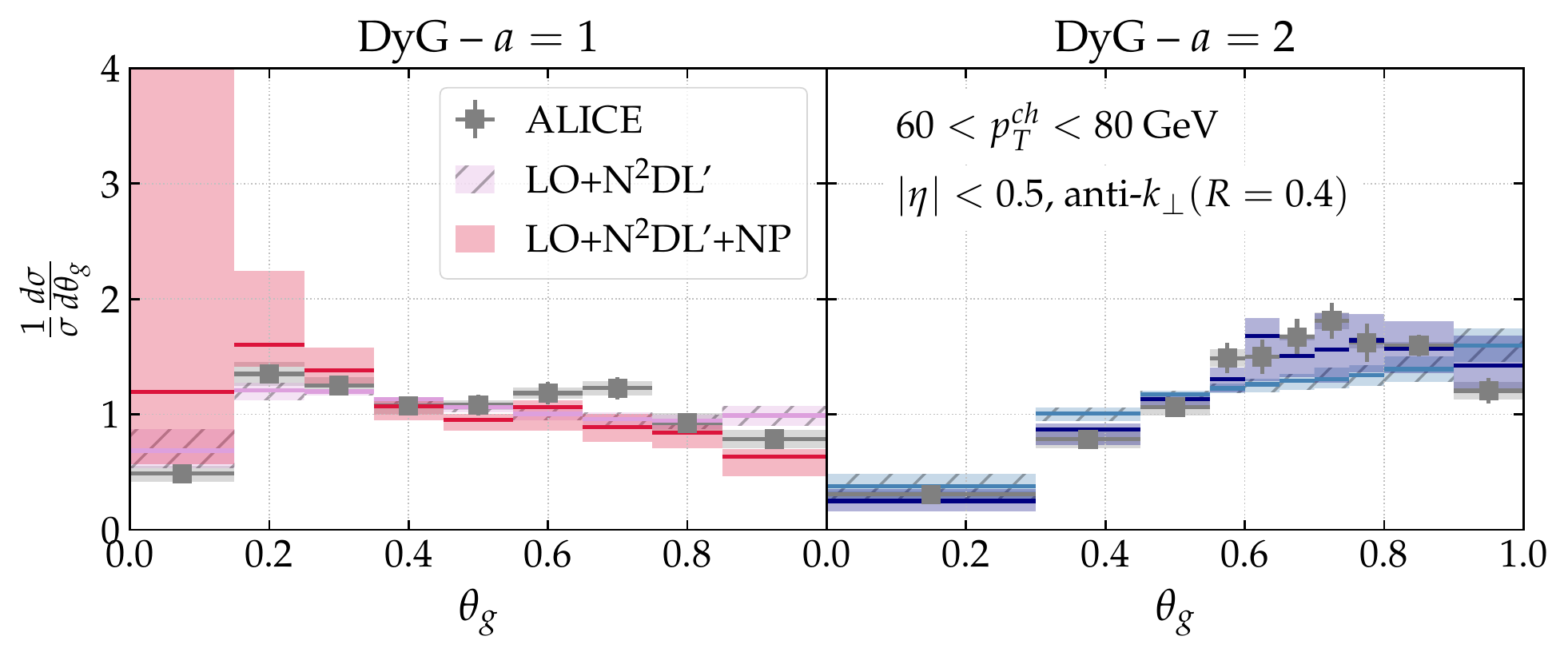}
  \caption{Same as Fig.~\ref{fig:ktg-theory-alice} for $\theta_g$. The experimental data was obtained from~\cite{Mulligan:2020cnp}.}
  \label{fig:thetag-theory-alice}
\end{figure}

Turning to $\theta_g$, represented in Fig.~\ref{fig:thetag-theory-alice}, we observe that the ad-hoc non-perturbative factor completely dominates the result in the first bin for $a\!=\!1$. The $a$-dependence of these results is also interesting from the point of view of missing terms in the analytic calculation. Indeed, we see a deficit in the analytic result for splittings with angles $\theta_g\!>\!0.6$ that disappears for the $a\!=\!2$. We have already mentioned that the $g_{nm}$ coefficients are inversely proportional to $a$ and thus higher-order terms would impact less the $a\!=\!2$ result than the $a\!=\!1$ one. In addition, the discrepancy appears in the region where the power-suppressed terms that we have neglected all along the calculation based on the narrow jet approximation, i.e.\ contributions of $\mathcal{O}(\theta_g^n)$, may matter. An obvious way to confirm this hypothesis would be either to include them or, alternatively, to make a jet radius scan of this observable on the experimental side. 

\begin{figure}
\centering
  \includegraphics[width=\textwidth]{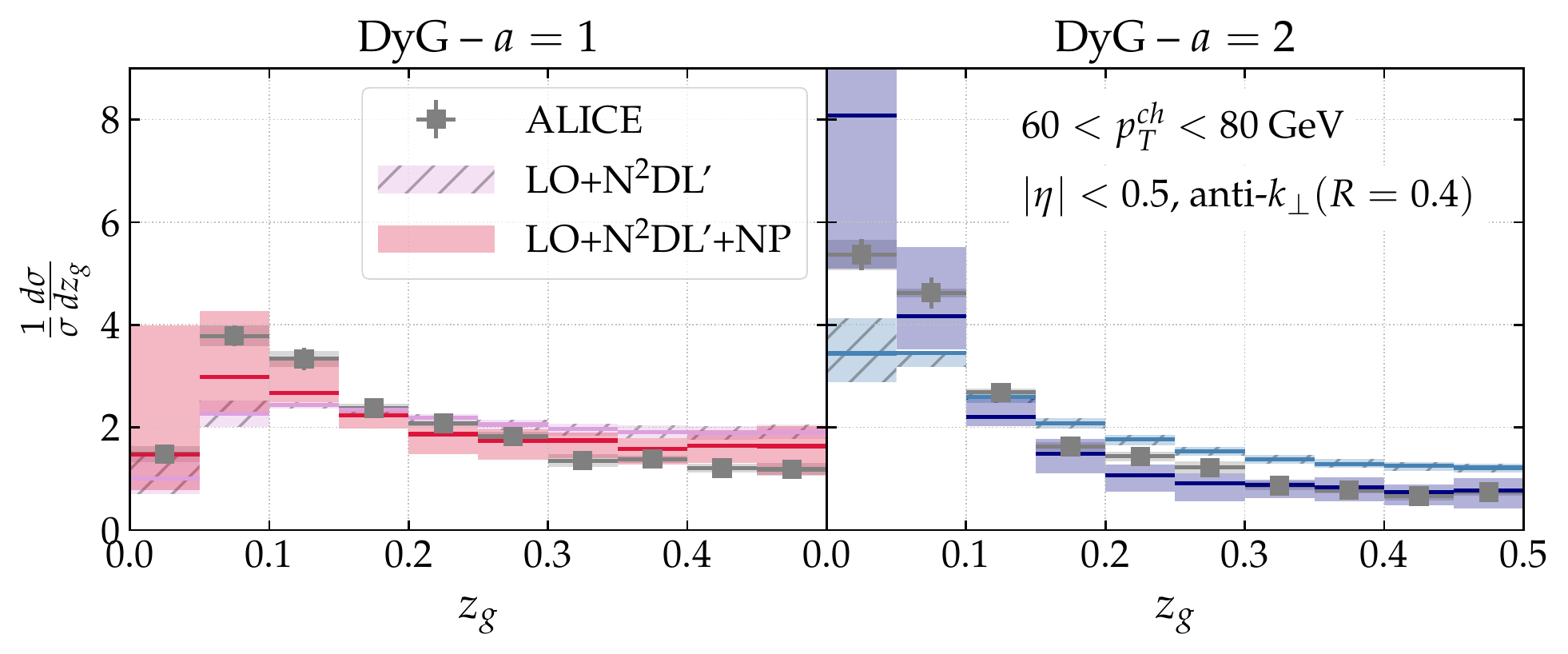}
  \caption{Same as Fig.~\ref{fig:thetag-theory-alice} for $z_g$.}
  \label{fig:zg-theory-alice}
\end{figure}

To end up this phenomenological section, we present the theory-to-data comparison for the momentum sharing fraction in Fig.~\ref{fig:zg-theory-alice}. In this scenario, the non-perturbative factor is prominent at small $z_g$, but has a relatively mild effect for $z_g\!>\!0.2$. In fact, in this interval both the LO+N$^2$DL' and LO+N$^2$DL'+NP results agree with the data within uncertainties. Clearly, this observable together with $k_{t,g}$ are the ones for which our theoretical calculation results provides the best description of the experimental measurements. This is a remarkable result given that the Sudakov nature of this observable complicates its theoretical analysis in many different ways, as we have discussed throughout the paper. The Monte-Carlo results are also consistent with the experimental measurement (see App.~\ref{sec:appendix-e}).

\section{Conclusions and outlook}
The work presented in this paper follows the current global effort towards a precise theoretical description of jet substructure observables that will help us to deepen our understanding of the space-time evolution of QCD jets, both in vacuum and in heavy-ion collisions. In particular, we have focused on three substructure observables defined on the splitting selected by the dynamical grooming method, that is, the hardest one in the jet tree. These three observables are the momentum sharing fraction $z_g$, the opening angle $\theta_g$ and the relative transverse momentum $k_{t,g}$. Out of the three, $k_{t,g}$ is the only infra-red and collinearly safe observable. Then, the definition of logarithmic accuracy for $z_g$ and $\theta_g$ is far from trivial and we extensively discuss a possible approach to tackle the problem that consists in defining the accuracy of the Sudakov safe observable in terms of the cumulative distribution of an IRC safe companion. In this way, we demonstrate that the resummation of dynamically groomed observables does not exponentiate, in general, and that a logarithmic counting at the level of the cumulative distribution is therefore better suited. Further, we present all the necessary ingredients to reach next-to-next-to-double logarithmic accuracy in the narrow jet limit. This includes: (i) the resummation of collinear logarithms arising from the running of the coupling and the hard-collinear correction to the splitting function, (ii) the contribution of non-global logarithms at $\mathcal O(\alpha_s^2)$ and (iii) the $\mathcal O(\alpha_s^2$) contribution of boundary logarithms in the case of $\theta_g$. Remarkably, neither clustering logarithms nor multiple emissions affect these dynamically groomed distributions. We make use of a matching scheme that naturally includes the $C_1$ term and allows us to recover the exact leading-order result, computed with MadGraph, for large values of $k_{t,g}$. On the Sudakov safe cases, we employ the full splitting function as a proxy for the fixed-order result and propose a dedicated matching scheme that depends on an ad-hoc cut-off. All in all, we achieve LO+N$^2$DL accuracy in our analytic computation.

The analytic framework is tested against three different parton-level Monte-Carlo simulations at high-$p_t$: Pythia8, angular-ordered, and dipole-style showers in Herwig. In the resummation dominated regime, we find a strong dependence of the Monte-Carlo results on the transverse momentum cut-off at which the parton shower stops. This value is smaller for Pythia ($\sim0.5$ GeV) than for Herwig ($\sim1$ GeV) by default and, as such, more radiation is allowed in the former case. The analytic result regulates the infra-red singularity through a freezing of the running coupling below $1$~GeV and, therefore, allows for emissions with all possible transverse momenta. Due to this fact, it is reasonable that the analytic result is closer to Pythia than to Herwig. Even in the region in which the analytic calculation reduces to the fixed-order contribution, the parton shower is fully responsible of the Monte-Carlo results, given that a leading-order matrix element is implemented by default. Despite this mismatch, an overall good agreement is found for the three jet substructure observables that we attribute to the use of the full splitting function in the parton showers that, as we have stated, generates the correct matrix element in the narrow jet limit. 

Our last step is to compare the analytic predictions against the preliminary ALICE data. To do so, we supplement the perturbative results with a non-perturbative factor extracted from Monte-Carlo simulations with Pythia and Herwig that accounts for the use of charged tracks, hadronisation and underlying event. This ingredient is particularly crucial in this experimental setup given that only low-$p_t$ jets, i.e.\ $p^{ch}_t\!\lesssim\!200$~GeV are measurable by the ALICE detector. In fact, it dominates the theoretical prediction in the lower bins of the $k_{t,g}, z_g$ and $\theta_g$ distributions. A quantitative description of $k_{t,g}$ and $z_g$ is achieved up to $5\!-\!10\%$ deviations in some bins. In the case of $\theta_g$, we find deviations of up to $15\%$ in the moderately large angle region. This is precisely the regime in which we have less confidence in our result considering that we have neglected all power suppressed terms of the type $\mathcal O (\theta_g^n)$.  

The natural extension of this work is to go beyond the small-$R$ limit and develop a numerical routine to account for the resummation of non-global and boundary logarithms. Notice that, as far as we are aware, the latter has yet never been achieved in the literature. On the collinear side of the resummation, we could include the recoil of the hard branch, make use of the NLO splitting function together with higher orders in the running coupling. Further, promoting our leading-order matching to NLO is straightforward from a conceptual point of view in the case of $k_{t,g}$ and would improve the agreement with data/parton level MC simulations both at low and high $p_t$ and reduce the theoretical uncertainties. Regarding the Sudakov safe observables, we would like to understand the feasibility of the dedicated matching scheme proposed in this paper and, specifically, quantify its dependence on the ad-hoc cut off. Finally, it would be insightful to compare these analytic results for dynamically groomed observables to the newly developed parton showers that aim at achieving perturbative control beyond leading double logarithmic accuracy and leading color~\cite{Dasgupta:2018nvj,Hamilton:2020rcu,Dasgupta:2020fwr}. From an experimental perspective, these theoretical efforts would highly benefit from a high-$p_t$ measurement where non-perturbative corrections are deemed to be milder.  

Beyond the possible improvements of the $p+p$ calculation, we would like to discuss two further extensions in terms of collision systems: $e+p$ (and eventually $e+A$), relevant for the future Electron Ion Collider and heavy-ion collisions. The former, despite the relatively low number of constituents per jet~\cite{Arratia:2019vju}, provides a cleaner environment with respect to $p+p$ given that both multi-parton interactions and the underlying event will have a residual effect. On the heavy-ion side, dynamically groomed observables can be used to characterise the properties of an in-medium parton shower. In particular, the $\theta_g$ distribution can be used to experimentally measure the critical resolution angle of the Quark-Gluon Plasma~\cite{us2}, while deviations at large $k_{t,g}$ from respect to the vacuum baseline could suggest rare, hard scatterings between the propagating parton and the medium. 

\section*{Acknowledgements}
We thank Leticia Cunqueiro, Raymond Ehlers and James Mulligan for clarifications on the experimental aspects of the measurements reported in~\cite{Mulligan:2020cnp,Ehlers:2020piz}. Further, we acknowledge Keith Hamilton, Silvia Ferrario-Ravasio, Yacine Mehtar-Tani, Gavin Salam, Gregory Soyez and Konrad Tywoniuk for many insightful discussions on different aspects of this calculation and their feedback on the manuscript. A.S.O.’s work was supported by the European Research Council (ERC) under the European Union’s Horizon 2020 research and innovation programme (grant agreement No. 788223, PanScales). A.T. is supported by a Starting Grant from Trond Mohn Foundation (BFS2018REK01), the MCnetITN3 H2020 Marie Curie Initial Training
Network, contract 722104, and wishes to thank the Institut de Physique Th\'{e}orique (IPhT) and Gregory Soyez for the hospitality. PC is also grateful to the IPhT for its support during the early stages of this work.

%%%%%%%%%%%%%%%%%%%%%%%%%%%%%%%%%
\appendix
\section{Details of analytic calculations at N$^2$DL'}
\label{sec:appendix-a}
%%%%%%%%%%%%%%%%%%%%%%%%%%%%%%%%%

The purpose of this section is to provide the analytic Sudakov form factor needed to achieve N$^2$DL' accuracy as explained in Sec.~\ref{sec:n2dl}. To that end, we need to perform the following integral:
\begin{equation}\label{eq:app-sudakov}
 -\ln(\Delta_i(\kappa|a))=\frac{2C_i}{\pi}\int_0^{e^{-B_i}}\frac{ \dd z}{z}\int_0^1\frac{\dd\theta}{\theta}\alpha_s(\mu_K z\theta Q)\Theta(z\theta^a-\kappa)\,,
 \end{equation}
with $
 \alpha_s(k_t)=\alpha_s^{2\ell}(k_t)\Theta(k_t-\mu_{\rm fr})+\alpha_s(\mu_{\rm fr})\Theta(\mu_{\rm fr}-k_t)$ and $Q=p_{t,\rm jet}R$. In the perturbative domain $k_t>\mu_{\rm fr}$, the two-loop running coupling $\alpha_s^{2\ell}(k_t)$ is given by Eq.~\eqref{eq:2loop-rc-def} with the reference $\alpha_s$ value taken at the renormalization scale $\mu_RQ$. 
 
 We define the following dimensionless variable: $\lambda_\kappa=2\alpha_s\beta_0\ln(\kappa)$, $\lambda_B=-2\alpha_s\beta_0 B_i$, $\lambda_K = 2\alpha_s\beta_0\ln(\mu_K/\mu_R)$ and $\lambda_{\rm fr}=2\alpha_s\beta_0\ln(\mu_{\rm fr}/(\mu_K Q))$, and the following functions:
\begin{align}
 W(x)&=-x+x\ln(x)\,,\\
 V(x)&=\ln(x)(2+\ln(x))\,.
\end{align}
Due to the presence of the constant $k_t=\mu_{\rm fr}$ line in the $(z,\th)$ phase space, the formulae depend on whether $a$ is larger or smaller than $1$. 

\paragraph{Case $a>1$.}
If $\lambda_\kappa\ge\lambda_{\rm fr}$:
\begin{align}
  -\ln\Delta_i(\kappa|a)=&\frac{C_i}{2\pi\alpha_s\beta_0^2}\left[W(1+\lambda_K+\lambda_B)+\frac{1}{a-1}\Big(W(1+\lambda_K +\lambda_\kappa)\right.\nonumber\\
  &\hspace{4.5cm}\left.\left.-aW\left(1+\lambda_K+\frac{a-1}{a}\lambda_B+\frac{\lambda_\kappa}{a}\right)\right)\right]\nonumber\\
  &+\frac{C_i\beta_1}{4\pi\beta_0^3}\left[V(1+\lambda_K+\lambda_B)+\frac{1}{a-1}\Big(V(1+\lambda_K +\lambda_\kappa)\right.\nonumber\\
  &\hspace{4.5cm}\left.\left.-aV\left(1+\lambda_K+\frac{a-1}{a}\lambda_B+\frac{\lambda_\kappa}{a}\right)\right)\right]\nonumber\\
  &-\frac{C_iK}{4\pi^2\beta_0^2}\left[\ln(1+\lambda_K+\lambda_B)+\frac{1}{a-1}\Big(\ln(1+\lambda_K +\lambda_\kappa)\right.\nonumber\\
  &\hspace{4.5cm}\left.\left.-a\ln\left(1+\lambda_K+\frac{a-1}{a}\lambda_B+\frac{\lambda_\kappa}{a}\right)\right)\right]\,.
\end{align}
If $\lambda_{\rm \kappa}\le a\lambda_{\rm fr}+(1-a)\lambda_B$:
\begin{align}
  -\ln\Delta_i(\kappa|a)=&\frac{C_i}{2\pi\alpha_s\beta_0^2}\left[-\lB+\lfr+(1+\lB+\lnu)\ln\left(\frac{1+\lB+\lnu}{1+\lfr+\lnu}\right)\right]\nonumber\\
  &+\frac{C_i\beta_1}{4\pi\beta_0^3} \left[V(1+\lB+\lnu)-\frac{2\lB-2\lfr+2(1+\lB+\lnu)\ln(1+\lfr+\lnu)}{1+\lfr+\lnu}\right.\nonumber\\
  &\hspace{9cm}+\ln^2(1+\lnu+\lfr)\Big]\nonumber\\
  &-\frac{C_iK}{4\pi^2\beta_0^2}\left[\frac{\lfr-\lB}{1+\lfr+\lnu}+\ln\left(\frac{1+\lB+\lnu}{1+\lfr+\lnu}\right)\right]\nonumber\\
  &+\frac{2C_R\alpha_s(\mu_{\rm fr})}{4\pi\alpha_s^2\beta_0^2}\left[\frac{(1-a)\lB^2+2a\lB\lfr-2\lB\lk-a\lfr^2+\lk^2}{2a}\right]\,.
\end{align}

If $a\lambda_{\rm fr}+(1-a)\lambda_B<\lambda_\kappa<\lambda_{\rm fr}$:
\begin{align}
  -\ln\Delta_i(\kappa|a)=&\frac{C_i}{2\pi\alpha_s\beta_0^2}\left[\frac{\lfr-\lk}{1-a}\ln(1+\lnu+\lfr)+W(1+\lnu+\lB)\right.\nonumber\\
  &\hspace{1cm}\left.\frac{1}{a-1}\left(W(1+\lambda_K +\lfr)-aW\left(1+\lambda_K+\frac{a-1}{a}\lambda_B+\frac{\lambda_\kappa}{a}\right)\right)\right]\nonumber\\
  &+\frac{C_i\beta_1}{4\pi\beta_0^3} \left[\frac{2(\lfr-\lk)}{1-a}\frac{\left(1+\ln(1+\lnu+\lfr)\right)}{1+\lambda_K+\lfr}+V(1+\lnu+\lB)\right.\nonumber\\
  &\hspace{1cm}\left.\frac{1}{a-1}\left(V(1+\lambda_K +\lfr)-aV\left(1+\lambda_K+\frac{a-1}{a}\lambda_B+\frac{\lambda_\kappa}{a}\right)\right)\right]\nonumber\\
  &-\frac{C_iK}{4\pi^2\beta_0^2}\left[\frac{\lfr-\lk}{1-a}\frac{1}{1+\lambda_K+\lfr}+\ln(1+\lnu+\lB)\right.\nonumber\\
  &\hspace{1cm}\left.\frac{1}{a-1}\left(\ln(1+\lambda_K +\lfr)-a\ln\left(1+\lambda_K+\frac{a-1}{a}\lambda_B+\frac{\lambda_\kappa}{a}\right)\right)\right]\nonumber\\
  &+\frac{C_R\alpha_s(\mu_{\rm fr})}{2\pi\alpha_s^2\beta_0^2}\frac{(\lfr-\lk)^2}{2(a-1)}\,.
\end{align}
\paragraph{Case a=1.} It is straightforward, albeit tedious, to take the limit $a\rightarrow1$ of the previous formulae. If $\lk>\lfr$,
\begin{align}
  -\ln\Delta_i(\kappa|1)=&\frac{C_i}{2\pi\alpha_s\beta_0^2}\left[-\lB+\lk+(1+\lB+\lnu)\ln\left(\frac{1+\lB+\lnu}{1+\lk+\lnu}\right)\right]\nonumber\\
  &+\frac{C_i\beta_1}{4\pi\beta_0^3} \left[V(1+\lB+\lnu)-\frac{2\lB-2\lk+2(1+\lB+\lnu)\ln(1+\lk+\lnu)}{1+\lk+\lnu}\right.\nonumber\\
  &\hspace{9cm}+\ln^2(1+\lnu+\lk)\Big]\nonumber\\
  &-\frac{C_iK}{4\pi^2\beta_0^2}\left[\frac{\lk-\lB}{1+\lk+\lnu}+\ln\left(\frac{1+\lB+\lnu}{1+\lk+\lnu}\right)\right]\,,
\end{align}
and if $\lk<\lfr$
\begin{align}
-\ln\Delta_i(\kappa|1)=&\frac{C_i}{2\pi\alpha_s\beta_0^2}\left[-\lB+\lfr+(1+\lB+\lnu)\ln\left(\frac{1+\lB+\lnu}{1+\lfr+\lnu}\right)\right]\nonumber\\
  &+\frac{C_i\beta_1}{4\pi\beta_0^3} \left[V(1+\lB+\lnu)-\frac{2\lB-2\lfr+2(1+\lB+\lnu)\ln(1+\lfr+\lnu)}{1+\lfr+\lnu}\right.\nonumber\\
  &\hspace{9cm}+\ln^2(1+\lnu+\lfr)\Big]\nonumber\\
  &-\frac{C_iK}{4\pi^2\beta_0^2}\left[\frac{\lfr-\lB}{1+\lfr+\lnu}+\ln\left(\frac{1+\lB+\lnu}{1+\lfr+\lnu}\right)\right]\nonumber\\
   &+\frac{C_R\alpha_s(\mu_{\rm fr})}{2\pi\alpha_s^2\beta_0^2}\left[(\lfr-\lk)(\lB-\lfr)+\frac{1}{2}(\lfr-\lk)^2\right]\,.
\end{align}

\paragraph{Case $a<1$.} For completeness, we provide also the formulae when $a<1$. For some values of $\lk$, they can be related to the expression in the $a>1$ case. If $\lk\ge a\lambda_{\rm fr}+(1-a)\lambda_B$, $\Delta(\kappa|a<1)$ is given by the expression of $\Delta(\kappa|a>1)$ when $\lk>\lfr$. In a similar way, when $\lk<\lfr$, $\Delta(\kappa|a<1)$ is given by the expression of $\Delta(\kappa|a>1)$ when $\lambda_{\rm \kappa}\le a\lambda_{\rm fr}+(1-a)\lambda_B$. In the remaining $\kappa$ domain, $\lfr\le\lk\le a\lambda_{\rm fr}+(1-a)\lambda_B$, one finds
 \begin{align}
  -\ln\Delta_i(\kappa|a)=&\frac{C_i}{2\pi\alpha_s\beta_0^2}\left[\frac{(1-a)\lB+a\lfr-\lk}{a-1}\ln(1+\lnu+\lfr)+W(1+\lnu+\lB)\right.\nonumber\\
  &\hspace{1cm}\frac{1}{a-1}(W(1+\lambda_K +\lk)-aW(1+\lambda_K+\lfr))\Big]\nonumber\\
  &+\frac{C_i\beta_1}{4\pi\beta_0^3} \left[\frac{2((1-a)\lB+a\lfr-\lk)}{a-1}\frac{\left(1+\ln(1+\lnu+\lfr)\right)}{1+\lambda_K+\lfr}+V(1+\lnu+\lB)\right.\nonumber\\
  &\hspace{1cm}\frac{1}{a-1}(V(1+\lambda_K +\lk)-aV(1+\lambda_K+\lfr))\Big]\nonumber\\
  &-\frac{C_iK}{4\pi^2\beta_0^2}\left[\frac{(1-a)\lB+a\lfr-\lk}{a-1}\frac{1}{1+\lambda_K+\lfr}+\ln(1+\lnu+\lB)\right.\nonumber\\
  &\hspace{1cm}\frac{1}{a-1}(\ln(1+\lambda_K +\lk)-a\ln(1+\lambda_K+\lfr))\Big]\nonumber\\
  &+\frac{C_R\alpha_s(\mu_{\rm fr})}{2\pi\alpha_s^2\beta_0^2}\left[\frac{((1-a)\lB+a\lfr-\lk)^2}{2a(1-a)}\right]\,.
\end{align}

\paragraph{The non-global term.} On top of Eq.~\eqref{eq:app-sudakov}, the Sudakov factor receives a contribution from soft non-global emissions of the form
\begin{align}
 -\ln(\Delta^{\rm NG}_i(\kappa|a))&=2C_iC_A\left(\frac{\alpha_s}{2\pi}\right)^2\frac{\pi^2}{3}\int_0^{e^{-B}}\frac{\dd z}{z}\ln\left(\frac{1}{\mu_K z}\right)\int_0^1\dd\th\,\Theta(z\theta^a-\kappa)\\
 &=C_iC_A\left(\frac{\alpha_s}{2\pi}\right)^2\frac{\pi^2}{3}\ln^2(\mu_K\kappa)+C_iC_A\left(\frac{\alpha_s}{2\pi}\right)^2\frac{\pi^2}{3}\Big[-\ln^2\left(\mu_K e^{-B}\right)\nonumber\\
 &\left.+2a^2\left(1-\kappa^{1/a} e^{B/a}\right)+2a\left(\ln(\kappa\mu_K)-\ln\left(e^{-B}\mu_K\right)\right)\kappa^{1/a} e^{B/a}\right]\,.
 \end{align}
In this expression, we have separated the single-log term from the pure $\alpha_s$ or power corrections in $\kappa$ which can be neglected to N$^2$DL accuracy.

%%%%%%%%%%%%%%%%%%%%%%%%%%%%%%%%%
%%%%%%%%%%%%%%%%%%%%%%%%%%%%%%%%%
\section{The size of finite $z$ corrections in the definition of $\kappa$}
\label{sec:appendix-b}
%%%%%%%%%%%%%%%%%%%%%%%%%%%%%%%%%
%%%%%%%%%%%%%%%%%%%%%%%%%%%%%%%%%
\begin{figure}
\centering
  \includegraphics[width=\textwidth]{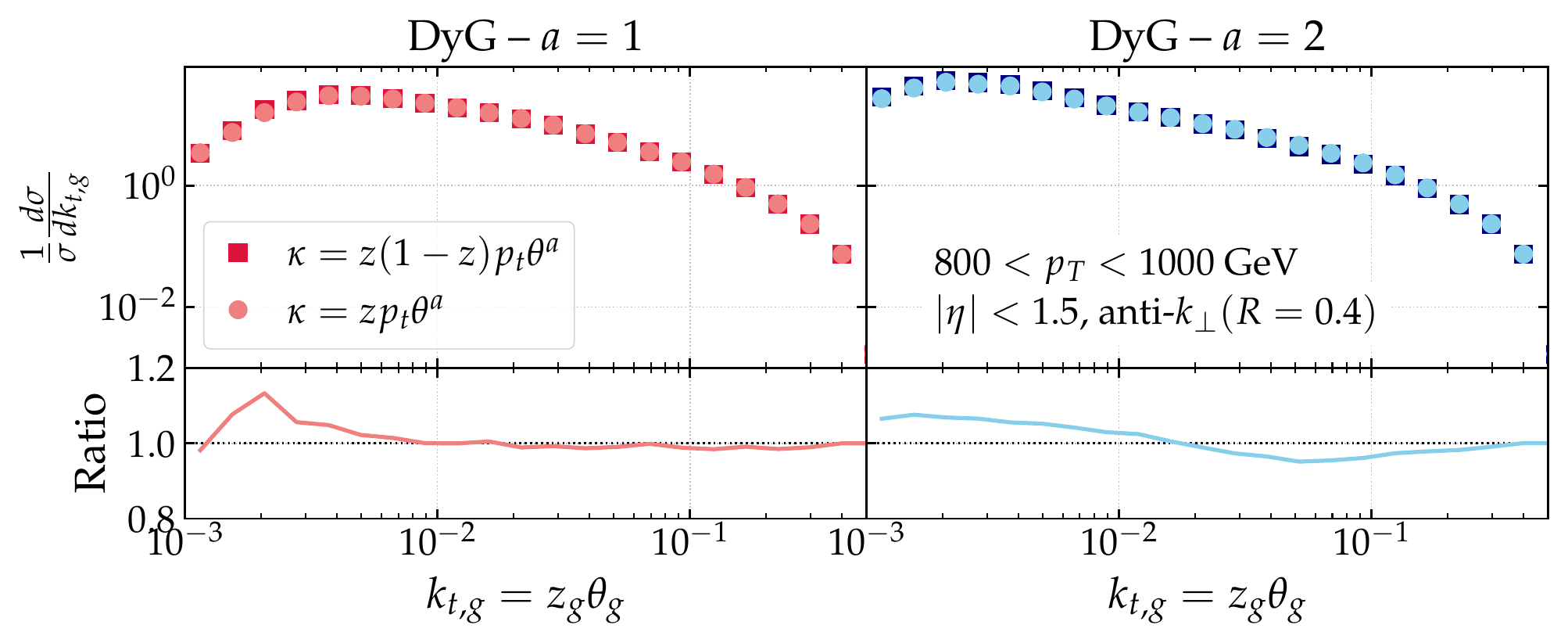}
    \includegraphics[width=\textwidth]{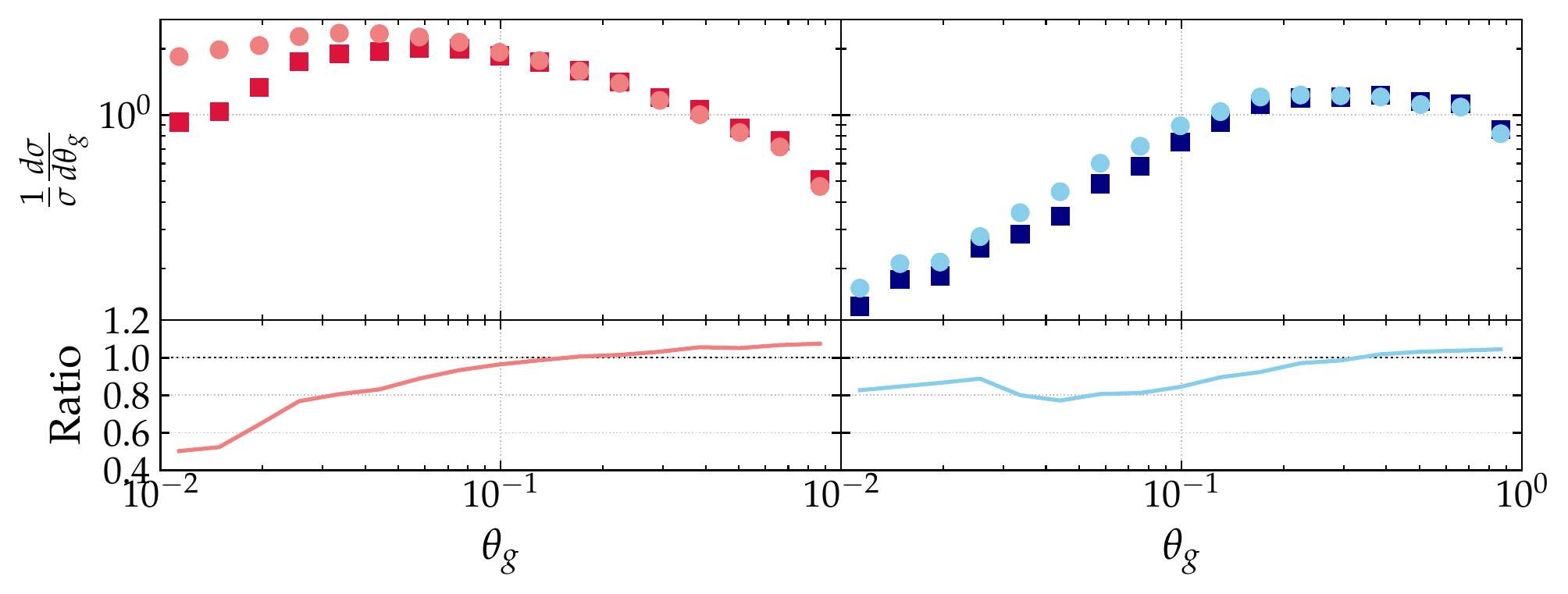}
  \includegraphics[width=\textwidth]{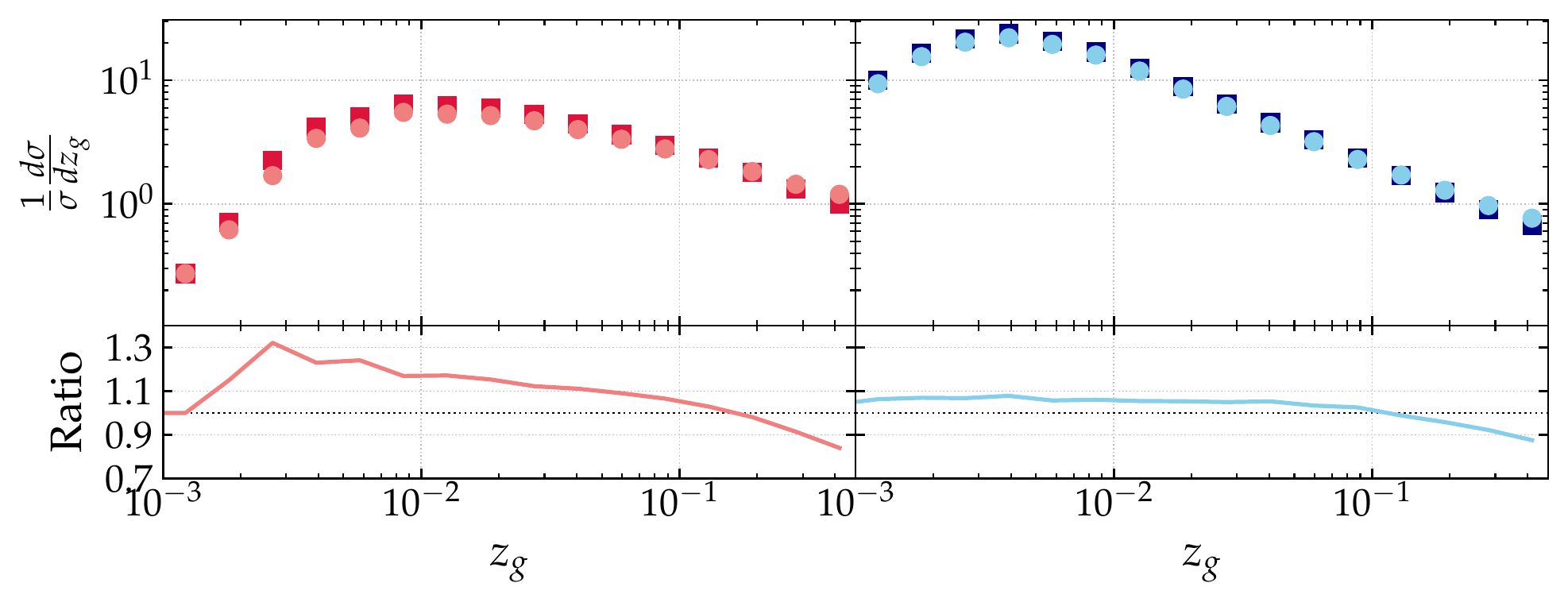}
  \caption{Impact of including or not the recoil factor in the hardness variable definition as a function of $a$ for dijet events at parton level in Pythia with $\sqrt s\!=\!13$~TeV for $k_{t,g}$ (top), $\theta_g$ (center) and $z_g$ (bottom).}
  \label{fig:pythia-recoil}
\end{figure}

In this appendix, we evaluate the impact of not neglecting the $1\!-\!z$ factor on the definition of $\kappa$ in Eq.~\eqref{eq:hardness}. From an analytic point of view, this is a sub-leading, non logarithmic correction and thus not needed to reach N$^2$DL. For example, at the level of the Sudakov $\Delta(\kappa|a)$, the $1-z$ in the definition of $k_\perp=z(1-z)p_T\theta$ for the running coupling scale induces a N$^3$DL correction of the form:
\begin{align}
\delta\ln(\Delta(\kappa|a))&=\frac{4\alpha_s^2\beta_0C_i}{\pi}\int_0^1\frac{\dd z'}{z'}\,\int_{0}^1\frac{\dd \theta'}{\theta'}\log(1-z')\Theta(z'\theta'^a-\kappa)\\
&=\frac{4\alpha_s^2\beta_0C_i}{\pi a}\left(\frac{\pi^2}{6}\ln(\kappa)+\zeta(3)+\mathcal{O}(\kappa)\right)
\end{align}
where we have used $\alpha_s(k_\perp)\simeq\alpha_s(zp_T\theta)(1-2\alpha_s\beta_0\ln(1-z))$ at our order of interest.
In the same way, one can determine the magnitude of the leading correction induced by $1-z$ in the definition of $\kappa = z(1-z)(\Delta R/R)^a$ by using the double logarithmic formula~\eqref{eq:sudakov_dla} for the Sudakov, with the veto constraint including the $1-z$ factor:
\begin{align}
\ln(\Delta(\kappa|a))&=-\frac{2\alpha_sC_i}{\pi}\int_0^{1}\frac{\dd z'}{z'}\,\int_0^{1}\frac{\dd \theta'}{\theta'}\Theta(z'(1-z')\theta'^a-\kappa)\\
 &=-\frac{\alpha_sC_i}{\pi a}\left(\ln^2(\kappa)-\frac{\pi^2}{3}+\mathcal{O}(\kappa)\right)
\end{align}
The correction to the double logarithmic result is therefore a sub-leading non logarithmic correction. 

That said, we would like to understand its impact on Monte-Carlo results in the experimental setups explored in this paper. The results are presented in Fig.~\ref{fig:pythia-recoil} while their implications are commented over the main text. 

%%%%%%%%%%%%%%%%%%%%%%%%%%%%%%%%%
\section{Impact of jet clustering algorithms}
\label{sec:appendix-c}
%%%%%%%%%%%%%%%%%%%%%%%%%%%%%%%%%
\begin{figure}
\centering
  \includegraphics[width=\textwidth]{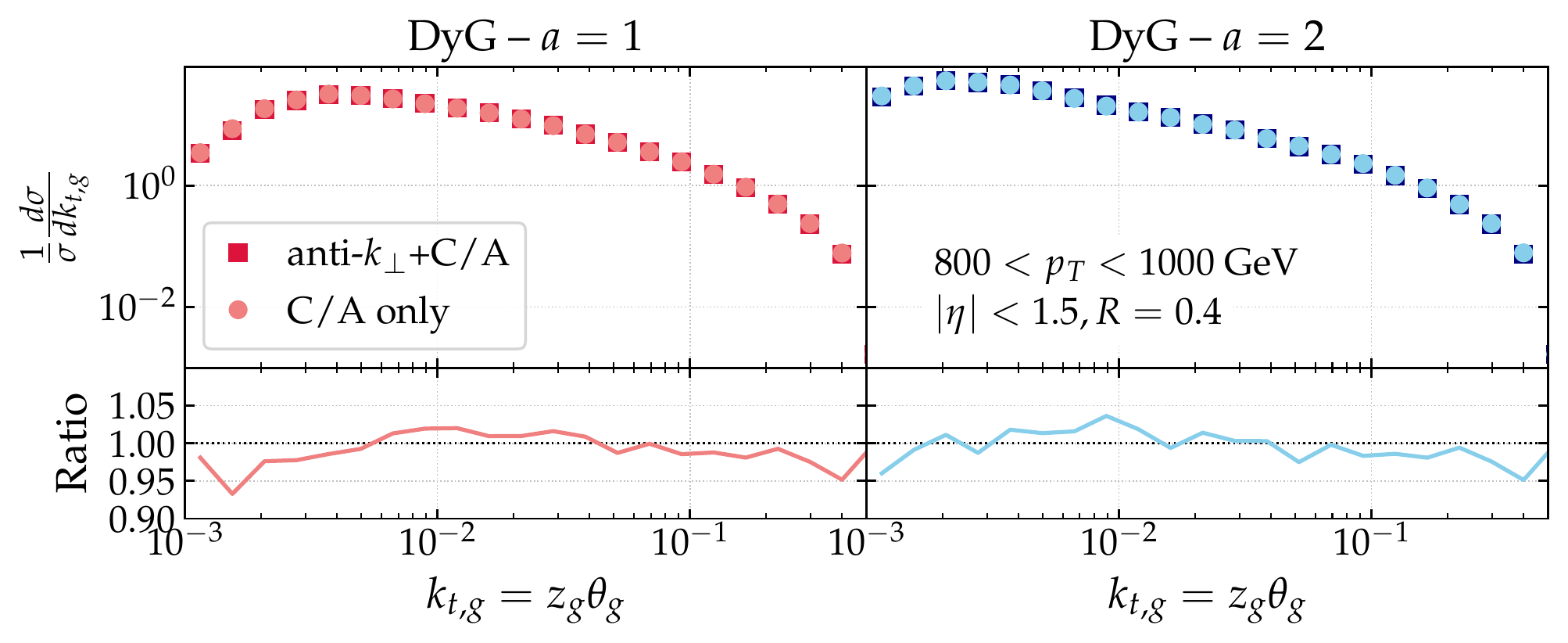}
   \includegraphics[width=\textwidth]{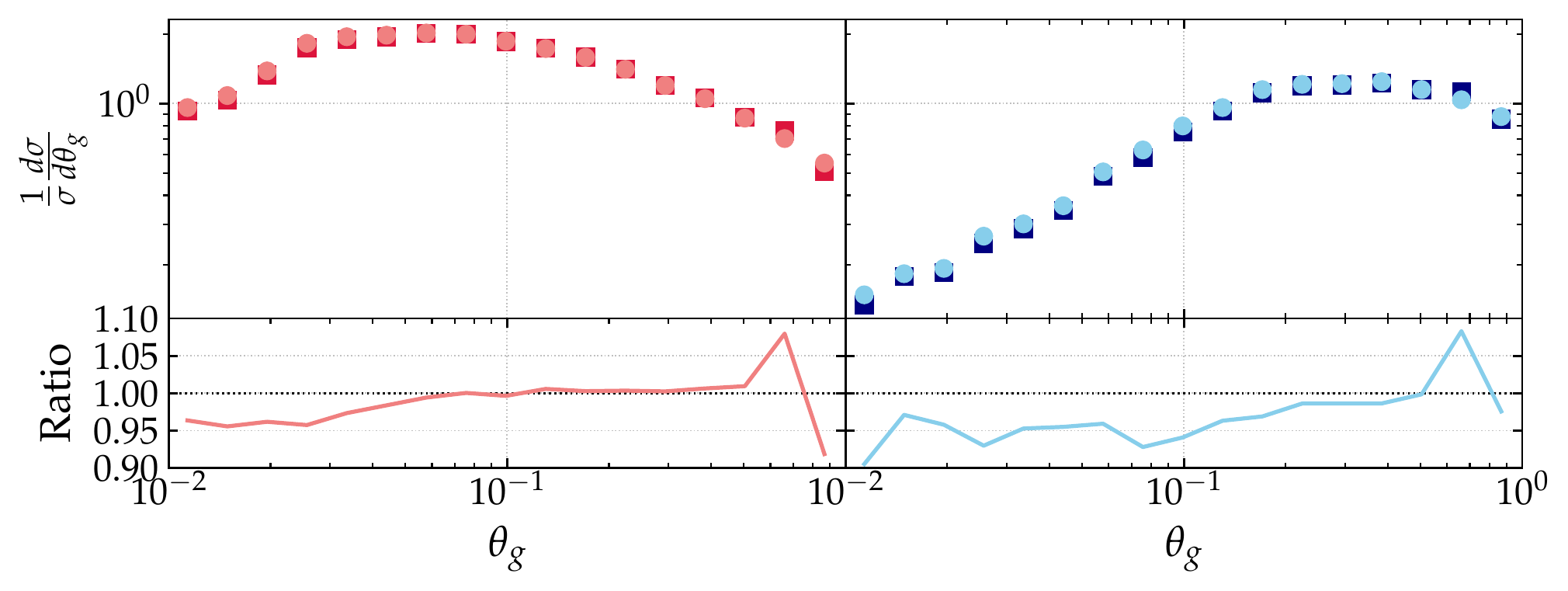}
     \includegraphics[width=\textwidth]{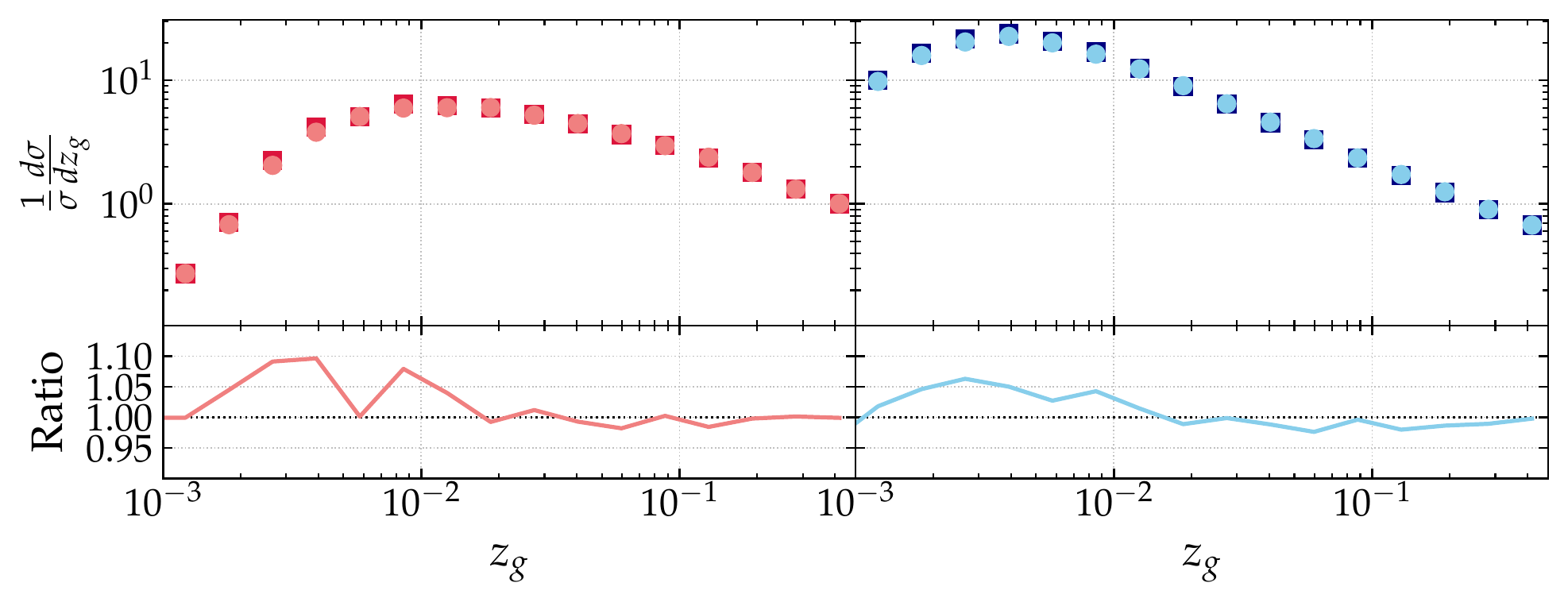}
  \caption{Impact of different clustering strategies in $k_{t,g}$ as a function of $a$ for dijet events at parton level in PYTHIA with $\sqrt s\!=\!13$~TeV for $k_{t,g}$ (top), $\theta_g$ (center) and $z_g$ (bottom).}
  \label{fig:pythia-ca}
\end{figure}

Throughout the main text we have determined dynamically groomed substructure observables for jets found with an initial anti-$k_t$ clustering and subsequently re-clustered with Cambridge/Aachen. This two-step process has advantages from an experimental point of view. However, from a theoretical perspective we have seen in Sec.~\ref{sec:boundary_logs} that this two-step process induces boundary logarithms in the calculation. In this Appendix, we would like to investigate at the Monte-Carlo level if the observables are modified when defining the jets only with C/A. This is shown in Fig.~\ref{fig:pythia-ca}. Interestingly, we observe how the bump at large angles whose origin we have discussed in the main text disappears when clustering with C/A. Nevertheless, the impact of these two jet clustering strategies is mild for all cases.
%%%%%%%%%%%%%%%%%%%%%%%%%%%%%%%%%
\section{Non-perturbative corrections with Pythia and Herwig}
\label{sec:appendix-d}
%%%%%%%%%%%%%%%%%%%%%%%%%%%%%%%%%
\begin{figure}
\centering
  \includegraphics[width=\textwidth]{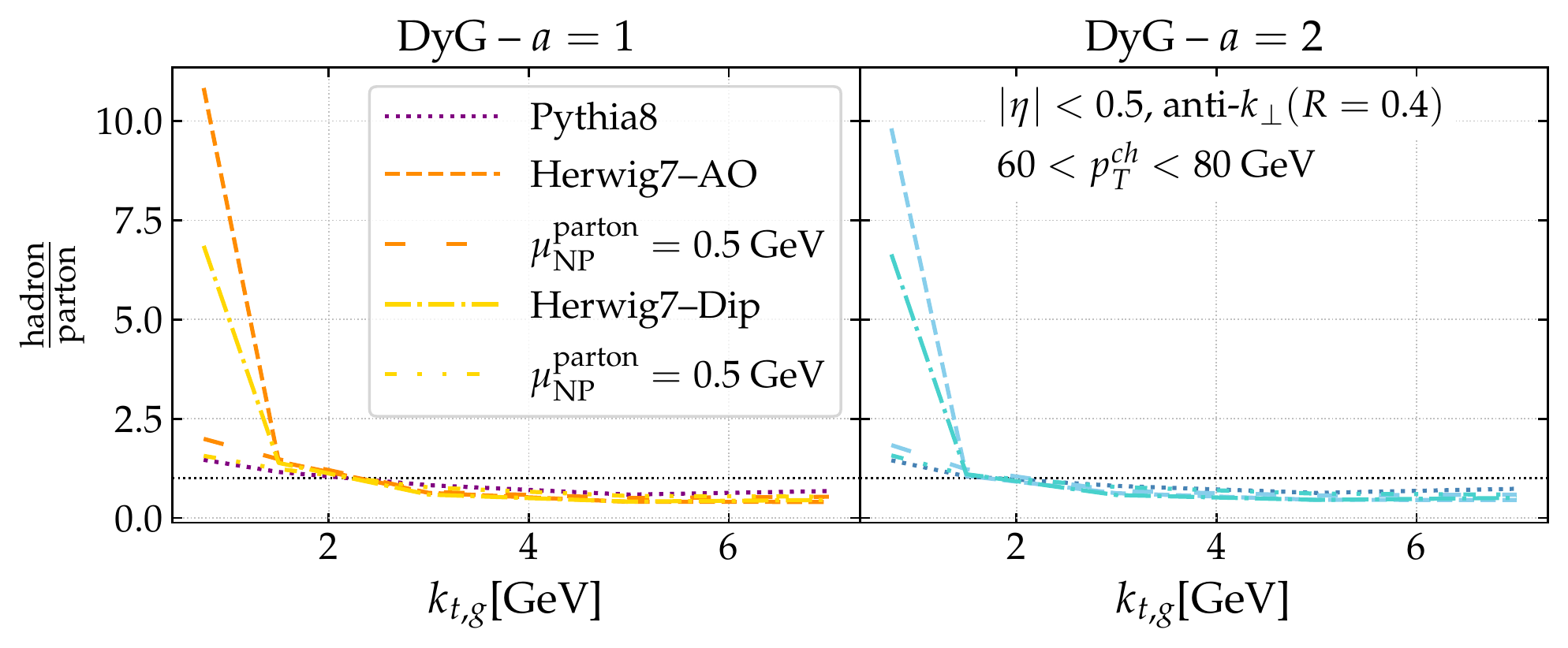}
   \includegraphics[width=\textwidth]{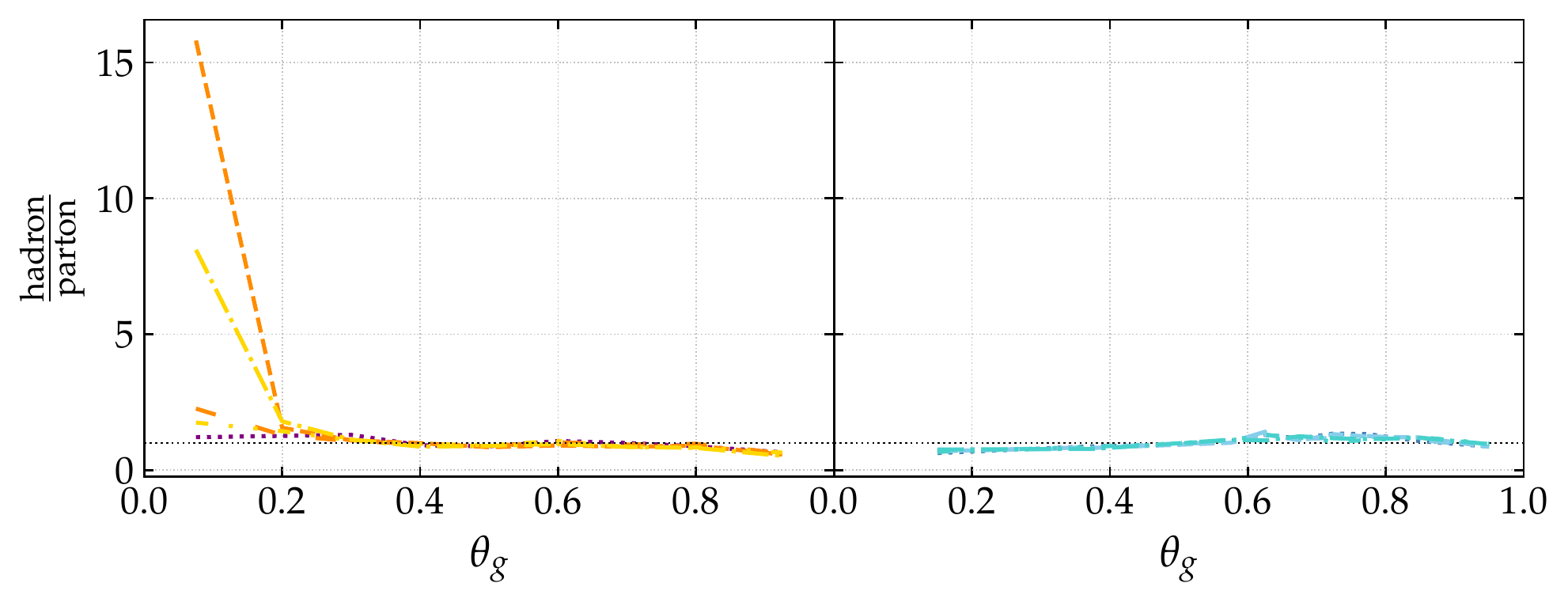}
   \includegraphics[width=\textwidth]{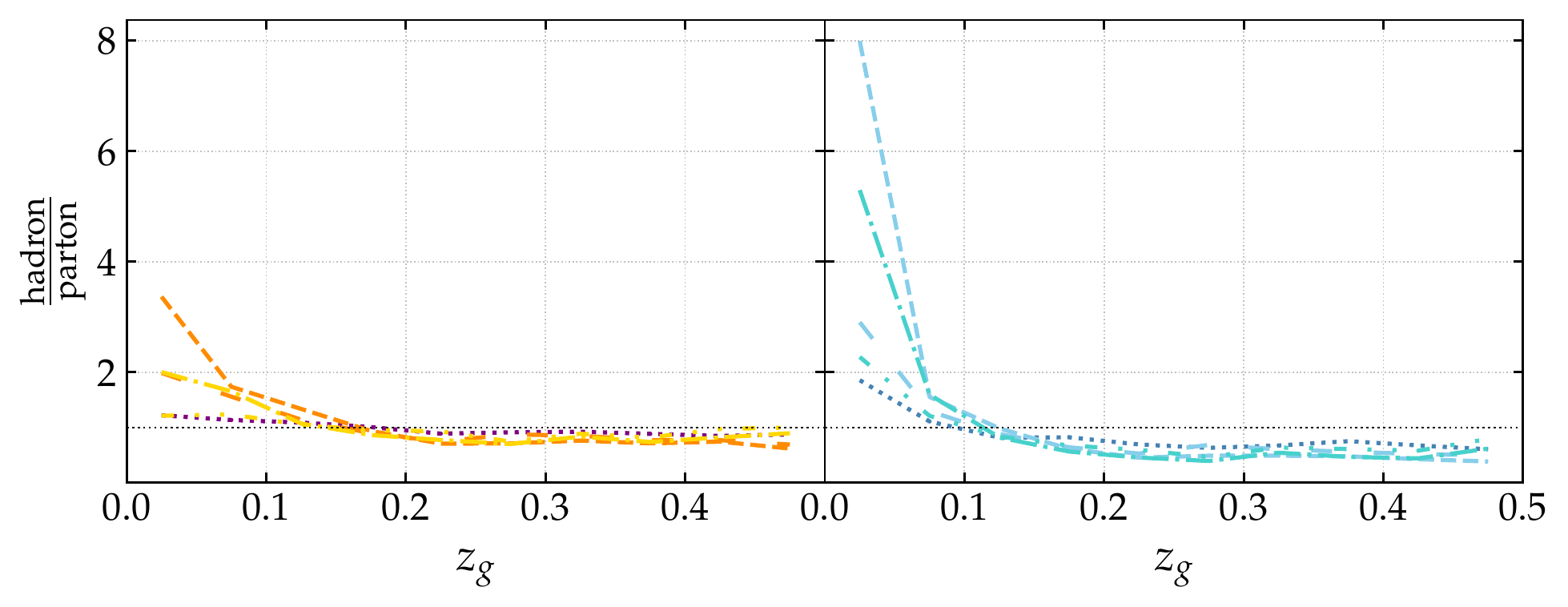}
  \caption{Ratio of hadron-to-parton level distributions for $k_{t,g}$ (top), $\theta_g$ (middle) and $z_g$ (bottom) with five different Monte-Carlo settings: Pythia8 (dotted, purple), Herwig7-AO with default parameters (orange, dashed) and with the shower cut-off set to $0.5$~GeV at parton level only (orange, loosely dashed), Herwig7-Dip with default parameters (gold, dotted dashed) and with the shower cut-off set to $0.5$~GeV at parton level only (gold, loosely dotted dashed). }
 \label{fig:np-factors}
\end{figure}

\begin{figure}
\centering
  \includegraphics[width=\textwidth]{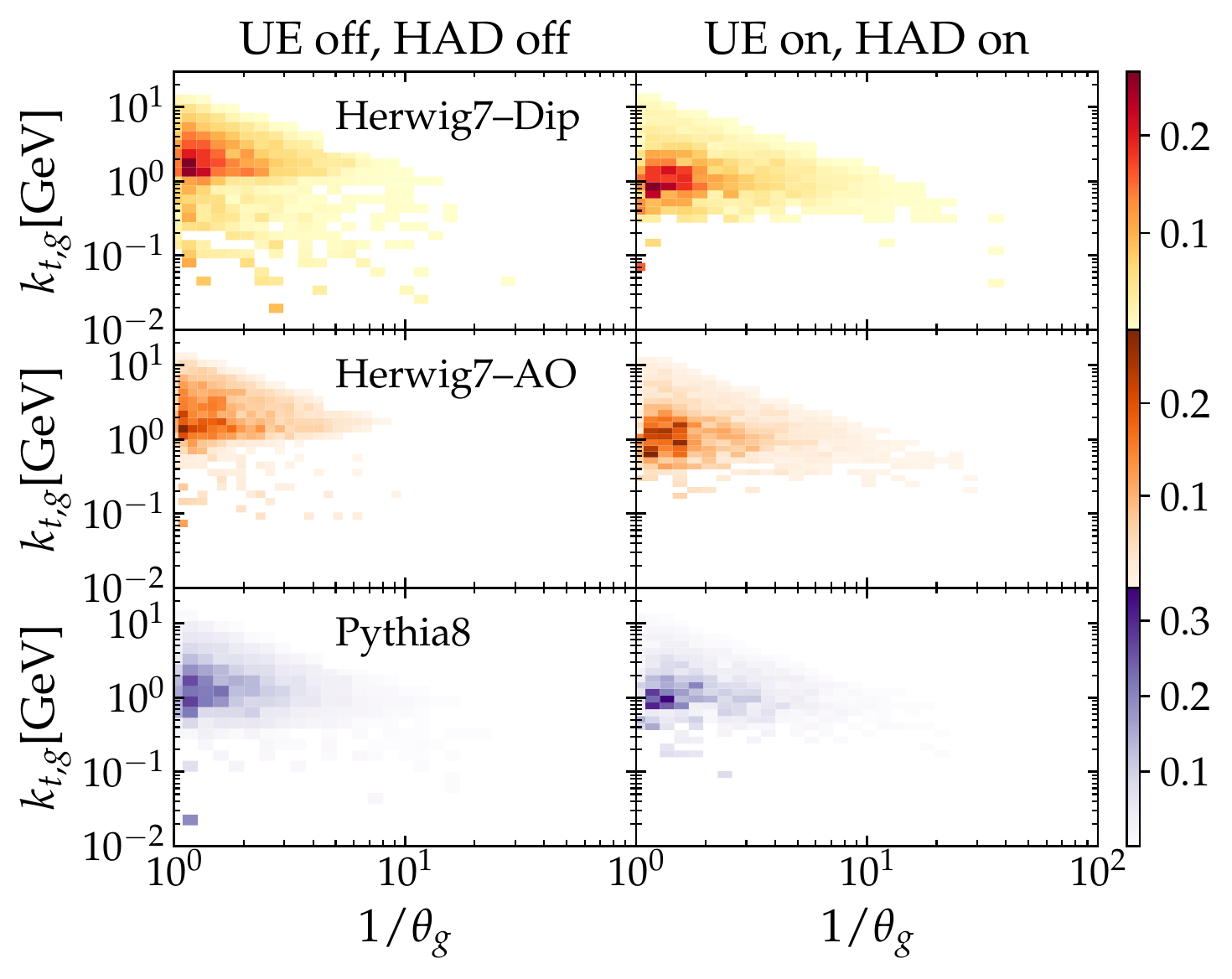}
  \caption{Lund planes generated by the three Monte-Carlo setups used in this work, Herwig7-Dip (top), Herwig-AO (center) and Pythia8 (bottom), with ALICE kinematics and $a\!=\!1$.}
  \label{fig:lp-mcs-alice}
\end{figure}

In order to compare our analytic predictions with ALICE's experimental data, we add non-perturbative effects through a single parameter extracted from Monte-Carlo simulations. This factor, thoroughly explained in Sec.~\ref{sec:data-to-theory}, is provided in Fig.~\ref{fig:np-factors}, where we took the ratio of MCs before and after hadronisation. Besides the default settings of Pythia and Herwig, we show two additional curves in which the parton shower cutoff, denoted as $\mu^{\rm parton}_{\rm NP}$, in Herwig is changed from its default value of $1$~GeV to the number used in Pythia where $\mu^{\rm parton}_{\rm NP}\!=\!0.5$~GeV. Note that changing this factor does not necessarily imply a one-to-one correspondence between the two event generators. The value of this factor, like any other hadronisation-related parameter, is tuned to data. Then, one cannot vary it when running the Monte-Carlo at hadron level because its predictive power would be negatively affected. Therefore, we only vary this factor for the parton level result, that is, for the denominator of our non-perturbative factor. 

The point of the variation of the parton shower stopping is to demonstrate the sensitivity of the dynamically groomed observables to that scale, and the limitations of this method for incorporating hadronisation corrections into analytic calculations. This is manifest in Fig.~\ref{fig:np-factors}, where the hadron-to-parton ratio varies from 0.5 to 2.5 for those settings that share the same value of $\mu^{\rm parton}_{\rm NP}$, while it explodes for the default Herwig-AO and Herwig-Dip in the limit of non-perturbative values of ($k_{t,g},z_g,\theta_g$). In the latter case, the reason for the rapid growth of the non-perturbative factor, e.g. in the low $k_{t}$ regime, is rooted in the fact that the parton-level shower does not generate splittings below $\mu^{\rm parton}_{\rm NP}$, while hadronization and underlying event populate this part of the phase-space. In terms of Lund planes, the area covered by the parton-level result and the hadron level one are clearly distinct in Herwig. This effect is less pronounced whenever $\mu^{\rm parton}_{\rm NP}$ is low, as in default Pythia. This is explicitly shown in Fig.~\ref{fig:lp-mcs-alice}.

As we have already mentioned, there is no preferred value of $\mu^{\rm parton}_{\rm NP}$ when running parton level simulations and the large variations encountered in the non-perturbative factor simply indicate that the parton-level results are out of their regime of applicability. Then, we decide to use the average of the Monte-Carlo generators with the same value of $\mu^{\rm parton}_{\rm NP}$ as the central value of the non-perturbative factor. The uncertainty band is obtained from the envelope of the five MC settings.

%%%%%%%%%%%%%%%%%%%%%%%%%%%%%%%%%
\section{Monte-Carlo description of ($z_g,\theta_g, k_{t,g}$) data}
\label{sec:appendix-e}
%%%%%%%%%%%%%%%%%%%%%%%%%%%%%%%%%
\begin{figure}
\centering
  \includegraphics[width=\textwidth]{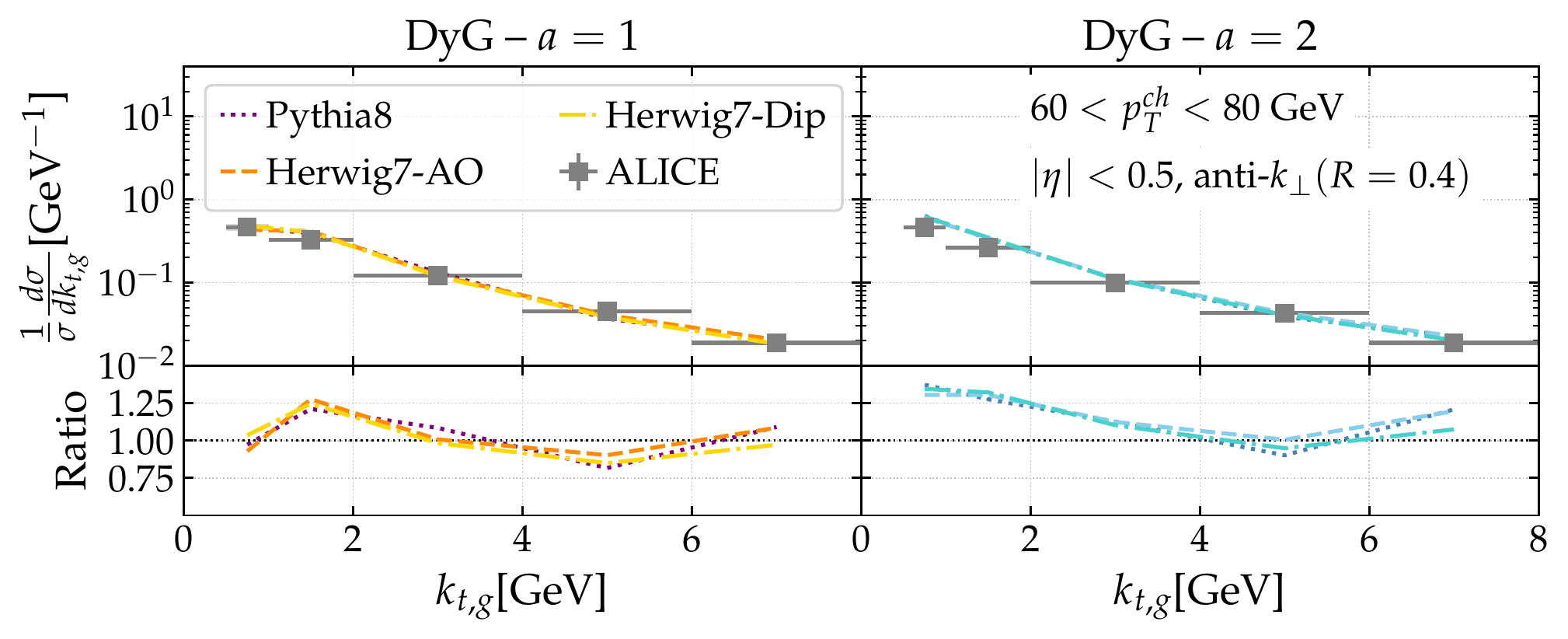}
   \includegraphics[width=\textwidth]{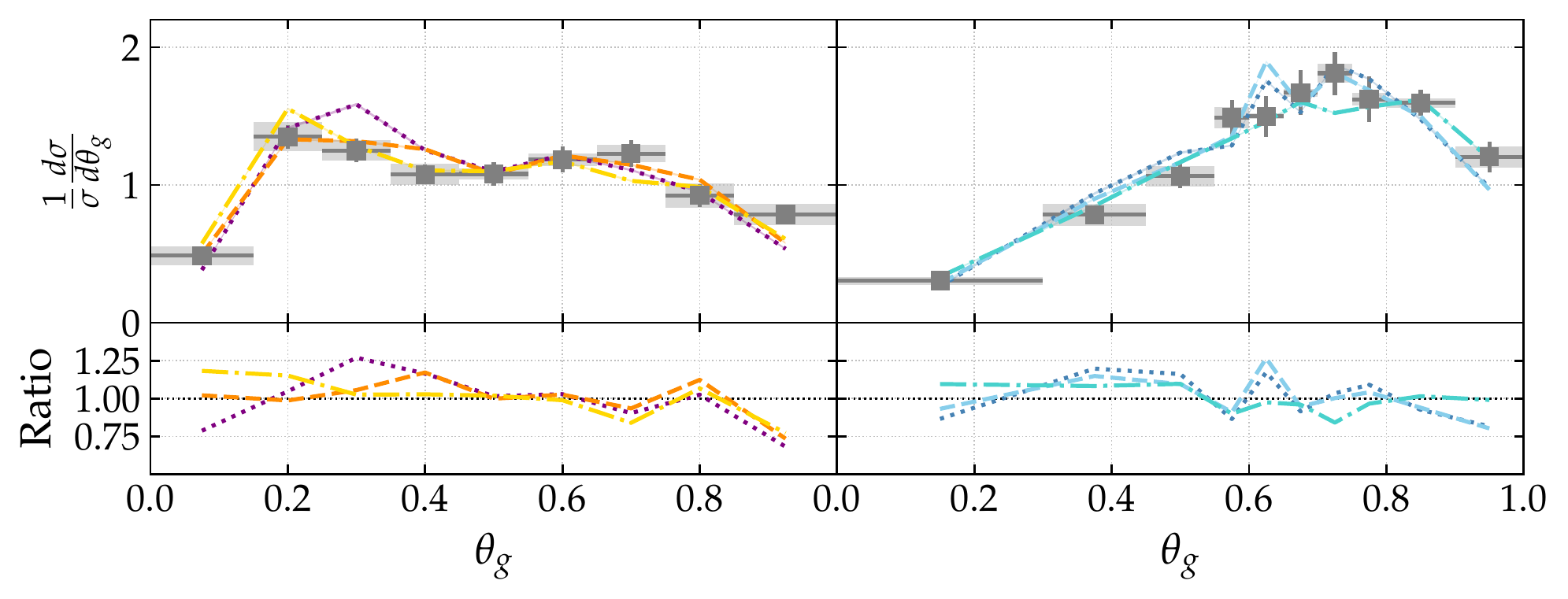}
    \includegraphics[width=\textwidth]{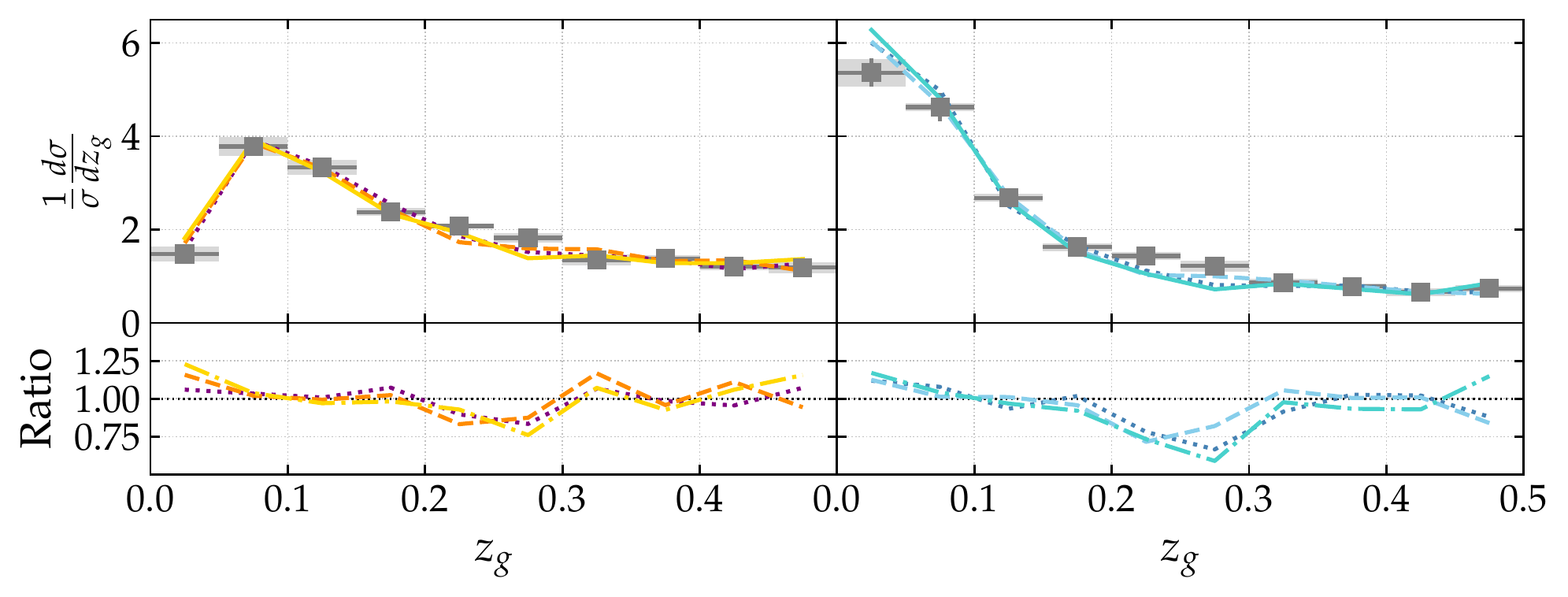} 
  \caption{Monte-Carlo to data comparison of $k_{t,g}$ (top), $\theta_g$ (middle) and $z_g$ (bottom) for $a\!=\!1$ (left) and $a\!=\!2$ (right) in the dynamical grooming condition, see Eq.\eqref{eq:hardness}. In the bottom panels, the theory-to-data ratios, computed using the same binning as the data, are presented.
  }
 \label{fig:mc-data}
\end{figure}

In this appendix we compare the three Monte-Carlo settings that we explore through this paper, i.e.\ Pythia8, Herwig7-AO and Herwig7-Dip, to the preliminary ALICE data. The results are shown in Fig.~\ref{fig:mc-data}. Notice that through these comparisons we are testing simultaneously the parton shower, i.e.\ dipole-style or angular-ordered, and the hadronization mechanism, i.e.\ Lund string or cluster models. In the case of $k_{t,g}$, no significant differences are observed among all Monte-Carlos. For $\theta_g$, Herwig7-Dip provides the best description of the data from small to large angles. All three Monte-Carlo settings are able to capture the data in the intermediate range of this measurement $0.4\!<\!\theta_g\!<\!0.7$  and differences only appear in the tails of Fig.~\ref{fig:mc-data}, where the hadronization mechanism seems to dominate for $\theta_g\!<\!0.4$. Finally, all Monte-Carlos show a significant depletion at $0.2\!<\!z_g\!<\!0.3$ that is ameliorated for $a\!=\!1$. Pythia achieves the best theory-to-data ratio, but its not obvious for this observable to disentangle between parton-shower dominated differences and hadronization mechanisms. 

%%%%%%%%%%%%%%%%%%%%%%%%%%%%%%%%%
\bibliographystyle{utcaps}
\bibliography{DyG-pp}

\end{document}